\documentclass[11pt,a4paper]{article}
\pdfoutput=1
\usepackage{setspace}
\usepackage{jheppub}
\usepackage{rotating}
\usepackage{array}
\usepackage{amsmath}
\usepackage[normalem]{ulem}
\usepackage{slashed}
\usepackage{booktabs}
\usepackage[pdftex,table,dvipsnames]{xcolor}
\usepackage{units}
\usepackage{multirow}
\usepackage{comment}
\usepackage{url}
\usepackage{xspace}
\usepackage{color}
\usepackage{float}
\usepackage{lipsum}
\usepackage[utf8]{inputenc}
\DeclareUnicodeCharacter{2032}{\ensuremath{^{\prime}}}
\usepackage{braket}
\usepackage{dsfont}

\newcommand{\me}{\mathrm{e}}

\newcommand{\GeV}{\,\mathrm{GeV}}
\newcommand{\eV}{\,\mathrm{eV}}
\newcommand{\alphaEM}{\alpha_{\scriptscriptstyle\text{EM}}}

\definecolor{Green}{rgb}{0.0, 0.5, 0.0}

\newcommand{\exclude}[1]{}
\def\beq{\begin{equation}}
\def\eeq{\end{equation}}
\newcommand{\C}[1]{\mathcal{#1}}

\preprint{TTK-21-02}

\title{Leading Logs in QCD Axion Effective Field Theory}

\author[1,2]{Gonzalo Alonso-{\'A}lvarez,}
\author[3]{Fatih Ertas,}
\author[1]{Joerg Jaeckel,}
\author[3]{Felix Kahlhoefer}
\author[1]{and Lennert J. Thormaehlen}

\affiliation[1]{Institut f\"ur theoretische Physik, Universit\"at Heidelberg, Germany}
\affiliation[2]{McGill University Department of Physics \& McGill Space Institute, \\
3600 Rue University, Montr\'eal, QC, H3A 2T8, Canada}
\affiliation[3]{Institute for Theoretical Particle Physics and Cosmology (TTK), RWTH Aachen University, \\ D-52056 Aachen, Germany}

\emailAdd{galonso@physics.mcgill.ca}
\emailAdd{ertas@physik.rwth-aachen.de}
\emailAdd{jjaeckel@thphys.uni-heidelberg.de}
\emailAdd{kahlhoefer@physik.rwth-aachen.de}
\emailAdd{l.thormaehlen@thphys.uni-heidelberg.de}

\abstract{
The axion is much lighter than all other degrees of freedom introduced by the Peccei-Quinn mechanism to solve the strong CP problem. It is therefore natural to use an effective field theory (EFT) to describe its interactions. Loop processes calculated in the EFT may however explicitly depend on the ultraviolet cutoff. In general, the UV cutoff is not uniquely defined, but the dimensionful couplings suggest to identify it with the Peccei-Quinn symmetry-breaking scale. 
An example are $K^+ \rightarrow \pi^+ + a$ decays that will soon be tested to improved precision in NA62 and KOTO and whose amplitude is dominated by the term logarithmically dependent on the cutoff.
In this paper, we critically examine the adequacy of using such a naive EFT approach to study loop processes by comparing EFT calculations with ones performed in complete QCD axion models.
In DFSZ models, for example, the cutoff is found to be set by additional Higgs degrees of freedom and to therefore be much closer to the electroweak scale than to the Peccei-Quinn scale.
In fact, there are non-trivial requirements on axion models where the cutoff scale of loop processes is close to the Peccei-Quinn scale, such that the naive EFT result is reproduced.
This suggests that the existence of a suitable UV embedding may impose restrictions on axion EFTs. We provide an explicit construction of a model with suitable fermion couplings and find promising prospects for NA62 and IAXO.
}


\begin{document}

\maketitle

\flushbottom

\newpage

\section{Introduction}

Effective Field Theories (EFTs) are extremely useful tools to test new physics scenarios in a model-independent way.
Their main ingredients are an often unknown cutoff energy scale and a set of low-energy fields and symmetries from which the operator expansion is constructed.
Some examples of EFTs include the Weak Effective Theory~\cite{Buchalla:1995vs,Aebischer:2017gaw,Jenkins:2017jig}, the Standard Model Effective Field Theory~\cite{Buchmuller:1985jz,Grzadkowski:2010es}, and the Higgs Effective Field Theory~\cite{Feruglio:1992wf,Alonso:2012px,Buchalla:2013rka}.

Nearly all phenomenological studies of axions and axion-like particles\footnote{Henceforth, we focus on the concrete case of (QCD) axions, though more general axion-like particles are usually motivated by similar symmetry considerations. Both particles are described by a generalized EFT that is based on the same construction principles, but for axion-like particles the requirement of addressing the strong CP problem of QCD is not enforced.} are performed within an EFT where one considers the Standard Model (SM) extended by a pseudoscalar particle -- the axion -- that interacts with the SM via dimension-5 operators involving either two gauge bosons or two SM fermions.

The existence of a light pseudoscalar boson with the aforementioned properties can have potentially observable consequences in a wide variety of setups.
Light axions may be copiously produced in astrophysical environments, allowing to search for them using helioscopes~\cite{Sikivie:1983ip,Anastassopoulos:2017ftl,Armengaud:2019uso}, cooling arguments~\cite{Raffelt:1996wa,Raffelt:2006cw}, or superradiance~\cite{Arvanitaki:2010sy,Arvanitaki:2014wva}.
Importantly, these light axions may constitute a fraction or the totality of the dark matter of the Universe~\cite{Preskill:1982cy,Abbott:1982af,Dine:1982ah}, in which case a plethora of direct~\cite{Sikivie:1983ip,Hagmann:1998cb,Horns:2012jf,Budker:2013hfa,Jaeckel:2013sqa,Chung:2016ysi,Kahn:2016aff,TheMADMAXWorkingGroup:2016hpc,Alesini:2017ifp,Melcon:2018dba,Du:2018uak,Braine:2019fqb} (see~\cite{Irastorza:2018dyq} for a recent overview) and indirect~\cite{Ressell:1991zv,Bershady:1990sw,Overduin:2004sz,Grin:2006aw,Boyarsky:2006fg,Boyarsky:2009ix,Vertongen:2011mu,Cadamuro:2011fd,Arias:2012az,Jaeckel:2014qea,Hook:2018iia,Caputo:2018ljp,Caputo:2018vmy,Foster:2020pgt,Wang:2021wae} detection experiments could be able to detect them.
Axions at the heavier end\footnote{See~\cite{Agrawal:2017ksf,Alves:2017avw,Gaillard:2018xgk,Gherghetta:2020keg} for some recent work motivating the possibility that QCD axions could be heavier than expected.} of the currently accessible mass range are best studied at the LHC~\cite{Jaeckel:2012yz,Mimasu:2014nea,Jaeckel:2015jla,Brivio:2017ije,Bauer:2017ris,Knapen:2016moh,Sirunyan:2018fhl,Aad:2020cje,Mariotti:2017vtv}
and $B$-factories~\cite{Freytsis:2009ct,Izaguirre:2016dfi,Dolan:2017osp,CidVidal:2018blh,Gavela:2019wzg,Merlo:2019anv,BelleII:2020fag}.
For the intermediate regime between the MeV and the GeV scales, reactors~\cite{AristizabalSierra:2020rom}, rare decay~\cite{Freytsis:2009ct,Dolan:2014ska,Izaguirre:2016dfi,Dobrich:2018jyi,Gavela:2019wzg,Beacham:2019nyx,Merlo:2019anv,Gori:2020xvq,MartinCamalich:2020dfe}, beam dump/fixed target~\cite{Bergsma:1985qz,Riordan:1987aw,Bjorken:1988as,Dobrich:2015jyk,Alekhin:2015byh,Dobrich:2019dxc,Darme:2020sjf,Kelly:2020dda,Brdar:2020dpr} experiments and long-lived particle detectors at the LHC~\cite{Feng:2018pew,Beacham:2019nyx,Aielli:2019ivi} compete with supernovae~\cite{Chang:2018rso,Carenza:2019pxu,Ertas:2020xcc} as the most favourable environments for probing axions (see~\cite{Beacham:2019nyx} for a recent review).
Of particular interest is the interplay between the various axionic couplings in different experiments and observations~\cite{Alonso-Alvarez:2018irt,Ertas:2020xcc} in these mass regimes -- see~\cite{Choi:2017gpf,Chala:2020wvs,Bauer:2020jbp} for the renormalization group (RG) equations that allow for a consistent comparison of constraints at different energy scales.

The rationale behind the employed EFTs is that the (QCD) axion is a pseudo-Goldstone boson arising from the Peccei-Quinn (PQ) solution to the Strong CP Problem~\cite{Peccei:1977hh,Peccei:1977ur,Weinberg:1977ma,Wilczek:1977pj}.  
It is expected that the underlying $U(1)_{\rm PQ}$ symmetry is broken spontaneously at a very high energy scale while the axion remains very light.
The pseudo-Goldstone boson nature of the axion then sets the guiding principles for the construction of its EFT.
Firstly, the shift symmetry $a\rightarrow a + \mathrm{const}$, resulting from the underlying $U(1)_{\rm PQ}$, restricts the interactions of the axion to be of purely derivative form.
Secondly, the $U(1)_{\rm PQ}$ symmetry must be anomalous under QCD in order for the strong CP problem to be solved. Non-perturbative effects then generate an axion potential and in particular a non-vanishing mass. Moreover, the QCD anomaly and similar anomalies under the other gauge groups result in characteristic couplings of the axion to two gauge bosons.
Finally, it seems natural to identify the cutoff scale of the EFT with the PQ symmetry-breaking scale, which is closely related to the axion decay constant, commonly denoted by $f_a$. This choice is suggested also by this scale appearing as the mass scale in the couplings of the dimension-5 derivative and two gauge boson interactions.

The adequacy of the EFT language to study axion phenomenology is based on the assumption that the presence of the axion is the only manifestation of the PQ construction at energies below the scale $f_{a}$.
At tree level, it is usually sufficient that any other particles 
are sufficiently heavy so that they cannot be produced at the energy scale relevant to the process under consideration. However, higher energy scales can become relevant at loop level where the contribution from off-shell particles in the loop needs to be included. In particular, when the EFT calculation yields divergent contributions, it may be more appropriate to identify the EFT cutoff scale with the mass of the lightest particle heavier than the axion instead of with $f_a$. 

In order to construct a suitable axion EFT it is therefore essential to check whether the underlying model features additional relevant degrees of freedom below $f_{a}$.
The goal of this paper is to do this in a number of full-fledged QCD axion models, thereby assessing the validity of studies performed using an EFT framework.
As a particularly pertinent example, we focus on loop-induced flavour-violating decays involving the axion.
Indeed, some of the one-loop diagrams involved in these processes are logarithmically divergent and therefore sensitive to the physics around the cutoff of the EFT.
We show that the \emph{leading log} prescription, where one focuses on the logarithmically divergent term and identifies the cutoff scale with $f_{a}$,
commonly employed 
in EFT calculations, produces results which can differ qualitatively and quantitatively from the ones obtained using full models.
The reason for this discrepancy, as will become clear, is that most popular (DFSZ- and KSVZ-type) QCD axion models~\cite{Kim:1979if,Shifman:1979if,Zhitnitsky:1980tq,Dine:1981rt} do not satisfy the assumptions that implicitly enter the EFT-based loop calculation. 
Our results are in line with the earlier work in a similar direction presented in~\cite{Choi:2017gpf}.
There, a leading-order RG evolution was employed to show that simple electroweak-complete axion models (e.g. DSFZ-type models explicitly including the two Higgs doublets) feature an absence of a large logarithm in the flavour-changing effects.
Our loop-based calculations are in agreement with these results.

In this situation, the question arises whether there exists an explicit field theoretic\footnote{The authors of~\cite{Choi:2017gpf} considered string-motivated UV models for which flavour effects are significantly suppressed.} UV-complete QCD axion model that allows for large logarithms in flavour-changing observables. 
A positive answer amounts to finding ways to generate tree-level couplings between the axion and SM fermions without introducing new states below the scale of PQ symmetry breaking.
We address this in two steps. First, we consider an effective model valid up to a scale $\gtrsim f_{a}$, which coincides with a specific charge assignment of the Lagrangian considered in ref.~\cite{Choi:2017gpf}, where it was however not further explored. Our effective model also features some similarities with the flavoured axion models of~\cite{Calibbi:2016hwq,Ema:2016ops}. Crucially however, the charge assignments in our case are not flavour-dependent and thus tree-level flavour-violating interactions are not present in our setup.
In the next step and drawing inspiration from Froggatt-Nielsen models~\cite{Froggatt:1978nt}, we construct a UV completion where the logarithms are explicitly calculable.
The result is a QCD axion model that features flavour-conserving SM fermion couplings without the need for an extended Higgs sector, which therefore provides an affirmative answer to the question posed above.

The logarithmic enhancement of the  $K^+\rightarrow \pi^+ + a$ decay rate in this model makes it a particularly good candidate to be tested at experiments like NA62~\cite{NA62:2017rwk} and KOTO~\cite{Ahn:2018mvc} as well as future experiments such as KLEVER~\cite{Ambrosino:2019qvz}.\footnote{For our concrete numerical examples we focus on charged kaon decays as measured by NA62.} 
Furthermore, the potential existence of an axion-electron coupling increases the production of axions in the solar interior.
This enhancement results in a larger expected axion flux in helioscopes like IAXO~\cite{Armengaud:2019uso}, which are projected to have sensitivity to the model in the $\mathcal{O}(10\,\mathrm{meV})$ mass range.

In essence, the goal of the present study is to asses whether a given combination of axion EFT and cutoff is consistent with an embedding into a more fundamental theory.
In the context of quantum gravity and string theory, the terms \emph{landscape} and \emph{swampland} have become popular to respectively denote the set of low energy effective theories that can and cannot consistently be incorporated into a theory of gravity~\cite{Vafa:2005ui}.
Although our considerations are purely field-theoretic (see~\cite{Freivogel:2019mtr} for a swampland conjecture that is independent of the Planck scale), the landscape/swampland analogy serves as a motivation to raise the question of whether there are additional constraints on embeddable axion EFTs that have not yet been fully appreciated.
In the case at hand, we find a promising avenue to construct the desired UV completion. That said, our model hints at some potentially non-trivial requisites. For example, achieving the absence of tree-level flavour-changing interactions in the EFT requires a non-trivial choice of the model parameters and potential tuning. Furthermore, realizing a sufficiently large top Yukawa coupling may
constrain the separation of scales between the UV cutoff and $f_a$, as was noted in~\cite{Choi:2017gpf}. This suggests that mild additional assumptions (e.g. the absence of fine-tuning) may give strong constraints on the possible embeddings of axion EFTs into full UV models.

This paper is structured as follows.
In section~\ref{sec:EFT}, we set up the QCD axion EFT under examination and calculate the expected rate for the $K^+\rightarrow\pi^+ + a $ decay within this framework.
This result is compared with detailed calculations in DFSZ- and KSVZ-type QCD axion models in section~\ref{sec:models}.
After establishing the discrepancy between the EFT prescription and existing models, in section~\ref{sec:new_model} we present a new QCD axion model that satisfies the assumptions under which the EFT is constructed.
The phenomenology of this model is explored in sections~\ref{sec:phenomenology} and~\ref{sec:constraintsandopportunities}, after which we conclude in section~\ref{sec:conclusions}. 
A discussion of CP violation in axion couplings and most of the technical details can be found in the appendices~\ref{app:CPVInt}--\ref{app:EDM}.

\newpage

\section{Loop-induced rare decays in QCD axion EFT}\label{sec:EFT}
The most general effective Lagrangian describing all possible interactions of an axion (or axion-like particle) with the Standard Model fields before electroweak (EW) symmetry breaking and involving dimension-$5$ operators can be written in a compact form as~\cite{Georgi:1986df,Bauer:2017ris} 
\begin{align}\label{eq:EFT_Lagrangian}
\mathcal{L} = -\frac{a}{f_a} \sum_F c_{FF} \frac{\alpha_F}{8\pi} F^i_{\mu\nu} \tilde{F}^{\mu\nu,i} +  \frac{\partial_\mu a}{f_a}  \sum_{\chi} \bar{\chi} C_{\chi} \gamma^\mu \chi\,,
\end{align}
where the first sum runs over SM gauge bosons with couplings $c_{FF}$.
In the second sum, the SM chiral fermion multiplets are summarized in $\chi=(Q_L,\,L_L,\,u_R,\,d_R,\,e_R)$ and each of the $C_{\chi}$ is a matrix in generation space that allows for flavour changing effects 
(see appendix~\ref{app:CPVInt} for a discussion of the CP properties of $C_{\chi}$).
It is also customary to define the gauge boson couplings $g_{aFF} = c_{FF} \alpha_F/(2\pi f_a)$, but for our purposes the dimensionless couplings $c_{FF}$ are more convenient.
The dimensionful quantity $f_a$, usually known as the axion decay constant, serves as the large scale for the expansion of the operators in the EFT.
In Eq.~\eqref{eq:EFT_Lagrangian}, there is an ambiguity in the definition of $f_a$ and the EFT coefficients.
As long as the coupling to gluons $c_{gg}$ is nonzero, as is necessarily the case for any QCD axion, we can solve this ambiguity by normalizing $f_a$ in such a way that $c_{gg}\equiv 1$, and we do so in the rest of this work.

The other relevant feature of the Lagrangian is that a potential for the axion is generated. For a nonzero coupling to gluons, non-perturbative QCD dynamics provide for a mass  term~\cite{Weinberg:1977ma,Shifman:1979if,diCortona:2015ldu,DiVecchia:1980yfw}
\begin{align}
m_a \simeq \sqrt{\frac{m_\pi^2 f_\pi^2}{f_a^2} \frac{m_u m_d}{(m_u + m_d)^2}}\simeq 5.7\,\mu\mathrm{eV} \left( \frac{10^{12}\,\mathrm{GeV}}{f_a} \right).
\end{align}

At energies below the electroweak scale, it is convenient to rewrite the axion-fermion couplings in Eq.~\eqref{eq:EFT_Lagrangian} using the mass eigenbasis for the SM fermions,
\begin{align}
\label{eq:EFT_Lagr_ferm}
    \mathcal{L} \supset \frac{\partial_\mu a}{f_a} \left( \sum_{f = u,d,\ell,\nu} \bar{f}_L\, c_{f,L}\, \gamma^\mu\, f_L + \sum_{f = u,d,\ell} \bar{f}_R\, c_{f,R}\, \gamma^\mu\, f_R\right),
\end{align} 
where the sum runs over the up, down, charged lepton, and neutrino flavour-triplets $u,\,d,\,\ell,$ and $\nu$. Each $f$ thus summarizes the three generations. The coupling matrices $c_{f,L}$ and $c_{f,R}$ for left- and right-handed fields are related to the initial $C_{\chi}$ matrices in Eq.~\eqref{eq:EFT_Lagrangian} by unitary matrices~\cite{Bauer:2017ris}.
In the left-handed quark sector, we have the additional relation \mbox{$c_{d,L} = V^\dagger\,c_{u,L}\, V$} involving the CKM matrix $V$, as both up- and down-type quarks are part of the same left-handed doublet. 
As neutrinos are massless in the SM, we can choose the unitary transformation of left-handed neutrino fields to equal the one of their charged-lepton counterparts so that $c_{\ell,L} = c_{\nu,L}$. Note that we could have also written down the interactions in a vector/axial-vector basis with the relevant matrices reading $c_{f,V} = (c_{f,R} + c_{f,L})/2$ and $c_{f,A} =  (c_{f,R}-c_{f,L})/2$, respectively.

The Lagrangian as written down in Eq.~\eqref{eq:EFT_Lagr_ferm} in principle allows for arbitrary flavour-coupling structures.  While some axion models~\cite{Ema:2016ops,Calibbi:2016hwq} feature flavour-dependent and even flavour non-diagonal interactions (see~\cite{MartinCamalich:2020dfe} for a summary of the associated phenomenology), they are not the main focus of this work. That said, as such couplings may also arise in our explicit model construction, we discuss them in some detail later.

\begin{figure}[t]
    \begin{minipage}{0.3\textwidth}
	\centering
	\includegraphics[width=1.0\linewidth]{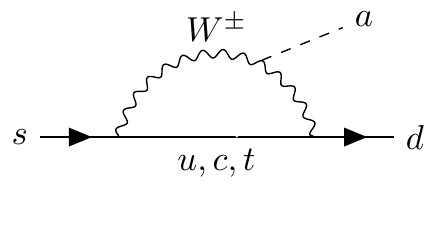}  \end{minipage}
	\hspace{0.2cm}
	\begin{minipage}{0.3\textwidth}
	\centering
	\vspace*{0.02cm}
	\includegraphics[width=1.0\linewidth]{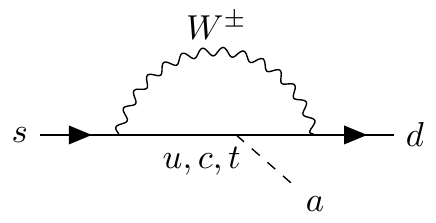}  \end{minipage}
	\hspace{0.2cm}
	\begin{minipage}{0.3\textwidth}
	\centering
	\vspace*{-0.32cm}
	\includegraphics[width=1.14\linewidth]{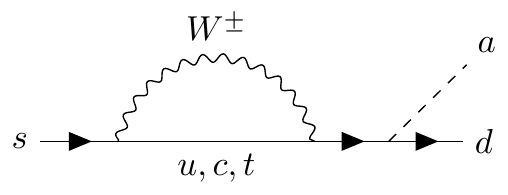}  
	\end{minipage}
	\vspace{-0.2cm}
	\caption{One-loop diagrams inducing the flavour-violating $s\rightarrow d + a$ transition. We include the self-energy contribution on the external strange-quark leg (analogously for the down-quark line) arising from the renormalization of quark fields, see appendix~\ref{app:Counterterms}. All Feynman diagrams throughout this work are drawn using the \texttt{TikZ-Feynman}~\cite{Ellis:2016jkw} package.}
	\label{fig:s2dALoop}
\end{figure}

For now, let us consider flavour-universal couplings in the quark and lepton sectors.
This is already sufficient to demonstrate our main point, namely that the leading logarithm in FCNCs obtained in the naive EFT does, in many cases, not agree with the result of more complete models.
In this case, we can remove the vectorial part of the quark and lepton couplings by performing universal vectorial phase rotations of the quark and lepton fields. These leave the mass terms and the charged current interactions with $W^\pm$ bosons invariant and therefore no scalar Yukawa couplings to $a$ appear.
The only possible effects are two-gauge-boson couplings arising for the chirally coupled electroweak fields, as noted in~\cite{MartinCamalich:2020dfe}, corresponding to the anomaly for the baryon and lepton number U(1) symmetries. The effect can therefore be absorbed in a redefinition of the corresponding two-gauge-boson coupling coefficient, most notably $c_{WW}$. As we discuss below, the contribution from $c_{WW}$ to the loop processes studied in this work is finite and therefore a shift in this coefficient does not play a role when evaluating the leading logarithm.
We thus only need to consider axial-vector couplings of the axion to quarks and leptons,\footnote{In DFSZ models, the couplings for up- and down-type quarks can actually differ even in absence of an explicit flavour structure due the the existence of an extended EW sector and mixing effects. In section~\ref{subsec:dfsz} we avoid this complication by adapting the calculation of~\cite{Freytsis:2009ct} which is done in the Yukawa basis.}
\begin{align}
\label{eq:AxQuarkFlavDiagUni}
     \mathcal{L} \supset \frac{\partial_\mu a}{2 f_a} \,c_q \sum_{f=u,d} \bar{f} \gamma^\mu \gamma_5 f  + \frac{\partial_\mu a}{2 f_a} \,c_l \sum_{f=\ell,\nu} \bar{f} \gamma^\mu \gamma_5 f  \,,
\end{align}
where the neutrino fields are understood to be purely left handed, $\nu=\nu_{L}$.
With these assumptions, no inter-generation fermion couplings and therefore no flavour-changing processes are present at tree level.
But even if these are absent, flavour-changing neutral currents (FCNCs) appear at one loop through the diagrams depicted in figure~\ref{fig:s2dALoop}.
Rare decays induced by such transitions have been exploited in the literature to test the axion paradigm~\cite{Freytsis:2009ct,Izaguirre:2016dfi,Gavela:2019wzg,Beacham:2019nyx,Merlo:2019anv,Gori:2020xvq,MartinCamalich:2020dfe}, and lepton-flavour violation has been studied in similar detail~\cite{Cornella:2019uxs,Escribano:2020wua}. 

In this work, we perform a detailed study of the FCNC transition in figure~\ref{fig:s2dALoop}, with a focus on the impact of the possible UV completions of the axion effective Lagrangian in Eq.~\eqref{eq:AxQuarkFlavDiagUni}.
For concreteness and also because it is of significant experimental relevance, we concentrate on the kaon decay $K^+\rightarrow \pi^+ + a$.
That said, our results are also applicable to other FCNC-induced processes.

The $s\rightarrow d + a$ transition can be described using the effective Hamiltonian
\begin{align}
\label{eq:EffLagrds}
\mathcal{H}_{s\rightarrow da} = \partial_\mu a\, \bar{d} \gamma^\mu (h^S_{ds} + h^P_{ds} \gamma_5) s + \text{h.c.}\,.
\end{align}
The hadronic matrix element for the $K^+\rightarrow \pi^+ + a$ decay\footnote{As already mentioned, we focus on the charged kaon decay as studied in NA62~\cite{NA62:2017rwk}, but a similar calculation can be done for neutral kaon decays (KOTO~\cite{Ahn:2018mvc} and KLEVER~\cite{Ambrosino:2019qvz}).} can then be parameterized as~\cite{Carrasco:2016kpy} 
\begin{equation}
\braket{\pi(p^\prime) | \bar{d} \gamma^\mu (h^S_{ds} + h^P_{ds} \gamma_5) s | K(p)} = h^S_{ds}\, P^\mu f_+(q^2) + h^S_{ds}\, q^\mu f_-(q^2)\,,
\end{equation}
where $P = p + p^\prime$ and $q = p - p^\prime$ so that the momentum transfer is $q^2 = m_a^2$.
For a sufficiently light axion, we only need the first form factor at $q^2\simeq 0$.
Recent Lattice QCD evaluations~\cite{Aoki:2019cca} give $f_+(0)=0.9706(27)$.
In this limit, the decay width can be expressed as
\begin{align}\label{eq:Kpia_width}
\Gamma (K^+ \rightarrow \pi^+ a) = \frac{|h^S_{ds}|^2}{16 \pi m_{K^+}^3} ( m_{K^+}^2 -  m_{\pi^+}^2 )^2 \lambda^{1/2}(m_{K^+}^2,\,m_{\pi^+}^2,\,m_{a}^2)\, f^2_+(m_a^2)\,,
\end{align}
where we have introduced the notation $\lambda(x,y,z) = x^2 + y^2 + z^2 - 2(xy +xz +yz)$.

Within the framework of Eq.~\eqref{eq:AxQuarkFlavDiagUni}, the coefficients $h^{S,P}_{ds}$ can be expressed in terms of the parameters of the axion EFT.
Given that, as we have seen, the QCD matrix element for $h^P_{ds}$ vanishes due to parity, we only need~\cite{Batell:2009jf,Izaguirre:2016dfi}
\begin{align}
\notag
h^S_{ds} = &-\frac{G_F}{16 \sqrt{2} \pi^2} \frac{1}{f_a} c_q\sum_{q=u,c,t}  V^*_{qd} V_{qs} m_q^2 \left( \log{\left(\frac{\Lambda^2}{m_q^2}\right)} - \frac{m_q^4 - 8 m_q^2 m_W^2 + 7 m_W^4 + 6 m_W^4 \log{\left(\frac{m_q^2}{m_W^2}\right)} }{2(m_q^2 - m_W^2)^2} \right)\\
\label{eq:effsdCoeff}
& -\frac{3 G_F^2 m_W^4}{\pi^2} \frac{c_{WW}}{32 \pi^2 f_a} \sum_{q=u,c,t} V^*_{qd} V_{qs} f(m_q^2/m_W^2)\,,
\end{align}
where we have substituted the divergence arising from the calculation of the diagrams in figure~\ref{fig:s2dALoop} by a leading logarithm depending on an a priori undetermined UV scale $\Lambda$. In an EFT sense, this identification can also be understood by considering a leading-order RG evolution between the high and the low scale~\cite{Choi:2017gpf,Bauer:2020jbp}.
We have also introduced the loop function~\cite{Izaguirre:2016dfi}
\begin{align}
 f(x) = \frac{x (1+x(\ln(x) -1))}{(1-x)^2}\,,
\end{align}
and expanded the amplitude at leading order in the external momenta (corresponding to dropping terms suppressed by $m_d/m_W$, $m_s/m_W$, and $m_a/m_W$) in order to determine the leading finite contribution. This expansion, however, has barely any effects compared to the ``leading log".
Finally, we have neglected the contribution from the gluon anomalous coupling, which can be computed based on mixing effects~\cite{Alves:2017avw,Bardeen:1986yb} but is subleading to the leading logarithmic term.

The result above has been computed in unitary gauge using the basis of Eq.~\eqref{eq:AxQuarkFlavDiagUni}, where the QCD axion couples only derivatively.
An axion-dependent rotation of the fermion fields allows to trade the derivative terms for pseudoscalar ones $\partial_\mu a \bar{f}\gamma^\mu\gamma_5 f \rightarrow 2m_f a \bar{f}i\gamma_5 f$, importantly along with anomaly terms.
One can check that the pseudoscalar interactions give the same leading log result as the axial-vector ones.
In Feynman gauge, a factor of $4$ difference between the axial-vector and pseudoscalar interaction has been reported (see footnote 3 in ref.~\cite{Dolan:2014ska}), but we have confirmed that the claimed discrepancy disappears when the EW Goldstone bosons are taken into account.

Combining the above equations, one can evaluate the decay rate as a function of the EFT parameters and, importantly, of the UV cutoff $\Lambda$.
This can in principle be used to place bounds on $f_a$ using the experimental constraint $\mathrm{Br}(K^+\rightarrow \pi^+ a) < 7.3\cdot 10^{-11}$ from the E787 and E949 experiments at BNL~\cite{Adler:2008zza}, which NA62 expects to improve by an order of magnitude by 2025~\cite{Fantechi:2014hqa,CortinaGil:2020fcx}. 
In practice, the reach of the constraint depends significantly on the choice of the value of $\Lambda$.
It is common to identify $\Lambda$ with the only other scale that is available in the EFT description, which is the decay constant $f_a$~\cite{Batell:2009jf,Gavela:2019wzg,Dolan:2014ska}. 
However, this is an \emph{ad hoc} choice which, as we will see, is not reproduced in the most popular UV completions of the axion EFT.
In order to clarify this issue, we now turn to study the $s\rightarrow d + a$ transition in the benchmark KSVZ and DFSZ QCD axion models.

\section{Loop-induced rare decays in QCD axion models}
\label{sec:models}

\subsection{DFSZ-type models}\label{subsec:dfsz}

Let us compute the flavour-violating decay of the kaon through the diagrams of figure~\ref{fig:s2dALoop} in DFSZ~\cite{Zhitnitsky:1980tq,Dine:1981rt} models.
Crucially, we use a complete DFSZ model explicitly including the two Higgs doublets in order to regularize the logarithmic divergence and identify the cutoff scale $\Lambda$.
The results are obtained from~\cite{Freytsis:2009ct}, where the $B\rightarrow K + a$ process was studied, by applying the obvious substitutions. Before going into the details, let us note that these calculations are performed in the basis where the axion interactions are of Yukawa and not of derivative type. As we briefly discuss below Eq.~\eqref{eq:KSVZChRot}, this avoids complications arising when separate chiral rotations of up- and down-type quarks are performed.
	
\begin{figure}[!t]
	\centering
	\includegraphics[width=0.49\textwidth]{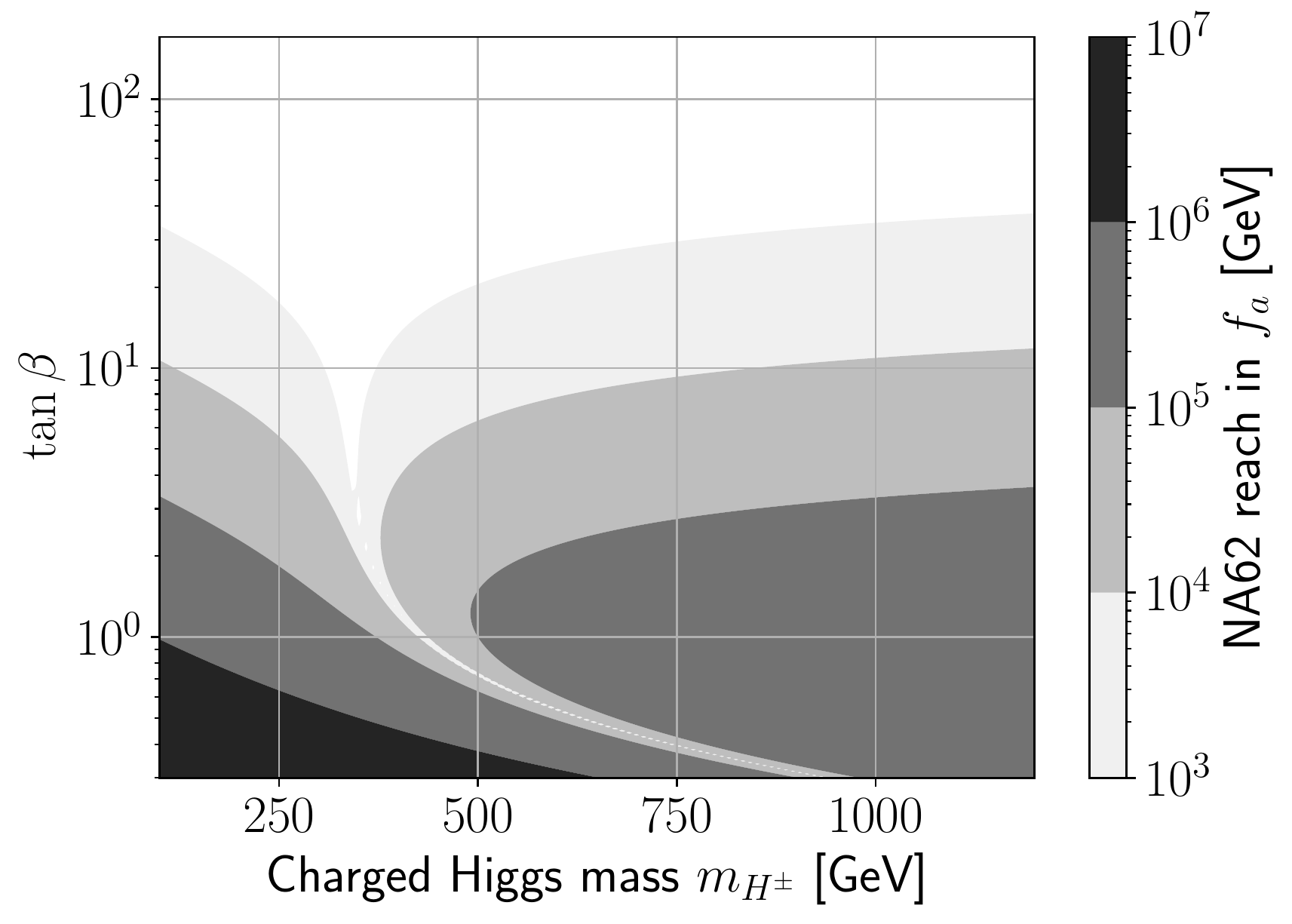}
	\hspace*{1cm}
	\caption{Expected NA62 reach on $f_a$ from $K^+\rightarrow \pi^+ + a$ in flavour-universal DFSZ models, in terms of the 2HDM parameters $m_{H^\pm}$ and $\tan\beta$. 
	We use the projection of~\cite{Fantechi:2014hqa} for $\mathrm{Br}(K^+\rightarrow\pi^++a)$.
	}
	\label{fig:K_pi_a_DFSZ}	
\end{figure}

Our starting point is the simplest DFSZ model, which is a two-Higgs-doublet model (2HDM) of Type II extended by a singlet complex scalar $\Phi$ that transmits its PQ charge via the operator 
\begin{equation}
\mathcal{L} \supset \Phi^2 H_u H_d.
\end{equation}
The PQ symmetry is broken when $\Phi$ acquires a vacuum expectation value (VEV) $f_a$, which we assume to be much larger than the EW VEV $v=(v_u^2 + v_d^2)^{1/2}$. As usual, we denote $\tan\beta = v_u / v_d$.
The axion inherits couplings to the SM quarks due to its mixing with the pseudoscalar Higgs, resulting in axion Yukawa interactions with the up $\sim (m_{u,i}/f_{a}) \cos^2\beta / 3$ and down $\sim (m_{d,i}/f_{a}) \sin^2\beta / 3$ quarks.
With this, the value of $h^S_{ds}$ in the effective Hamiltonian Eq.~\eqref{eq:EffLagrds} describing the $s\rightarrow d + a$ transition is found to be 
\begin{equation}\label{eq:hds_DFSZ}
h^S_{ds} = - \frac{G_F}{16\pi^2} \frac{\cos^2\beta}{3f_a} \sum_{q=u,c,t} V_{qd}^* V_{qs} m_q^2 \left( X^q_1 + X^q_2 \cot^2\beta \right).
\end{equation}
The sum runs over all the up-type quark flavours and the (finite) loop functions are given by~\cite{Freytsis:2009ct}
\begin{align}
X^q_1 = 2 &+ \frac{m_{H^\pm}^2}{m_{H^\pm}^2-m_q^2} - \frac{3m_W^2}{m_q^2-m_W^2}
+ \frac{3m_W^4 \left( m_{H^\pm}^2 + m_W^2 - 2m_q^2 \right)}{\left( m_{H^\pm}^2 - m_W^2 \right) \left( m_q^2 - m_W^2 \right)^2} \log \frac{m^2_q}{m^2_W}\notag \\
&+ \frac{m_{H^\pm}^2}{m_{H^\pm}^2 - m_q^2} \left( \frac{m_{H^\pm}^2}{m_{H^\pm}^2-m_q^2} - \frac{6m_W^2}{m_{H^\pm}^2 - m_W^2} \right) \log \frac{m_q^2}{m_{H^\pm}^2}, \\
X^q_2 = \,\,\,\, &- \frac{2m_q^2}{m_{H^\pm}^2-m_q^2} \left( 1 + \frac{m_{H^\pm}^2}{m_{H^\pm}^2 - m_q^2}\log \frac{m_q^2}{m_{H^\pm}^2} \right),
\end{align}
in terms of the charged Higgs boson mass $m_{H^\pm}$. The different sign of the leading logarithmic term in the limit of large $m_{H^\pm}$ compared to eq.~\eqref{eq:effsdCoeff} is due to a negative $c_q$ in the DFSZ model.
We can then use Eq.~\eqref{eq:Kpia_width} to compute the expected $K^+\rightarrow\pi^+ + a$ branching ratio in DFSZ models. 
The expected NA62 constraints~\cite{Fantechi:2014hqa} as a function of $m_{H^\pm}$ and $\tan\beta$ are shown in figure~\ref{fig:K_pi_a_DFSZ}. 
Note that for each value of $\tan\beta$, there is a value of $m_H^\pm$ for which $h_{ds}^S$ in Eq.~\eqref{eq:hds_DFSZ} changes sign, which causes the funnel in figure~\ref{fig:K_pi_a_DFSZ} along which the bound disappears. 

With this computation, it becomes explicit that the value of the cutoff scale of the leading logarithm for the DFSZ model is the charged Higgs boson mass, $\Lambda_{\mathrm{DFSZ}} = m_{H^\pm}$, which is orders of magnitude smaller than $f_a$ in the interesting region of parameter space. Moreover, cancellations in the decay rate can occur for specific parameter points, an effect that is not captured within the EFT approach. 
Finally, FCNCs can also be suppressed by large values of $\tan\beta$~\cite{Choi:2017gpf}.
Overall, this means that the $K^+\rightarrow \pi^+ + a$ branching ratio in DFSZ models is much smaller than expected based on the EFT computation.
As a consequence, the actual bounds and projections from experiments such as NA62 on the DFSZ axion parameter become correspondingly weaker.
This comparison can be seen in figure~\ref{fig:K_pi_a_EFT} in section~\ref{sec:constraintsandopportunities}, where the dark region represents the branching ratio for a benchmark value of $m_{H^\pm}=800\GeV$ and $1\leq\tan\beta \leq 5$, roughly representative of the experimentally allowed 2HDM parameter space (see e.g.~\cite{Oda:2019njo,Kling:2020hmi}).
We also display a lighter region which corresponds to only requiring perturbativity of the Yukawa couplings, $0.25\leq\tan\beta\leq 170$~\cite{DiLuzio:2020wdo}.

\subsection{KSVZ-type models}
The computation above shows that the EFT calculation with the naive leading-log approximation is not appropriate for DFSZ axion models.
The reason for this is that DFSZ axion models introduce new degrees of freedom (the second Higgs doublet) at a scale below the naive EFT cutoff $f_a$.
In KSVZ~\cite{Kim:1979if,Shifman:1979if} models, one expects that the EFT leading log term should not appear either.
The reason is that in KSVZ models no tree-level fermion couplings are present, and even if they can be generated by an axion-dependent chiral rotation of fermion fields, the logarithmically divergent contribution from the 1-loop process involving the derivative coupling $\partial_\mu a /(2 f_a) \bar{f}\gamma^\mu\gamma^5 f$ exactly cancels with the one coming from the pseudoscalar coupling $ia/f_a m_f \bar{f}\gamma^5 f$, so that only a finite piece remains. From a different perspective, this cancellation reflects the equivalence of the linear and polar representation used for the complex scalar field containing the axion~\cite{Quevillon:2019zrd}.

To check this statement, we start with
\begin{equation}
\mathcal{L} \supset - \frac{a}{f_a} \frac{\alpha_s}{8\pi} G^{a,\mu\nu} \tilde{G}^a_{\mu\nu}\,,
\end{equation}
and perform a chiral rotation~\cite{Bauer:2017ris} on \textit{all} SM quarks, i.e.~$q=(u,\,d,\,s,\,c,\,b,\,t)^{T}$, with
\begin{equation}
q \rightarrow \exp\left( i \kappa_q \frac{a}{2 f_a}\gamma_5\right) q\,,
\end{equation}
where $\kappa_q$ is a $6\times 6$ diagonal matrix. Taking, e.g., $\text{Tr}[\kappa_q] = 1$ removes the gluon coupling completely~\cite{Bauer:2017ris}. As the one-loop process for $s \rightarrow d + a$ is $\mathcal{O}(g^2)$, we have to collect all terms that contribute at the same order. The relevant terms before the rotation are 
\begin{equation}
\begin{aligned}
\mathcal{L} \supset \bar{q} i \gamma^\mu \partial_\mu q - \bar{q} M_q q - \frac{g}{\sqrt{2}} \bar{u}_L \gamma^\mu W^+_\mu V d_L + \text{h.c.}\,,\\
\end{aligned}
\end{equation}
yielding after the chiral rotation (expanded at LO in the ALP field)
\begin{equation}
\begin{aligned}
\label{eq:KSVZChRot}
\mathcal{L} \supset - \frac{\partial^\mu a}{2 f_a} \bar{q} \kappa_q \gamma_\mu \gamma_5 q - \frac{i a}{f_a} \bar{q}\, \kappa_q M_q  \gamma_5 \,q - \frac{i a g}{2\sqrt{2} f_a} \bar{u}_L \gamma^\mu W_\mu^+ \left(\kappa_u V -  V \kappa_d \right) d_L + \text{h.c.}\,,
\end{aligned}
\end{equation}
where $M_q$ is the quark mass matrix and $\kappa_u$ and $\kappa_d$ are the $3\times3$ submatrices of $\kappa_q$ including only the elements for the up-type and down-type quarks, respectively. After tracing the influence of these terms on $s \rightarrow d + a$ at one loop, one notices that the UV-divergent terms cancel exactly. Note that in this context the third term in Eq.~\eqref{eq:KSVZChRot} provides crucial contributions by axion emissions off the $W-q-q'$ vertices, which do not vanish unless $\kappa_q$ is fully flavour-universal.\footnote{Intriguingly, this means that for DFSZ-like models such terms can in general not be neglected upon performing a chiral rotation from the Yukawa to the derivative basis.} We can therefore robustly conclude that a leading log does not appear in KSVZ models at the one-loop level.

\section{An EFT-inspired QCD axion model}\label{sec:new_model}

\subsection{Low-energy effective model}\label{sec:effectivemodel}

Because the naive EFT result is not recovered in common axion models like KSVZ or DFSZ, one wonders if there is any UV completion of the axion EFT where the leading-log result is actually applicable, meaning that the full result of the kaon decay width is also enhanced by as large a logarithm (or an even larger one) as in~\eqref{eq:effsdCoeff}. From our discussion above, we can identify two key requirements for such a model. First, it should not contain any additional new states below the new-physics scale $\Lambda$ apart from the axion (unlike the DFSZ model). And second, it should feature physical couplings to quarks that cannot be removed by a chiral rotation (unlike the KSVZ model).

A general class of effective QCD axion models that satisfy both of these requirements is given by the Lagrangian\footnote{As noted before, this model is a member of the class of effective Lagrangians considered in~\cite{Choi:2017gpf} based on electroweak symmetry considerations. 
Ref.~\cite{Choi:2017gpf} also provided a string-related axion-like particle realisation to give a large logarithm. However, in that setup the FCNC is parametrically suppressed by $1/(16\pi^2) \sim 10^{-2}$ compared to the field theoretic construction presented below.
}
\begin{equation} \label{eq:new_model}
\mathcal{L} \supset -\frac{\Phi}{\Lambda^u_{ij}}  \bar{Q}_{Li}\tilde{H} u_{Rj} - \frac{\Phi}{\Lambda^d_{ij}} \bar{Q}_{Li}H  d_{Rj} + \mathrm{h.c.}\, , 
\end{equation}
where $Q_L$, $u_R$, $d_R$, and $H$ denote the usual SM chiral fermion fields and Higgs doublet before EW symmetry breaking, and the indices $i,j$ label the three generations. 

The complex scalar $\Phi = \phi/\sqrt{2}\,\me^{ia/\braket{\phi}}$ contains the axion as its angular degree of freedom, and we have therefore normalized its PQ charge to $\chi_\Phi = 1$. The Higgs doublet has no PQ charge\footnote{The model does not change if the Higgs is allowed to carry charge. The reason is that the five charges of $Q_L$, $u_R$, $d_R$, $H$, and $\Phi$ are restricted by the two interaction terms in Eq.~\eqref{eq:new_model} to three conserved global U(1) symmetries which can be identified as hypercharge, baryon number, and the PQ symmetry. In this three-dimensional space, a direction for the PQ symmetry not involving the Higgs field can always be found.}
and the charges of the quarks therefore satisfy $\chi_{Q_{L_i}} - \chi_{u_{R_i}} = \chi_{Q_{L_i}} - \chi_{d_{R_i}}=1$. 

In order to avoid having to impose any restrictions on the coupling matrices $\Lambda^{u,d}$, we assume that the PQ charges are flavour independent,
\begin{equation}
	\chi_{L} =	 \chi_{Q_{Li}}  \quad  \mathrm{and} \quad \chi_{R}  =	\chi_{d_{Ri}} = \chi_{u_{Ri}}  \quad \textrm{for all generations.}  
\end{equation}
Note that this is the essential difference to the models considered in~\cite{Calibbi:2016hwq,Ema:2016ops}, where the charges were assumed to be flavour-dependent in order to explain the observed flavour structure of the Standard Model Yukawa couplings. Such a choice generically leads to tree-level flavour-violating interactions for the axion. As our focus is on the loop-induced flavour effects, we choose flavour-independent charges to avoid these complications.

The QCD anomaly induced by the SM quarks is
\begin{equation}
N = \sum_f  (\chi_{f_L}-\chi_{f_R}) T(R_f)=3,
\label{eq:anomalycoeffN}
\end{equation}
as in the usual DFSZ models. The electromagnetic anomaly coefficient,
\begin{equation}
\label{eq:anomalycoeffE}
E = \sum_f  (\chi_{f_L}-\chi_{f_R}) Q_f^2,
\end{equation}
can either be $8$ or $5$ depending on whether the leptons are PQ-charged or not (leptons can also couple to $\Phi^*$, resulting in a negative contribution to $E$). Once the PQ scalar acquires a VEV $\braket{\Phi}=\braket{\phi}/\sqrt{2}$, we can write
\begin{equation}
	\mathcal{L} \supset -Y_{ij}^u \me^{i \frac{a}{\braket{\phi}}} \bar{Q}_{Li}\tilde{H}  u_{Rj} - Y_{ij}^d \me^{i \frac{a}{\braket{\phi}}}\bar{Q}_{Li} H  d_{Rj} + \mathrm{h.c.}, \label{eq:mediumenergy_Lag}
\end{equation}
where the SM Yukawa couplings are $Y_{ij}^{u,d} = \braket{\phi} /(\sqrt{2} \Lambda^{u,d}_{ij})$.
Because the axion is automatically aligned with the Yukawas, there are no tree-level flavour-violating axion couplings in this model.

After EW symmetry breaking and CKM diagonalization, the usual pseudoscalar couplings of the axion to quarks,
\begin{equation}
\mathcal{L} \supset -\sum_q m_q \bar{q}\, \me^{i \frac{a}{\braket{\phi}}\gamma^5}q, \label{eq:lowenergy_coup}
\end{equation}
are recovered, meaning that the couplings are strictly proportional to the fermion masses.

The presence of axion-quark couplings in this model implies the existence of a divergent contribution to flavour-changing amplitudes at one loop from diagrams as the ones shown in figure~\ref{fig:s2dALoop}. When replacing this divergence by a leading logarithm, the only possibilities are $\log(\braket{\phi}^2/m_q^2)$ or $\log(\Lambda_{ij} / m_q^2)$. Since all of these scales are at least as large as the PQ breaking scale, we therefore necessarily obtain the large logarithmic enhancement of flavour-changing processes anticipated from the axion EFT.

Since we started from a non-renormalizable Lagrangian in Eq.~\eqref{eq:new_model}, it is not possible to obtain an exact result for the leading logarithm and as such one also lacks its physical interpretation. It is therefore instructive to understand how the effective interactions discussed above can be embedded in a UV-complete model, which allows to calculate the flavour-changing processes in detail.

\subsection{UV completion of the model}

In order to embed the effective model introduced above into a renormalizable UV completion, we add three generations of heavy coloured up- and down-type fermions, $F^u_i$ and $F^d_i$, which resemble the messenger fields in the Froggatt-Nielsen mechanism~\cite{Froggatt:1978nt}.
For simplicity, we first study the case of a single generation, and later generalize it to three generations.

For one generation, the relevant renormalizable Lagrangian involves the terms
\begin{align}
\begin{split}
	\mathcal{L} \supset \quad &-\alpha^{u} \bar{Q}_{L} \tilde{H} F^u_{R} - \beta^{u} \bar{F}^u_{L} \Phi u_{R}   + \mathrm{h.c.}\\
	&- \alpha^{d} \bar{Q}_{L} H F^d_{R} - \beta^{d} \bar{F}^d_{L} \Phi d_{R}   + \mathrm{h.c.}\;,
\label{eq:highenergy_Lag}
\end{split}
\end{align}
where $\alpha^{u,d}$ and $\beta^{u,d}$ denote the couplings in the up- and down-type sectors.\footnote{The only other dimension-4 term that is invariant under all symmetries considered is $\Phi \Phi^* H H^\dagger$. If included, such a term would generate a large contribution to the Higgs mass unless it comes with an extremely small prefactor. This is nothing but the usual naturalness problem that is common to all axion models, and which we do not seek to address in this work.}

More generally, one can allow the possibility to use either $\Phi$ or $\Phi^*$.
This leads to slightly different models, which are described in appendix~\ref{sec:app_models}. We focus on the example of the model as specified in Eq.~\eqref{eq:highenergy_Lag}, but the other options can be treated in a similar way.
For all terms to be gauge invariant, $F^u$ and $F^d$ must transform under the SM gauge groups like $u_R$ and $d_R$, respectively, and left- and right-handed $F$ fields (we omit the $u/d$ superscript when referring to any of the two fields) must transform identically.
The terms in Eq.~\eqref{eq:highenergy_Lag} always enforce the conditions $\chi_{Q_L}-\chi_{F_R}=0$ and $\chi_{F_L}-\chi_{q_R}= 1$.
It is precisely the sum of these equations that appears in the anomaly coefficients $N$ and $E$ in Eqs.~\eqref{eq:anomalycoeffN} and \eqref{eq:anomalycoeffE}, as up- and down-type $F$ and SM quarks couple similarly to gluons and photons.  Hence, the additional SU(3) quarks do not change the anomaly coefficients $N=3$ and $E=$~8~or~5.
Furthermore, a bare mass term for the $F$ quarks,
\begin{equation}
\mathcal{L} \supset - \lambda \frac{\braket{\Sigma}}{\sqrt{2}} \bar{F}_L F_R + \mathrm{h.c.},
\label{eq:F_massS}
\end{equation}
can be added.
For easier extension to the multi-generation case, we have parameterized the mass in terms of the VEV of a real spurion field $\Sigma$ for which $\chi_\Sigma=0$.
This imposes the additional condition $\chi_{F_L}-\chi_{F_R}=0$.
With this, all axial charges of all combinations of SM and $F$ quarks have been fixed and the only remaining freedom is a shift of all charges by an arbitrary constant. 

It is clear that after PQ and EW symmetry breaking, the terms in Eq.~\eqref{eq:highenergy_Lag} induce mass mixing between the different fields involved.
In unitary gauge, we can write the mixing in matrix form as 
\begin{equation}
\mathcal{L}\supset -\begin{pmatrix}
\bar{q}_{L} & \bar{F}_{L}
\end{pmatrix}
M
\begin{pmatrix}
q_{R} \\
F_{R}
\end{pmatrix} + \mathrm{h.c.},
\end{equation}
where $M$ depends on the VEVs of all scalar fields in our theory ($H$, $\Phi$, $\Sigma$), the corresponding coupling matrices ($\alpha$, $\beta$, $\lambda$), and the axion field $a$.

\subsection{Light- and heavy-quark mass matrix diagonalization}
\label{sec:diagonalization}

In order to recover the quark masses and couplings, we proceed exactly as in the Standard Model and diagonalize the mass matrices $M^u$ and $M^d$ by unitary transformations of left- and right-handed fields.
Only the main steps and results are mentioned here: we refer to appendix~\ref{app:diagonalization} for detailed calculations.
Throughout this and the next subsection, we drop the labels $u/d$ whenever the procedure is identical for both types of quarks.
We consider at every point all three generations of SM and $F$ quarks even though the generation indices are also omitted. 

The full mass matrix $M$ is given by
\begin{equation}
\label{eq:MassMatr}
M = \frac{\braket{\Sigma}}{\sqrt{2}} \begin{pmatrix} 0 & \epsilon \epsilon'\alpha\\ \epsilon' \beta e^{i a/\langle \phi \rangle}& \lambda\end{pmatrix},
\end{equation}
where we have introduced two parameters 
\begin{equation}
    \epsilon = \frac{v}{\braket{\phi}}
    \quad \mathrm{and} \quad \epsilon' = \frac{\braket{\phi}}{\braket{\Sigma}}\,,
\end{equation}
with the Higgs VEV $v = 246\,\mathrm{GeV}$, which highlight the hierarchies of scales in our model. 
We can safely assume that the parameter $\epsilon$ is much smaller than one. In contrast, $\epsilon'$ may but does not need to be smaller than one for a general axion model, where extra degrees of freedom other than $\Phi$ may appear close to the PQ scale. Importantly, such additional degrees of freedom close to the PQ scale can lead to sizable corrections to our desired effective model in Eq.~\eqref{eq:mediumenergy_Lag}, introducing e.g. tree-level flavour-violating axion couplings. In order to avoid this complication, we assume that the additional degrees of freedom at hand (the $F$ quarks) are heavy enough so that these effects are sufficiently suppressed. In practice, this means that we require 
\begin{equation}
   \epsilon'' = \frac{\braket{\phi}}{\min m_{F_i}} \ll 1,
\end{equation}
where $\min m_{F_i}$ is the mass of the lightest fermion that is not part of the SM.
More details, such as quantifying how small $\epsilon''$ has to be in the most general case as well as how tree-level flavour violation can be avoided even for sizeable $\epsilon''$, can be found in appendix~\ref{app:diagonalization}.

Continuing with the diagonalization, we define $U$ to be a unitary matrix which diagonalizes the hermitian product $MM^\dagger$,
\begin{equation}
U^\dagger MM^\dagger U =   \Lambda^2 =  \begin{pmatrix} M_q^2 & 0\\ 0 & M_F^2\end{pmatrix}.
\end{equation}
Here, $M_q$ and $M_F$ are the diagonal mass matrices of $Q$ and $F$ quarks.
Transforming the left-handed quark fields with $U$ and the right-handed ones with $S= M^\dagger U\Lambda^{-1}$,
\begin{align}
\begin{pmatrix}q_L\\F_L\\\end{pmatrix} \rightarrow U  \begin{pmatrix}q_L\\F_L\\\end{pmatrix} \, , \quad
\begin{pmatrix}q_R\\F_R\\\end{pmatrix} \rightarrow S  \begin{pmatrix}q_R\\F_R\\\end{pmatrix}\, ,
\label{eq:DiagTrafo}
\end{align}
exactly gives us a basis of fields in which the mass matrix is diagonal and independent of $a$. Note that to lighten the notation we are using identical symbols for fields in the original Lagrangian, Eq.~\eqref{eq:highenergy_Lag}, and for the mass eigenstates after the diagonalization.

By expanding $U$ in $\epsilon$ and $\epsilon''$, we can express the quark masses in terms of the UV parameters. At leading order, the masses are
\begin{align}
    M_q^2 &= \frac{v^2}{2}\,\epsilon'^2 U_\delta^\dagger(\alpha\lambda^{-1}\beta\beta^\dagger\lambda^{\dagger -1} \alpha^\dagger) U_\delta, \label{eq:SMquarkmasses}\\
    M_F^2 &= \frac{\braket{\Sigma}^2}{2}  U_\xi^\dagger(\lambda \lambda^\dagger ) U_\xi,\label{eq:Fquarkmasses}
\end{align}
where $U_\delta$ and $U_\xi$ are unitary matrices which diagonalize the hermitian matrices in parentheses. 
It is important to keep in mind that all matrices in \eqref{eq:SMquarkmasses} and \eqref{eq:Fquarkmasses} exist for up- and down-type particles, giving a total of 12 different quark masses.
Any physical realization of the coupling matrices needs to reproduce the masses of SM quarks in Eq.~\eqref{eq:SMquarkmasses}. Since the mass of the top quark is close to $\braket{H}=v/\sqrt{2}$, it may be difficult or impossible to recover such a high value when all $F$ quarks are much heavier than the PQ scale and when simultaneously requiring perturbativity of all Yukawa couplings. 
A similar observation was made in~\cite{Choi:2017gpf}, where it was concluded that the cutoff of a Lagrangian as the one in \eqref{eq:new_model} can at most be of order $f_a$. 
However, in our explicit UV completion the masses of the six $F$ quarks can have a large intrinsic hierarchy, which can in principle lead to logarithmic enhancements of flavour-violating effects by factors larger than $\log(f_a^2/m_q^2)$ and possibly also tree-level flavour violation, and simultaneously reproduce the observed top Yukawa. These effects are part of the non-trivial flavour properties of our model, whose detailed investigation goes beyond the scope of this work. In the main part of this text, we work in the limit $\epsilon''\ll 1$ while remaining agnostic about the exact flavour structure that may realize this hierarchy. As a proof of principle, in appendix~\ref{app:diagonalization} we give an explicit example with $\epsilon''=0.2$ for which all of our calculations apply and the SM is reproduced. 

\subsubsection*{Axion couplings to gauge bosons}
\label{subsec:AnomTerms}
The matrices $U$ and $S$ introduced in the previous section generally contain axion-dependent phases.
As a consequence, the quark-field transformations in Eq.~\eqref{eq:DiagTrafo}, which are required to diagonalize the mass matrix, source axion interaction terms.
To understand these better, it is helpful to split the diagonalization procedure into two subsequent field redefinitions, of which only the first one depends on the axion field.

Looking at the full UV Lagrangian in \eqref{eq:highenergy_Lag}, it is clear that the $a$ dependence of the mass-mixing terms can be absorbed into the quark fields by the redefinitions
\begin{align}
\label{eq:ChiralRotAx}
u_R\rightarrow e^{-\frac{i a}{\braket{\phi}}} u_R \, ,\quad d_R\rightarrow e^{-\frac{i a}{\braket{\phi}}} d_R\,.
\end{align}
This transformation removes the axion field $a$ from $M$ in Eq.~\eqref{eq:MassMatr}, while the quark kinetic terms generate derivative couplings of the axion to right-handed quarks.
Hence, both diagonalization matrices $U$ and $S$ can subsequently be chosen to be independent of the axion field.
The resulting derivative interactions are explicitly written out below in terms of the mass-diagonal quark fields (see Eq.~\eqref{eq:AxDerivInt}).

In addition to the kinetic terms, the path integral measure is not invariant under the transformation in \eqref{eq:ChiralRotAx} and anomalous interaction terms between the axion and gauge bosons arise.
The $U(1)_{A}\times G\times G$ anomaly, with $G$ being either the strong or the EW gauge group, sources the interaction terms~\cite{Adler:1969gk,Bell:1969ts,Bardeen:1969md,Peccei:1977hh,Peccei:1977ur}
\begin{align}
\begin{split}
\label{eq:AnomGluPhot}
 \mathcal{L} \supset &- 2N\cdot\frac{\alpha_s}{16\pi \braket{\phi}}\, a\, \epsilon^{\mu\nu\alpha\beta} G^a_{\mu\nu} G^a_{\alpha\beta} - E \cdot \frac{\alphaEM}{8\pi \braket{\phi}}\,a\,\epsilon^{\mu\nu\alpha\beta} F_{\mu\nu} F_{\alpha\beta}\\
 & + E \cdot \frac{\alphaEM}{4\pi} \frac{s_W}{c_W} \frac{a}{\braket{\phi}} \epsilon^{\mu\nu\alpha\beta} F_{\mu\nu} Z_{\alpha\beta} - E \cdot \frac{\alphaEM}{8\pi} \frac{s_W^2}{c_W^2} \frac{a}{\braket{\phi}} \epsilon^{\mu\nu\alpha\beta} Z_{\mu\nu} Z_{\alpha\beta}\,,
 \end{split}
\end{align}
where $s_W$ and $c_W$ denote the sine and cosine of the Weinberg angle $\theta_W$. Note that there are no anomalous $W$-couplings because we are only transforming right-handed fields, which are singlets under $SU(2)$.
From here, it is customary to define
\begin{equation}
    f_a \equiv  \frac{1}{N_{\rm DW}}\braket{\phi},
\end{equation}
where $N_\text{DW}\equiv 2N=6$ is the domain-wall number counting the number of inequivalent vacua in the QCD-induced axion potential. The consequences of $N_\text{DW} \ne 1$ in our model are briefly discussed in section~\ref{sec:conclusions}.
The definition of $f_a$ results in the conventional normalization of the axion couplings,
\begin{align}
\begin{split}
\mathcal{L} \supset &- \frac{\alpha_s}{16\pi f_a}\, a\, \epsilon^{\mu\nu\alpha\beta} G^a_{\mu\nu} G^a_{\alpha\beta} - \frac{E}{N} \cdot \frac{\alphaEM}{16\pi f_a}\,a\,\epsilon^{\mu\nu\alpha\beta} F_{\mu\nu} F_{\alpha\beta} \\
 & + \frac{E}{N} \cdot \frac{\alphaEM}{8\pi} \frac{s_W}{c_W} \frac{a}{f_a} \epsilon^{\mu\nu\alpha\beta} F_{\mu\nu} Z_{\alpha\beta} - \frac{E}{N} \cdot \frac{\alphaEM}{16\pi} \frac{s_W^2}{c_W^2} \frac{a}{f_a} \epsilon^{\mu\nu\alpha\beta} Z_{\mu\nu} Z_{\alpha\beta}\,.
 \label{eq:axion_boson_coup}
\end{split}
\end{align}
By comparison to the EFT Lagrangian defined in Eqs.~\eqref{eq:EFT_Lagrangian} and \eqref{eq:EFT_Lagr_ferm}, we can identify $c_{gg}$, $c_{WW}$, $c_{\gamma\gamma}$, $c_{ZZ}$, and $c_{\gamma Z}$ as listed in table~\ref{tab:EFT_coefficients}. If leptons are also charged under the PQ symmetry, an analogous axial rotation has to be performed in the leptonic sector, resulting in an additional contribution to the electromagnetic anomaly. In any case, Eq.~\eqref{eq:axion_boson_coup} is equally applicable with the corresponding value of $E/N$.

\subsubsection*{Axion couplings to light and heavy quarks}\label{subsec:quark_couplings}

After absorbing the axionic phase of the mass matrix in the $q_R$ fields, $M$ becomes independent of $a$.
Hence, $U$ and $S$ can also be chosen to be independent of $a$ when proceeding with the subsequent steps in the diagonalization as described in Eq.~\eqref{eq:DiagTrafo}. After the diagonalization is completed, the derivative axion-quark couplings induced by the transformation Eq.~\eqref{eq:ChiralRotAx} can be expressed in terms of the mass-diagonal fields as
\begin{align}
\mathcal{L} &\supset 
	 \frac{\braket{\phi}}{2} \begin{pmatrix}\bar{q}_R & \bar{F}_R \\\end{pmatrix} (\slashed{\partial}a)\, \Lambda^{-1}\begin{pmatrix}
	(\epsilon\epsilon')^2 \mathcal{ABA^\dagger} & \epsilon\epsilon' \mathcal{AB}\\ \epsilon\epsilon' \mathcal{BA^\dagger} & \mathcal{B}\\
	\end{pmatrix}\Lambda^{-1} \begin{pmatrix} q_R \\ F_R\\ \end{pmatrix}
	\label{eq:AxDerivInt_init}\\
& = \frac{1}{\braket{\phi}} \bar{q}_R \,(\slashed{\partial}a)\, q_R  
+\frac{\braket{\phi}}{2} \bar{F}_R(M_F^{-1} \mathcal{B} M_F^{-1} ) \,(\slashed{\partial}a)\, F_R  
+ \frac{v}{2}\, \epsilon'\,\bar{q}_R (M_q^{-1}\mathcal{A}\, \mathcal{B} M_F^{-1})\,(\slashed{\partial}a)\, F_R + \text{h.c.} ,
\label{eq:AxDerivInt}
\end{align}
where $\mathcal{A}$ and $\mathcal{B}$ are coupling matrices which, to leading order in $\epsilon''$, are given by
\begin{equation}
    \mathcal{A} = U^\dagger_\delta \alpha \lambda^{-1} U_\xi \quad \textrm{and}\quad \mathcal{B} = U_\xi^\dagger  \beta \beta^\dagger U_\xi\; .
\end{equation}
Note that the reason why the axion only couples to right-handed quarks is that we chose to only rotate these chiral components in Eq.~\eqref{eq:ChiralRotAx}.
This is clearly an arbitrary choice: we can perform an axion-dependent vector (non-axial) rotation in the quark field to include derivative couplings to left-handed quarks.

\bigskip

The field basis in which axions only couple derivatively is convenient for the calculation of loop processes as the one in section~\ref{sec:Kpia_new_model}. But if we want to recover the effective terms in Eq.~\eqref{eq:lowenergy_coup}, we have to rewrite the first term in Eq.~\eqref{eq:AxDerivInt} by 
performing a rotation of the right-handed fields (neglecting interactions with the Higgs and CP-conserving anomalous axion terms),
\begin{align}
\label{eq:Noether2}
 \frac{1}{\braket{\phi}} \bar{q}_R \,(\slashed{\partial}a)\, q_R &\to-\frac{a}{\braket{\phi}} \bar{q} M_q i \gamma_5 q\,.
\end{align}
Hence, we exactly arrive at purely pseudoscalar couplings to SM quarks which are proportional to the quark masses, as anticipated in Eq.~\eqref{eq:lowenergy_coup}. Note that this implies that the right-handed flavour-diagonal coupling structure conserves CP, which can equally be shown by applying the CP transformation explicitly, see appendix~\ref{app:CPVInt}.

In contrast, applying the same steps to the second term in Eq.~\eqref{eq:AxDerivInt} yields
\begin{align}
 & \frac{\braket{\phi}}{2} \bar{F}_R (M_F^{-1} \mathcal{B} M_F^{-1})\,(\slashed{\partial}a)\, F_R \nonumber\\
     \to&- \frac{\braket{\phi}}{2} a \left[ \bar{F} \frac{ \mathcal{B} M_F^{-1} - M_F^{-1}\, \mathcal{B}}{2} i F + \bar{F} \frac{ \mathcal{B} M_F^{-1} + M_F^{-1}\, \mathcal{B}}{2} i \gamma_5 F\right]\,.
     \label{eq:axionFFcouplings}
\end{align}
This demonstrates that the coupling of heavy $F$-quarks is not proportional to their masses, which is due to the fact that their mass is (mostly) generated by the spurion field $\Sigma$, whose VEV does not break the PQ symmetry.
Furthermore, the interaction with $F$ quarks is not of purely pseudoscalar form, which, as is discussed in section~\ref{sec:CPviolation}, has important implications for CP violation in our model.
 
Finally, for the last term in \eqref{eq:AxDerivInt} we make use of the fact that $q_R$ and $F_R$ only differ in their respective masses but transform identically under the SM gauge group, and therefore
\begin{align}
 \partial_\mu\,(\bar{q}_{Ri} \gamma^\mu F_{Rj}) &\to \bar{q}_i \frac{(M_q)_{ii} - (M_F)_{jj}}{2} i F_{j} + \bar{q}_i \frac{(M_q)_{ii} + (M_F)_{jj}}{2} i \gamma_5 F_{j}\,,
\end{align}
where the indices $i$ and $j$ do not imply a sum but only a specific quark. This allows us to rewrite the last term in \eqref{eq:AxDerivInt} as
\begin{align}
     &\frac{v}{2}\,\epsilon' \bar{q}_R (M_q^{-1}\mathcal{A}\, \mathcal{B} M_F^{-1})\,(\slashed{\partial}a)\, F_R \nonumber\\
     \to&-\frac{v}{2}\,\epsilon' a \left[ \bar{q} \frac{\mathcal{A}\, \mathcal{B} M_F^{-1} - M_q^{-1}\mathcal{A}\, \mathcal{B}}{2} i F + \bar{q} \frac{\mathcal{A}\, \mathcal{B} M_F^{-1} + M_q^{-1}\mathcal{A}\, \mathcal{B}}{2} i \gamma_5 F\right]\,.
     \label{eq:axionqFcouplings}
\end{align}
Once more, the interaction contains both pseudoscalar and scalar parts which can in general be CP violating.

\subsection{Low-energy effective couplings}

To conclude this section, let us summarize how the model introduced above maps onto the generic axion EFT. In this context, it is important to note that the low-energy effective model does not contain any free parameter other than the axion decay constant and the mass of the heavy $F$-fermions, the latter only being relevant as a cut-off scale for loop processes (see the next section).
All the couplings of the axion to SM particles are therefore fully determined once $f_a$ is fixed. 
There is only one discrete choice, namely whether the SM leptons are charged under the PQ symmetry or not, leading to the two variations of the model.
Table~\ref{tab:EFT_coefficients} summarizes the values that the EFT coefficients introduced in Eqs.~\eqref{eq:EFT_Lagrangian} and~\eqref{eq:EFT_Lagr_ferm} take for both variations.

\begin{table*}[ht]
	\centering
	\small
	\begin{tabular}{| c || c | c | c |}
	\hline
	EFT coefficient & UV parameter & EFT model & EFT-$\ell$ model \\ \hline \hline
	$f_a$ & $\braket{\phi}/(2N)$ & $\braket{\phi}/6$ & $\braket{\phi}/6$ \\ \hline
	$c_{gg}$ & $1$ & $1$ & $1$ \\ \hline
	$c_{\gamma\gamma}$ & $E/N$ & $5/3$ & $8/3$ \\ \hline
	$c_{ZZ}$ & $\tan(\theta_W)^2\ E/N$ & $\tan(\theta_W)^2\ 5/3$ & $\tan(\theta_W)^2\ 8/3$ \\ \hline
    $c_{\gamma Z}$ & $-2 \tan(\theta_W)\ E/N$ & $-2 \tan(\theta_W)\ 5/3$ & $-2 \tan(\theta_W)\  8/3$ \\ \hline
	$c_{WW}$ & $W/N$ & $0$ & $0$ \\ \hline
	$(c_{q,R}-c_{q,L})$ & $-(\chi_{q_R}-\chi_{q_L})/(2N)$ & $ 1/6-0$ & $1/6-0$ \\ \hline
	$(c_{\ell,R} - c_{\ell,L})$ & $-(\chi_{\ell_R}-\chi_{\ell_L})/(2N)$ & $0$ & $1/6-0$ \\ \hline
	\end{tabular}
	\caption{Summary of couplings of the model introduced in the present section, expressed in the EFT basis introduced in Eqs.~\eqref{eq:EFT_Lagrangian} and~\eqref{eq:EFT_Lagr_ferm}.
	The general dependence on the high-energy parameters is shown, together with the explicit value for each of the two variations of the model, the one with and the one without tree-level couplings to leptons. In the three cases $c_{\gamma\gamma}$, $c_{\gamma Z}$, and $c_{ZZ}$, the coupling term should be written as in \eqref{eq:axion_boson_coup} with the electromagnetic fine structure constant $\alphaEM$. Further, the fermion coupling matrices in the last two rows are proportional to the unit matrix and $1/6-0$ indicates that only right-handed fields couple at tree-level. 
		\label{tab:EFT_coefficients}
}
\end{table*}

\section{Flavour and CP effects in the EFT-inspired QCD axion model}\label{sec:phenomenology}

In the previous section, we have introduced a QCD axion model with tree-level couplings to SM fermions and without any additional new states below the scale $f_a$ other than the axion itself. As a consequence, the model fully reproduces the phenomenological expectations based on the generic EFT description introduced in section~\ref{sec:EFT}, even when one-loop processes are involved.
To confirm this, in this section, we delineate the main phenomenological features of this QCD axion, including a calculation of the $K^+\rightarrow\pi^+ + a$ decay rate in the full model.

\subsection{\texorpdfstring{$K^+\rightarrow \pi^+ + a$ decay rate}{K to pi a decay rate}} \label{sec:Kpia_new_model}

In the previous section, we have computed the effective interactions between the axion and the SM quarks, which are summarized in table~\ref{tab:EFT_coefficients}. That said, our initial motivation for deriving the model in section~\ref{sec:new_model} was to obtain an EFT-like model with no ambiguity in the log-enhanced FCNC contribution. If the interactions listed in table~\ref{tab:EFT_coefficients} were the only ones emerging in our UV model, there would be no additional contributions to the $K^+ \rightarrow \pi^+ + a$ amplitude at one loop and we would again be confronted with the same problem of UV divergences as before. Crucially, however, there are indeed further relevant interactions arising in our UV model that are not contained in table~\ref{tab:EFT_coefficients} and which render the loop computation finite. These operators are, as expected, related to the heavy $F$ quarks and are hence not captured by the EFT approach.

\begin{figure}
\hspace*{2cm}
\vspace*{-0.2cm}
\begin{minipage}{0.3\textwidth}
    \vspace*{0.45cm}
	\centering
	\includegraphics[width=1.1\linewidth]{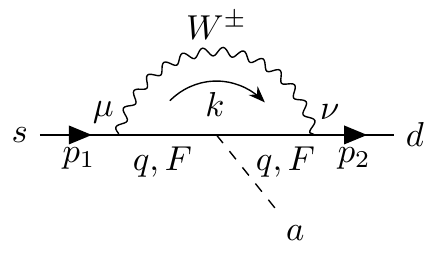}
	\end{minipage}
\hspace*{2cm}
\begin{minipage}{0.3\textwidth}
	\centering
	\includegraphics[width=0.9\linewidth]{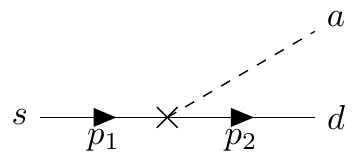}
	\end{minipage}
\caption{Relevant one-loop contributions to the $s\rightarrow d + a$ process in our UV model. These consist of axion emissions through the quark couplings (left) as well as counterterm contributions (right). The symbol $\times$ denotes a counterterm insertion. }
	\label{fig:diagramssketch}	
\end{figure}

To obtain these additional vertices, it is important to discuss further implications of the unitary transformation of the quark fields that we performed in Eq.~\eqref{eq:DiagTrafo}. As in the SM, this rotation generates new flavour-dependent interactions of neutral and charged hadronic currents with the weak gauge bosons. The detailed calculations are included in appendix~\ref{app:diagonalization}.
The leading-order $W$ interactions in the mass-diagonal basis can be written as 
\begin{align}
	\mathcal{L} \supset \frac{-g}{\sqrt{2}} \begin{pmatrix}\bar{u}_L \\ \bar{d}_L \\ \bar{F}^u_L \\ \bar{F}^d_L\end{pmatrix}^T \gamma^\mu
	&\left[    W^+_\mu \begin{pmatrix}
	0 &  V  & 0 &-\epsilon\epsilon'V\mathcal{A}_d  \\
    0 & 0 & 0 & 0 \\
	0 &  -\epsilon \epsilon' \mathcal{A}_u^\dagger V\phantom{^\dagger}  & 0 & \phantom{^\dagger}(\epsilon \epsilon')^2 \mathcal{A}^\dagger_u V \mathcal{A}_d \\
	0 & 0 & 0 & 0 \\
	\end{pmatrix}\right.\nonumber \\
 + & \phantom{\Bigg[}\left.  W^-_\mu\begin{pmatrix}
	0 & 0 & 0 & 0 \\
	V^\dagger & 0 & -\epsilon\epsilon' V^\dagger\mathcal{A}_u & 0 \\
	0 & 0 & 0 & 0 \\
	-\epsilon\epsilon'\mathcal{A}_d^\dagger V^\dagger & 0 &  (\epsilon \epsilon')^2 \mathcal{A}^\dagger_d V^\dagger \mathcal{A}_u & 0 \\
	\end{pmatrix} \right]
	\begin{pmatrix}
	u_L \\ d_L \\ F^u_L \\  F^d_L \\
	\end{pmatrix},
	\label{eq:quark-W-coupling}
\end{align}
where we have reintroduced the labels for up- and down-type quarks as well as coupling matrices, and $V$ denotes the CKM matrix as before. At leading order, $V$ is given by $V=U_\delta^{u\dagger}U_\delta^d$ and turns out to be unitary up to terms of the order $(\epsilon\epsilon'')^2$. 
Similarly, the leading-order $Z$ interactions become
\begin{equation}
\label{eq:ZBosInt}
    \mathcal{L} \supset\begin{pmatrix}\bar{q} \\ \bar{F}\end{pmatrix}^T \gamma_\mu Z^\mu \frac{-g}{\cos(\theta_W)} \left[
	\pm\frac{1}{2}\begin{pmatrix} \mathds{1} & -\epsilon\epsilon'\mathcal{A} \\-\epsilon\epsilon'\mathcal{A}^\dagger & (\epsilon\epsilon')^2\mathcal{A}^\dagger \mathcal{A} \\\end{pmatrix}P_L
	-Q \sin^2(\theta_W)   \begin{pmatrix}\mathds{1} & 0 \\ 0 & \mathds{1} \\\end{pmatrix}
	\right] \begin{pmatrix}
	q \\ F\\
	\end{pmatrix}.
\end{equation}
Here, the upper (lower) sign refers to up- (down-) type quarks, $P_L$ denotes the projector onto left-handed fields and $Q$ is the electromagnetic charge of each field. We find that only the left-handed coupling structure is affected by the transformation to the mass-diagonal basis. 

With these additional interactions at hand, we can move forward to perform the computation of the loop processes shown in figure~\ref{fig:diagramssketch}. The first diagram in figure~\ref{fig:diagramssketch} represents the sum of four individual contributions because each internal fermion propagator can either be a SM or an $F$ quark. To calculate these diagrams, we make use of the axion-quark interactions as given in Eq.~\eqref{eq:AxDerivInt_init} and the $W$ couplings to hadronic currents in Eq.~\eqref{eq:quark-W-coupling}.
Expressing the interactions in this way means that the $W$ boson and the axion only interact with left-handed and right-handed fields, respectively. Hence, we need mass insertions by each of the internal propagators. The corresponding factors of fermion masses are however cancelled by the inverse $\Lambda$ matrices in the axion interaction. Putting everything together, it is easy to see that all four diagrams have the same flavour structure and are of the same order in $\epsilon$ and $\epsilon''$.
We can write their combined contribution to the $s\rightarrow d + a$ amplitude in the compact form
\begin{align}
\label{eq:sdaUVAmpl}
	i\mathcal{M}=&i\frac{g^2}{2}(\epsilon\epsilon')^2  \frac{\braket{\phi}}{2}\sum_{i,j,k,l}  V^\dagger_{di}(\mathcal{A}_u)_{ij}(\mathcal{B}_u)_{jk}(\mathcal{A}^\dagger_u)_{kl}V_{ls}\ \int\frac{d^4k}{(2\pi)^4}\bar{u}_d(p_2)\left[\gamma_\nu\gamma_\rho(p_2-p_1)^\rho\gamma_\mu P_L\right]u_s(p_1)\nonumber\\
	&\times\left( \frac{1}{k^2-m_W^2}\left(g^{\mu\nu}-\frac{k^\mu k^\nu}{m_W^2}\right)\right)\nonumber\\
	&\times\bigg(\frac{1}{((p_2-k)^2+m_{Q_i}^2)((p_1-k)^2+m_{Q_l}^2)}-\frac{1}{((p_2-k)^2+m_{F_j}^2)((p_1-k)^2+m_{Q_l}^2)}\nonumber\\
	&\phantom{\Bigg[}-\frac{1}{((p_2-k)^2+m_{Q_i}^2)((p_1-k)^2+m_{F_k}^2)}+\frac{1}{((p_2-k)^2+m_{F_j}^2)((p_1-k)^2+m_{F_k}^2)}\bigg),
\end{align}
where $m_Q$ refers to the up-type quarks. We evaluate the loop integral using \texttt{Package-X}~\cite{Patel:2015tea} and under the simplifying assumption that all $F$ quarks have equal masses. Only considering the leading-order terms in the down and strange fermion masses, we arrive at
\begin{align}\label{eq:K_pi_a_new_model}
i \mathcal{M} =  h_{ds} \times \bar{u}_d(p_2) \left(  (m_s-m_d) +(m_s+m_d) \gamma_5 \right)u_s (p_1),
\end{align}
with 
\begin{eqnarray}
\label{eq:hdsUVCoeffWqF}
	h_{ds}\!\!&=&\!\!- \frac{G_F}{16\sqrt{2}\pi^2} \frac{1}{\braket{\phi}}
	\\\nonumber
	&&\times\sum_{q=u,c,t} V_{qd}^* V_{qs} m_q^2 \left(\log\left(\frac{m_F^2}{m_q^2}\right)-\frac{2m_q^4-7m_q^2m_W^2+5m_W^4+3m_W^4\log\left(\frac{m_q^2}{m_W^2}\right)}{(m_q^2-m_W^2)^2}\right),
\end{eqnarray}
where we expanded the finite contributions in terms of $1/m_F$. This gives a contribution to $h^S_{ds} = h_{ds}$ as defined in the effective interaction Hamiltonian in \eqref{eq:EffLagrds}. The logarithmically enhanced term is identical to the EFT result for $c_q=\frac{1}{6}$ and $\Lambda = m_F$. Note that the finite term is different from the EFT result as the loops with internal $F$-quarks also yield relevant contributions.
\newpage
In principle, we also need to include diagrams exchanging the $W$ boson in the first diagram in figure~\ref{fig:diagramssketch} with a $Z$ boson. However, even if the couplings of the $Z$ boson do have non-diagonal entries, they do not induce $s\rightarrow d + a$ at one-loop at the order in $\epsilon''$ that we are considering. This is due to the matrix structure $\mathcal{A}\mathcal{B}\mathcal{A}^\dagger$ appearing in the relevant amplitudes, which is flavour-diagonal as it is identical to the matrix that determines the quark masses in Eq.~\eqref{eq:SMquarkmasses}.

Finally, there are also counterterm contributions~\cite{Hall:1981bc} from the renormalization of quark fields as depicted on the right of figure~\ref{fig:diagramssketch}. Performing the calculations, however, one notices that the relevant contributions to Eq.~\eqref{eq:K_pi_a_new_model} cancel each other at linear order in the down- and strange-quark masses (see appendix~\ref{app:Counterterms} for more details). We can therefore also discard these counterterm diagrams and work with Eq.~\eqref{eq:hdsUVCoeffWqF} as our effective coefficient.

\subsection{CP violation in axion interactions}
\label{sec:CPviolation}

From Eq.~\eqref{eq:AxDerivInt}, we observe that the axion has a right-handed coupling structure to all quarks in our theory. As discussed in appendix~\ref{app:CPVInt}, CP violation in flavour-violating couplings occurs when the corresponding coupling is complex valued. The axion interactions in Eq.~\eqref{eq:AxDerivInt} can therefore induce CP violation since $\mathcal{A}$ and/or $\mathcal{B}$ can have imaginary entries. The fact that the interactions of the QCD axion can be CP violating may seem worrisome at first, since the QCD axion is introduced precisely to eliminate the CP violation in the strong sector.
As shown in this section, however, the effects of this kind of CP violation on the electric dipole moment (EDM) of the neutron are comparable or even smaller to those induced by the phase of the CKM matrix, and are therefore not a threat to the axion solution of the strong CP problem.
We point the reader to refs.~\cite{OHare:2020wah,DiLuzio:2020oah} for recent discussions on CP-violating axion interactions.

To see this, firstly note that this CP violation is of explicit nature, like the CKM phase in the SM, and not due to any spontaneous breaking.
This can be understood considering the full UV theory in Eq.~\eqref{eq:highenergy_Lag}.
If the couplings $\alpha$, $\beta$, and $\lambda$ are real-valued, no CP-violating terms are generated before or after EW and PQ symmetry breaking. This is because in that case, $U_\delta$ and $U_\xi$ appearing in Eqs.~\eqref{eq:SMquarkmasses} and \eqref{eq:Fquarkmasses} can be chosen to be orthogonal and real because the matrices that they diagonalize are real and symmetric to begin with.
This then results in $\mathcal{A} = U^T_\delta \alpha \lambda^{-1} U_\xi$ and $\mathcal{B} = U_\xi^T  \beta \beta^T U_\xi$ being real-valued as well.
Hence, Eq.~\eqref{eq:AxDerivInt} is CP conserving in this case.
However, starting from complex-valued $\alpha$, $\beta$, or $\lambda$ couplings generically makes $\mathcal{A}$ and/or $\mathcal{B}$ complex valued as well.
Therefore, the CP violation present in the axion interactions of Eq.~\eqref{eq:AxDerivInt} is a reflection of potential explicit CP-breaking in the UV model.
As a matter of fact, given that the CKM matrix, which is constructed from $\alpha$, $\beta$, and $\lambda$, is experimentally confirmed to be complex valued, there is no reason to expect the coefficients $\mathcal{A}$ and $\mathcal{B}$ to be real without invoking an \emph{ad hoc} cancellation.

The relevant question is therefore whether this CP violation is problematic, that is, whether it is in conflict with any observations.
To answer this, we consider a very sensitive observable to CP violation: the neutron EDM $d_n$.

As shown in appendix~\ref{app:EDM}, one-loop processes do not generate quark EDMs in our UV model.
Hence, a neutron EDM based on free quark EDMs may occur at the earliest at the two-loop level.
Without delving into involved two-loop calculations, we instead derive a very conservative upper bound based on the expected scaling of the contributions.
Assuming that at least two new physics vertices appear (such as internal emission and absorption of the axion), each of which carries a coupling suppression of $f_a$, we get the rough estimate
\begin{align}
\label{eq:NEDMestimate}
    d_n^\text{\,UV} \lesssim \frac{e}{(16 \pi^2)^2} \frac{m_n}{f_a^2} \approx 5\cdot10^{-32}\,e\cdot\mathrm{cm}\,.
\end{align}
Here, $m_n$ denotes the neutron mass, which we use as the characteristic scale, and we have inserted $f_a = 4\cdot10^6\,\mathrm{GeV}$ as a lower-end value for the range of axion decay constants of interest.
The above estimate is already very conservative in the sense that we have not included any suppression due to new physics contributions to $W^\pm$ and/or $Z$ gauge boson interactions and/or the heavy $F$-quark mass scale $\braket{\Sigma}$, since at least one of the three has to participate to generate a quark EDM.
Moreover, we have not invoked any additional electric/weak coupling insertion or Yukawa/light quark mass suppression, the latter being expected to appear given the discussion above regarding the origin of the CP-violating interactions.

Interactions between the quark constituents within the neutron have also been shown to contribute to the neutron EDM (see ref.~\cite{Dar:2000tn} for a review).
That said, these effects are already below present sensitivity for purely SM-related interactions.
This does not change in our UV model, as the new physics couplings/particles appearing would only lead to a stronger suppression.
We can therefore safely neglect such contributions.

All in all and despite being very conservative, the estimate in Eq.~\eqref{eq:NEDMestimate} is still much smaller than the current bounds~\cite{Abel:2020gbr} $d_n^\text{\,exp} \lesssim 10^{-26}\,e\,\mathrm{cm}$.
We therefore conclude that the resulting explicit CP violation is too small to affect near-future experiments measuring CP-sensitive observables and does not pose a threat to the axion solution to the strong CP problem in the model considered here.

\section{Phenomenology of the model and discovery opportunities}
\label{sec:constraintsandopportunities}

Because our QCD axion model was constructed to reproduce the general features of the EFT setup, one may be sceptical that the model could posses any particular phenomenological feature that may serve as a handle to probe it.
But interestingly, what at first seems like a lack of features does in fact have interesting consequences for experimental searches, which we now describe.

\subsection{Astrophysical limits and helioscope searches}

\begin{figure}[t]
	\centering
	\begin{tabular}{cc}
		\includegraphics[width=0.5\textwidth]{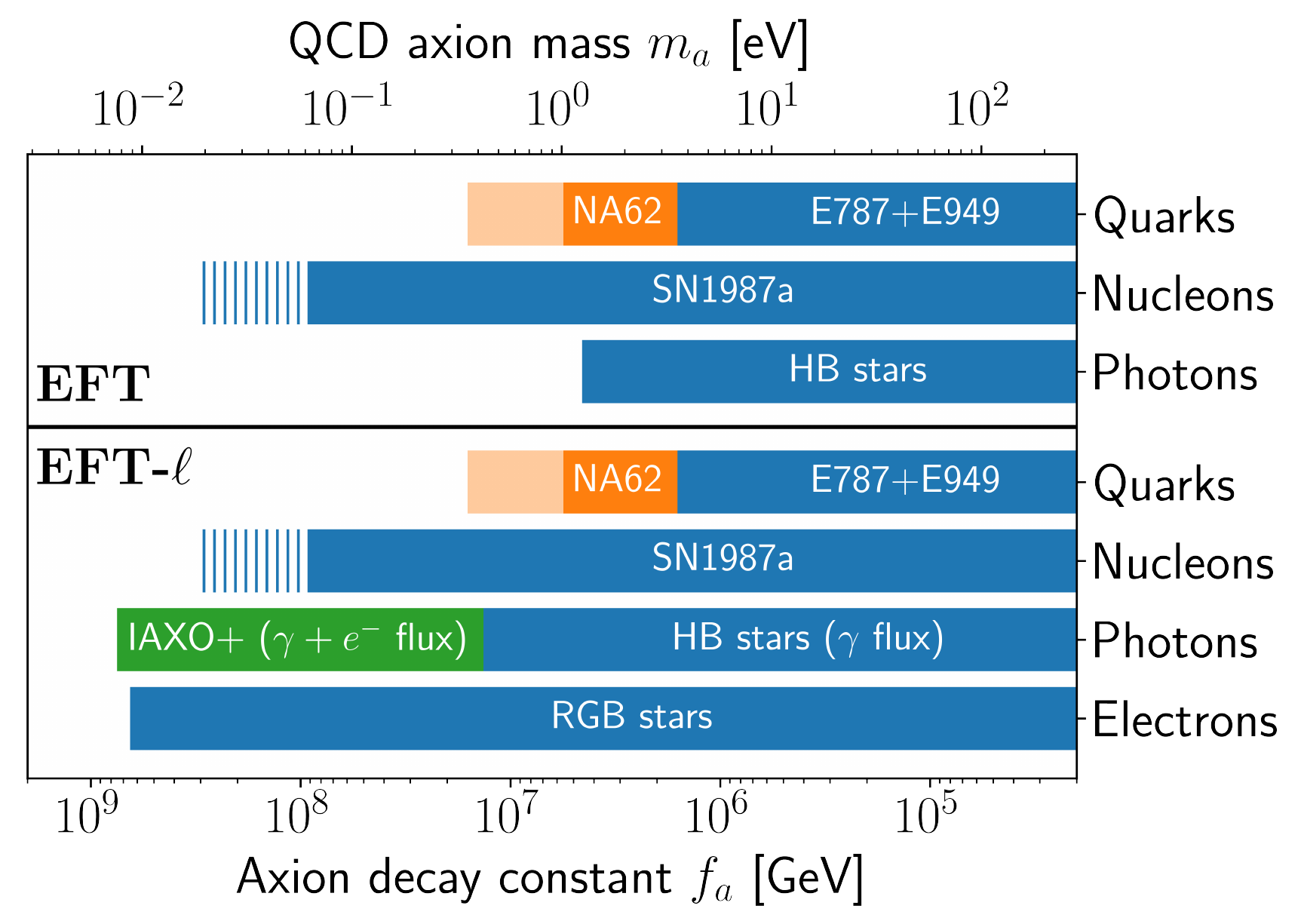}
	&
		\includegraphics[width=0.45\textwidth]{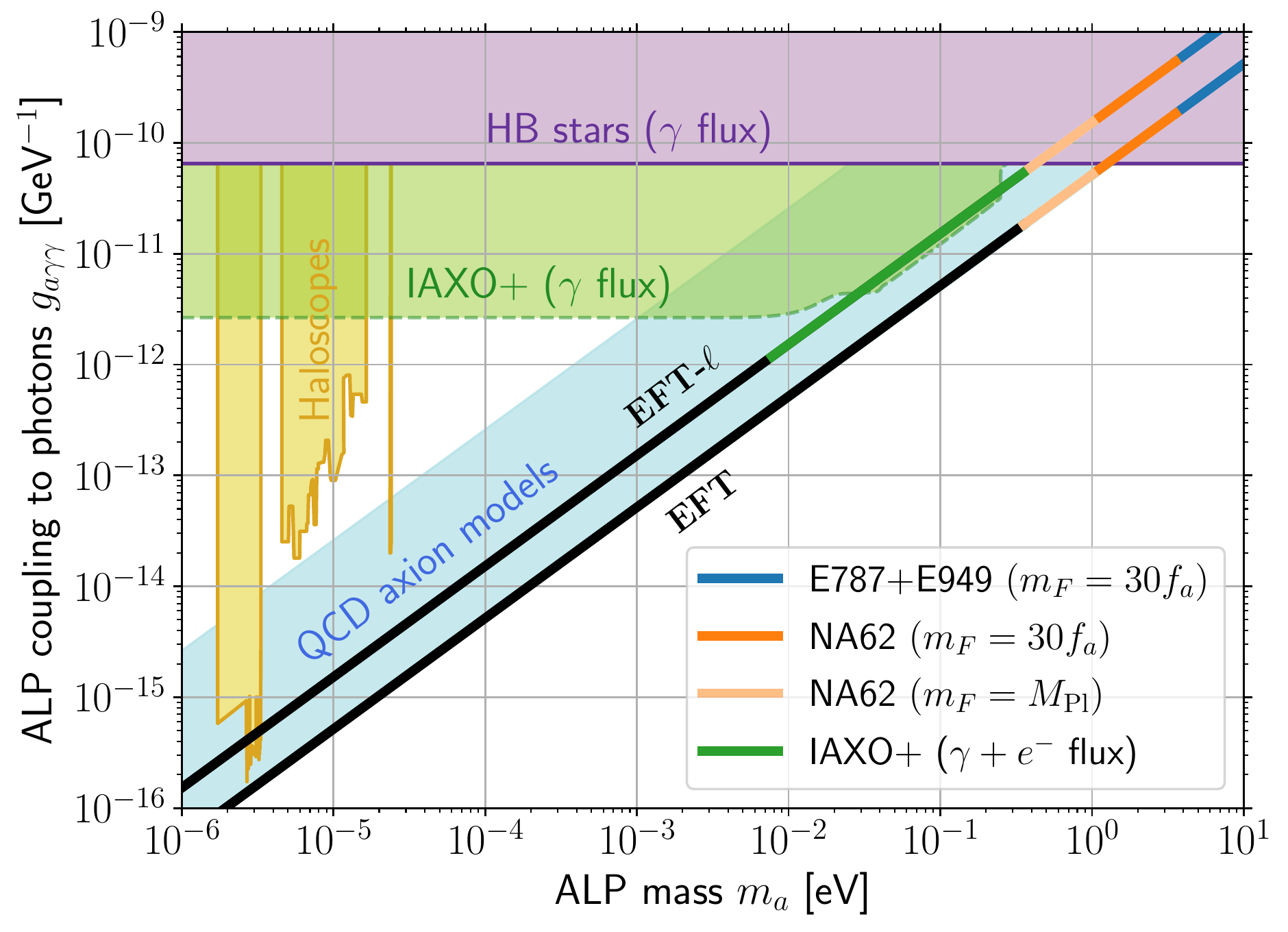}
	\end{tabular}
	\vspace{-0.4cm}
	\caption{Constraints on the two variations of the model introduced in section~\ref{sec:new_model}, which we denote EFT-$\ell$ and EFT depending on whether tree-level axion-lepton couplings are present or not.
	On the left, we present the bounds as a function of the QCD axion decay constant.
	For the $K^+\rightarrow\pi^++a$ constraints, the lighter shading of orange represents the range of possible values of the cutoff scale for the leading log, $\Lambda\in(30f_a,\,M_{\rm Pl})$ as discussed in the main text. The vertical hatching indicates uncertainties in the supernova limit.
	On the right, we show the models predictions in the usual mass vs.~photon coupling parameter space.
	The band of QCD axion models corresponds to $E/N\in(44/3,\,5/3)$ as defined in~\cite{DiLuzio:2017pfr}.
	Note that IAXO becomes particularly sensitive to the EFT-$\ell$ model due to solar axion production through axion-electron interactions. Hence, the predicted IAXO sensitivity in dark green is based on the interactions with both electrons and photons. }
	\label{fig:parameter_space}
\end{figure}

As any other QCD axion model, the one presented in this work is subject to constraints coming from its couplings to photons, electrons and nucleons.
Figure~\ref{fig:parameter_space} summarizes the bounds as a function of $f_a$ and in the usual $m_a$-$g_{a\gamma\gamma}$ parameter space.

The axion-photon coupling is tested by a combination of laboratory experiments and astrophysical observations.
The negative results from the solar axion search performed by CAST~\cite{Anastassopoulos:2017ftl}, together with the absence of exotic cooling in Horizontal Branch (HB) stars in globular clusters~\cite{Ayala:2014pea}, constrain the photon coupling to be $g_{a\gamma\gamma}\lesssim 7\times 10^{-11}$.
This translates into a bound on $f_a$ which is different for the leptonic and the non-leptonic models due to their different values of $E/N$ (note that in the leptonic case, this bound only considers the axion flux generated by interactions with photons and not electrons; it is therefore overly conservative and only illustrative).
Furthermore, the proposed successor of CAST, the IAXO solar telescope~\cite{Armengaud:2019uso}, is expected to improve this limit by over an order of magnitude in its upgraded version IAXO+.
Unfortunately, the loss of sensitivity for axion masses above $\sim 10^{-2}\eV$, combined with an accidental cancellation\footnote{The low-energy QCD axion-photon coupling is a combination of the model-dependent anomaly coefficient $E/N$ and a model-independent piece coming from hadronic contributions~\cite{diCortona:2015ldu}.} in the photon coupling, renders the non-leptonic model out of reach for IAXO+, as can be seen in the right panel of figure~\ref{fig:parameter_space}.
The model with tree-level couplings to SM leptons, however, is within reach of the sensitivity forecast, as can be appreciated in the left panel of figure~\ref{fig:parameter_space}.
The reason for this is the enhanced production of axions in the solar interior caused by the existence of a large electron coupling.
Finally, haloscopes searches have an exquisite sensitivity to axions in the $1-100\,\mu\mathrm{eV}$ range, provided these particles were to make up the dark matter of the Universe.
The currently excluded region, as compiled in~\cite{Irastorza:2018dyq,Beacham:2019nyx} and including new ADMX data~\cite{Du:2018uak,Braine:2019fqb}, is shaded in yellow in the right panel of Fig.~\ref{fig:parameter_space}.

Axions coupling to electrons induce additional cooling mechanisms in stars.
Currently, the strongest limits come from observations of the brightness of the tip of the red-giant branch (RGB) in globular clusters, which exclude values of $f_a/c_e\geq 3.9\times 10^9\GeV$, as derived in~\cite{Capozzi:2020cbu}.\footnote{Slightly weaker limits come from measurements of the R parameter in globular clusters~\cite{Giannotti:2015kwo,Hoof:2018ieb} and from observations of white dwarfs, which exclude values of $f_a/c_e\geq 1.9\times 10^9\GeV$, as derived in~\cite{Bertolami:2014wua}, although the data seems to prefer some amount of extra cooling compatible with a non-vanishing electron (and possibly photon) coupling~\cite{Giannotti:2015kwo,Giannotti:2017hny}.}
The axion-electron coupling is present at tree level in the leptonic variant of the model, and only loop induced (and therefore suppressed) in the non-leptonic one.
As a consequence, the previously mentioned bound is only competitive for the former variation of the model, and suppressed by a loop factor~\cite{Srednicki:1985xd,DiLuzio:2020wdo} $\sim \alphaEM^2/(\pi^2)\log \sim 10^{-4}$ in the latter.
Indeed, in the presence of tree-level leptonic couplings, RGB observations currently place the strongest constraints on the model.
As mentioned above, these are only expected to be improved once IAXO has reached its full sensitivity.

Finally, we discuss the constraints arising from the effective coupling of axions to nucleons, which arises at low energies from the interactions with gluons and quarks.
It is customary to define it in a way analogous to the other fermionic couplings as
\begin{equation}
    \mathcal{L}\supset \frac{\partial_\mu a}{2f_a}\sum_{N=p,n} c_N\,\bar{N}\gamma^\mu\gamma_5 N\,.
\end{equation}
The expression for the coefficients $c_N$ in terms of the quark couplings can be found in~\cite{diCortona:2015ldu}, which shows good agreement with a recent reevaluation~\cite{Vonk:2020zfh}.
For the model at hand, they are $c_p = -0.39875$ and $c_n = 0.05125$. These couplings are best tested by studying their impact on the extreme dynamics of the proto-neutron star that forms in the course of a core-collapse supernova.
Building on the seminal reference~\cite{Raffelt:1996wa}, the most recent limits are given in~\cite{Chang:2018rso,Carenza:2019pxu,Ertas:2020xcc}. 
While there is overall agreement given the uncertainties, the bound of~\cite{Carenza:2019pxu} on $f_a$ is a factor of $\sim 3.5$ stronger than the most conservative ones in~\cite{Chang:2018rso,Ertas:2020xcc}.
In figure~\ref{fig:parameter_space}, we choose to use vertical hatching to showcase the spread in the different evaluations of the limit.

\subsection{\texorpdfstring{Constraints and opportunities from $K^+\rightarrow \pi^+ + a$}{Constraints and opportunities from K to pi a}}

\begin{figure*}[t]
	\centering
	\includegraphics[width=0.6\textwidth]{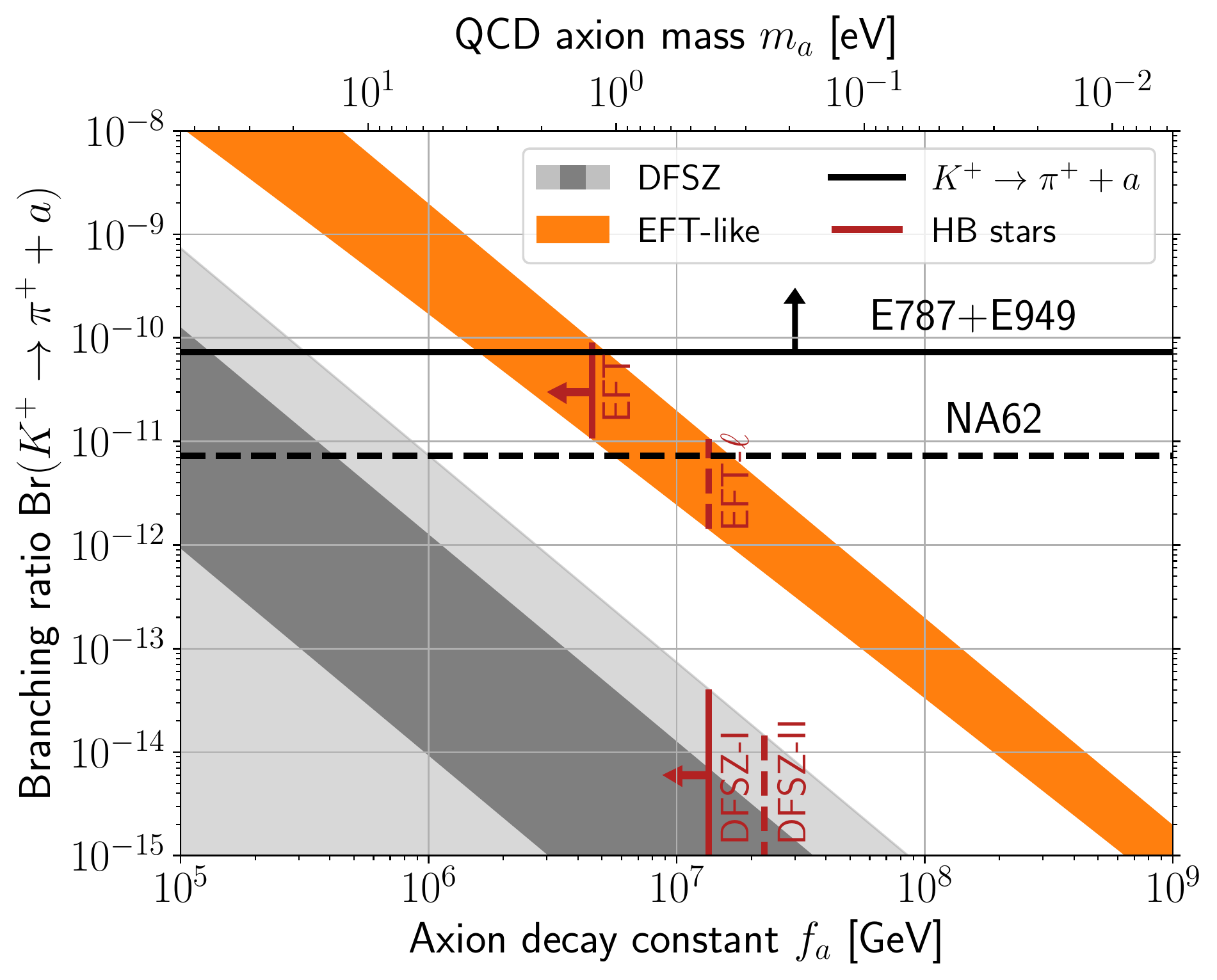}
	\vspace{-0.4cm}
	\caption{Comparison of the branching ratio of $K^+\rightarrow\pi^+ + a$ predicted in the EFT-like model (orange band) and in the DFSZ model (grey bands).
	For the EFT-like model, we assume a universal quark coupling $c_q = 1/6$ and different values of the cutoff scale $\Lambda$ (between $30f_a$ and $M_\text{Pl}$ as discussed in the text).
	The DFSZ bands are obtained for a charged Higgs boson mass $m_{H^\pm}=800\GeV$ and $1<\tan\beta<5$ (darker band) or $0.25<\tan\beta<170$ (lighter band).
	The horizontal black lines show the existing constraint by the E787 and E949 experiments~\cite{Adler:2008zza} (solid line) and the expected reach of NA62~\cite{Fantechi:2014hqa} (dashed line).
	The vertical red lines describe the bounds set by HB stars, which depend on the photon coupling and therefore differ for the EFT/EFT-$\ell$ and DFSZ-I/II models.}
	\label{fig:K_pi_a_EFT}
\end{figure*}

Perhaps the most important feature of the model at hand is that quark-flavour violating transitions $q\rightarrow q^\prime + a$, which are induced at the one-loop level, are enhanced by a large logarithm as discussed in the previous sections.
This means that processes involving such transitions are particularly suitable for testing this QCD axion model.
Among all such processes, the $s\rightarrow d + a$ decay is the one that offers the best experimental perspectives, thanks to the very precise measurements that can be performed at kaon facilities.

The predicted decay rate for $K^+ \rightarrow \pi^+ + a$ is given in Eq.~\eqref{eq:K_pi_a_new_model}, where we can see that the leading-log piece matches the EFT result with the identification $\Lambda=m_F$.
The corresponding branching ratio thus depends on the value of $m_F$.
In principle, the only requirement for the mass of the heavy $F$-fermions is for them to be above the axion decay constant, $m_F\geq f_a$.
However, given that SM Yukawa couplings are proportional to $\braket{\phi}/m_F$, it is possible to obtain some rough limits on the largest possible values of $m_F$ that reproduce the SM quark masses.
Assuming perturbativity of all dimensionless couplings, reproducing the top quark Yukawa requires at least one of the $F$ fermions to have a mass $m_{F_j}\lesssim 10^3 f_a$ for some $j$. However, the other $F$ fermions could be much heavier, and without a complete construction of the $3$-dimensional coupling matrices $\alpha$, $\beta$, and $\gamma$, it is not possible to point down the contribution of each of the $F$ fermions to $s\rightarrow d + a$ and therefore to the loop cutoff $\Lambda$.
Therefore, and although it is reasonable to expect that $\Lambda\lesssim 10^3 f_a$, one cannot rule out the possibility of larger values.

Taking into account the discussion above, we can derive the bounds and prospects for this model in past and present kaon facilities.
The $K^+\rightarrow \pi^+ + a$ branching ratio as a function of $f_a$ for different values of $\Lambda$ corresponds to the orange lines in figure~\ref{fig:K_pi_a_EFT}.
A robust exclusion can be placed by assuming 
a conservative value for the cutoff, $\Lambda=30\,f_a$. Note that this choice is equivalent to $\epsilon'' \simeq 0.2$ and hence corresponds to a slight scale separation between $\braket{\phi}$ and $\braket{\Sigma}$. This choice may appear in tension with the assumption $\epsilon'' \ll 1$ employed in our discussion so far. However, as shown in appendix~\ref{app:diagonalization} for an explicit choice of coupling matrices, our expression for the relevant effective $s \to d + a$ coefficient derived in eq.~\eqref{eq:hdsUVCoeffWqF} still holds to a very good degree in this case.
With this, the E787 and E949 experiments rule out $f_a < 1.6\times 10^{6}\GeV$, and NA62 could push this up to $f_a < 5.6\times 10^{6}\GeV$ (left panel of figure~\ref{fig:parameter_space}, dark orange) assuming an order of magnitude better sensitivity in the branching ratio.
That said, given that $\Lambda$ could be much larger, the discovery potential for these experiments extends somewhat further. Only in this way, by identifying $m_F$ as the scale appearing in the logarithm and treating it as an independent parameter from the PQ breaking scale $\Braket{\phi}$, can the NA62 sensitivity go beyond the robust limits set by the cooling of HB stars for the EFT scenario.
For the most extreme value of $\Lambda=M_{\mathrm{pl}}$, the potential reach of NA62 extends to $f_a\lesssim 1.6\times 10^{7}\GeV$, which as can be seen in the left panel of figure~\ref{fig:parameter_space} (light orange) comes closer to the SN1987a limit.

\section{Conclusions and Discussion}\label{sec:conclusions}
Effective Field Theory tools are a cornerstone of studies of the QCD axion and general axion-like particles. 
However, the validity of an EFT is limited by the existence of a cutoff. This is evident in log-divergent loop processes, where the cutoff explicitly appears in the final result. It is tempting to infer the cutoff scale from the scale set by the dimensionful coupling constants of the higher dimensional operators in the axion EFT. Yet, one may wonder if this choice is validated by embeddings into simple axion models with renormalizable couplings.
In this work, we have compared calculations made using an EFT setup with those obtained using complete QCD axion models.
Our results show that matching EFT predictions to full models is far from trivial when loop processes are involved.

Owing to its experimental relevance, we have chosen to focus on the rare decay $K^+\rightarrow \pi^+ + a$ for our study.
We have shown that the $K^+\rightarrow \pi^+ + a$ decay rate in the most popular QCD axion models can differ by orders of magnitude from naive EFT predictions, similar to what was found in~\cite{Choi:2017gpf} based on effective models that respect electroweak symmetry.
The origin of this discrepancy lies in the UV sensitivity that is induced by the logarithmic divergence of this loop-induced decay.
For DFSZ models, it is the existence of new degrees of freedom much below the PQ symmetry-breaking scale which violates the assumptions on which the EFT loop calculation is based.
On the other hand, for KSVZ (or hadronic) QCD axion models, which do not have extra degrees of freedom below the PQ scale, suitable tree-level couplings to SM fermions are absent.

The above considerations appear to question the existence of a complete axion model which can successfully reproduce the EFT result, i.e.~the large logarithmic enhancement found in the EFT calculation for loop-induced rare decays involving an axion.
We have addressed this issue by explicitly constructing such a model, whose low-energy dynamics are dictated by the Lagrangian in Eq.~\eqref{eq:new_model}.
Crucially, this new model allows for tree-level axion-SM fermion couplings without introducing any extra degree of freedom below the PQ symmetry-breaking scale other than the axion itself.

While very minimal at energies below $f_a$, the EFT-inspired QCD axion model constructed in this work presents rich dynamics above the PQ scale.
In addition to the PQ complex scalar, three generations of heavy coloured fermions have to be introduced in order to obtain a renormalizable model.
Through mixing with SM quarks, these heavy fermions induce modifications of the low-energy axion couplings.
Ultimately, it is the presence of these additional states which tames the logarithmic divergence of the $K^+\rightarrow \pi^+ + a$ rate, therefore providing a physically-motivated cutoff as opposed to the artificial one that is employed in the EFT.

The implementation that we have followed in this work is only one of the possibilities to UV complete the low-energy Lagrangian Eq.~\eqref{eq:new_model}.
Some other scenarios are delineated in appendix~\ref{sec:app_models}.
These may feature qualitatively different dynamics for the interactions of the QCD axion with the SM fields.
One example that is briefly described in appendix~\ref{app:diagonalization} is the alternative that the mass of the heavy quarks lies close (or even below) the PQ-breaking scale.
In this case, the QCD axion generically enjoys tree-level flavour-violating interactions with the SM quarks, which significantly change the flavour phenomenology (see, e.g.~\cite{MartinCamalich:2020dfe}).
Other interesting possibilities, like potential CP-violating axion interactions or connections between the axion dynamics and the flavour puzzle are interesting avenues of further research.
Finally, the large logarithms appearing together with the top-quark Yukawa suggests upgrading our loop calculations with RG improvements along the lines of~\cite{Choi:2017gpf,Chala:2020wvs,Bauer:2020jbp} to achieve better accuracy in the phenomenological limits.

The EFT-inspired QCD axion model has a rich phenomenology, the main features of which are summarized in figure~\ref{fig:parameter_space}. 
As is also the case in other axion models, astrophysical observables currently place the strongest constraints on axions with sub-eV masses.
In particular, the duration of the neutrino burst in the supernova 1987a excludes decay constants $f_a\lesssim 10^8$~GeV.
However, owing to the logarithmic enhancement of the $K^+\rightarrow\pi^++a$ rate, $K^+\rightarrow\pi^+ + \mathrm{inv}$ searches at the NA62 experiment have the potential to supersede all astrophysical bounds except for the SN1987a one.
Given the uncertainties in the modelling of axion emission in supernovae,\footnote{Some authors~\cite{Bar:2019ifz} have proposed an alternative scenario for the SN1987a explosion, which would render the SN bound on axions invalid. However,~\cite{Cigan:2019shp,Page:2020gsx} argue that recent hints towards the existence of a neutron star remnant in the site of the explosion favour the standard hypothesis.} a confirmation of the exclusion from a terrestrial experiment is highly desirable.
Additionally, the presence of additional colour-charged fermions with PQ charge could act to decrease the value of the QCD anomaly coefficient $N$, in which case the SN1987a limit would be weakened compared to the NA62 one.

The possibility to couple the axion to the SM leptons in a similar way as we have done for quarks offers even more exciting possibilities.
Due to the particular balance between the photon and the electron couplings in our model, the IAXO helioscope (in its upgraded version IAXO+) is expected to become the most sensitive probe of sub-eV mass QCD axions, with a reach close to $10^9$~GeV in $f_a$ (corresponding to a mass as small as $0.01$~eV).
This would supersede all astrophysical limits, including the previously mentioned SN1987a one and the one derived from observations of the brightness of the tip of the red giant branch in globular clusters.

The dark matter phenomenology of the investigated models is similar to those in the DFSZ and KSVZ benchmark models, with two notable differences. One is that the photon coupling can be quite small due to an electromagnetic anomaly coefficient of $5/3$. Second, in the models that we have investigated the domain-wall number is greater than one, favouring the scenario where the PQ symmetry is broken during inflation and not restored afterwards. Alternatively, one may have an additional explicit breaking of the discrete $Z_{N}$ symmetry. While this can be useful to allow for the correct dark matter abundance at smaller values of the PQ scale $f_{a}$~\cite{Ringwald:2015dsf}, potentially accessible with experiments such as NA62, it may also lead to a certain degree of tuning.

The findings of this work showcase the necessity of developing a detailed understanding of the strengths and limitations of EFT tools in axion studies.
We have seen that it is far from trivial to construct a UV embedding of an axion EFT coupled to fermions that agrees with the leading log predictions for relevant experimental processes. More generally, this leads to the question of whether a generic axion EFT with a cutoff given by the (inverse of) the dimension-5 coupling constants can be obtained from a UV model. While our explicit construction answers this question in an affirmative manner for the specific case at hand, it also highlights potential issues such as the quite complicated structure of the model, the potential appearance of additional flavour-changing couplings as well as potentially non-perturbative Yukawas being required to reproduce the top-quark mass.\footnote{Problems in obtaining the top-quark mass while retaining perturbativity were already noted in~\cite{Choi:2017gpf}.} It therefore remains an intriguing question whether even mild extra assumptions may severely restrict the space of axions EFTs that are the low-energy manifestations of reasonable UV models, possibly giving rise to an \emph{ALPine swampland}.

\section*{Acknowledgements}
We thank Susanne Westhoff for helpful discussions. GA acknowledges support by a ``la Caixa" postgraduate fellowship from the Fundaci\'on ``la Caixa", by a McGill Trottier Chair Astrophysics Postdoctoral Fellowship, and by NSERC (Natural Sciences and Engineering Research Council, Canada).
FE and FK are funded by the Deutsche Forschungsgemeinschaft (DFG) through the Emmy Noether Grant No.\ KA 4662/1-1. LT is funded by the Graduiertenkolleg \textit{Particle physics beyond the Standard Model} (GRK 1940).

\appendix
\section{CP violation in pseudoscalar interactions}
\label{app:CPVInt}
In this appendix, we investigate under what conditions CP violation arises in the couplings studied in this work.
In particular, we focus on axion-quark, $W$-quark, and $Z$-quark interactions, where quark generally refers to a quark of any type, i.e.~SM or BSM.
The flavour-diagonal structure of photon and gluon interactions is not altered in our UV model compared to the SM and hence does not require any further comments.

\subsection*{Axion-quark interactions}
We begin with derivative axion-quark interactions, differentiating between flavour-violating and flavour-conserving operators. The 
latter can in full generality be written as 
\begin{align}\label{eq:app1_aq_diagonal}
(\partial_\mu a) \,\bar{q} \gamma^\mu (h^S + h^P \gamma_5) q\,.
\end{align}
Hermiticity of this operator restricts $h^S$ and $h^P$ to be real-valued.
A CP transformation leaves~\eqref{eq:app1_aq_diagonal} unchanged and hence this interaction does never violate CP.
On the other hand, the flavour-violating version of this operator and its hermitian conjugate read
\begin{align}
(\partial_\mu a) \,\bar{q} \gamma^\mu (h^S + h^P \gamma_5) q' + (\partial_\mu a) \,\bar{q}' \gamma^\mu (\left(h^S\right)^* + \left(h^P\right)^* \gamma_5) q\,.
\end{align}
Applying the CP operation on the first term, we obtain
\begin{align}
    (\partial_\mu a) \,\bar{q} \gamma^\mu (h^S + h^P \gamma_5) q'  \stackrel{\text{CP}}{\longrightarrow} (\partial_\mu a) \,\bar{q}' \gamma^\mu (h^S + h^P \gamma_5) q\,,
\end{align}
and hence CP violation is closely related to the imaginary parts of $h^S$ and $h^P$. Importantly, CP violation does not arise even if the axion simultaneously enjoys scalar and pseudoscalar couplings in the derivative basis, as long as both coupling constants remain real.
For example, in the EFT scenario in Eqs.~\eqref{eq:EffLagrds} and~\eqref{eq:effsdCoeff}, the CP violation is fully determined by the imaginary part of the CKM matrix elements.

\subsection*{Gauge boson-quark interactions}
We start with the $Z$ bosons-quark couplings.
For flavour-conserving interactions, we have
\begin{align}
    \bar{q} \gamma_\mu (V + A \gamma_5)Z^\mu q\;, 
\end{align}
where hermiticity requires both couplings $V$ and $A$ to be real-valued.
As a consequence, invariance under a CP transformation is automatically guaranteed.
In contrast, flavour-violating couplings involve
\begin{align}
    \bar{q} \gamma_\mu (V + A \gamma_5)Z^\mu q' + \bar{q}' \gamma_\mu(V^* + A^* \gamma_5)Z^\mu q \;.
\end{align}
A CP transformation on the first operator gives 
\begin{align}
   \bar{q} \gamma_\mu (V + A \gamma_5)Z^\mu q'  \stackrel{\text{CP}}{\longrightarrow} \bar{q}' \gamma_\mu (V + A \gamma_5)Z^\mu q\;,
\end{align}
and hence CP violation is again closely related to complex-valued couplings, i.e.~phases of $V$ and/or $A$.
In our case the coupling structure for flavour-violating interactions of the $Z$ bosons is always left-handed (see Eq.~\eqref{eq:Zint}) and hence we have $V = -A$. CP violation therefore also amounts here to a global phase attached to the operator $\bar{q} \gamma_\mu Z^\mu P_L q'$.

The situation of $W$ bosons is simpler given that in general only left-handed fields couple to it.
The coupling structure are always of the type 
\begin{align}
    \bar{q} (V-A) \gamma_\mu W^\mu P_L q' + \text{h.c.}\;,
\end{align}
As is familiar from the CKM matrix in the SM, CP violation manifests itself when $(V-A)$ is complex-valued.

\section{Variations of the model}
\label{sec:app_models}
Instead of the specific UV completion in Eq.~\eqref{eq:highenergy_Lag}, we could have allowed either or both of the up- and down-type fields to couple to $\Phi^*$ instead of $\Phi$. In total, there are four possible ways to choose between $\Phi$ and $\Phi^*$. The most general Lagrangian reads
\begin{align}
\begin{split}
	\mathcal{L} \supset \quad &-\alpha^{u} \bar{Q}_{L} \tilde{H} F^u_{R} - \beta^{u} \bar{F}^u_{L} \Phi^{(*)} u_{R}   + \mathrm{h.c.}\\
	&- \alpha^{d} \bar{Q}_{L} H F^d_{R} - \beta^{d} \bar{F}^d_{L} \Phi^{(*)} d_{R}   + \mathrm{h.c.}\;.
\label{eq:highenergy_Lag_appendix}
\end{split}
\end{align}
Since $\Phi$ is a SM scalar, this choice does not influence the charges of any fields under the SM gauge groups. However, each case results in different relations between PQ charges of the different fermions.
The terms in Eq.~\eqref{eq:highenergy_Lag_appendix} enforce the conditions
\begin{align}
    \chi_{Q_L}-\chi_{F_R^u}&=\chi_{Q_L}-\chi_{F_R^d}=0\, , \\
    \chi_{F_L^u}-\chi_{u_R}&=\pm 1\, , \\
    \chi_{F_L^d}-\chi_{d_R}&=\pm 1 \, ,
\end{align}
where the upper sign corresponds to $\Phi$ and the lower sign to $\Phi^*$ in the Lagrangian~\eqref{eq:highenergy_Lag_appendix}.
Because all $q$ and $F$ quarks are in the fundamental representation of SU(3), the sum of the four equations above appears in the anomaly coefficient,
\begin{equation}
N = \sum_f  (\chi_{f_L}-\chi_{f_R}) T(R_f) \, .
\end{equation}
Hence, $|N|=3$ as long as we choose either $\Phi$ or $\Phi^*$ for both up- and down-type quarks. 
Otherwise, the two contributions cancel, there is no QCD anomaly and $a$ is not a QCD axion.
We therefore do not further pursue that last possibility.
Therefore, PQ charges of up- and down-type fields are necessarily identical, and we drop the corresponding labels in what follows.

After this, the only other choice is the PQ charge assignment of the $F$ quarks.
This determines the possible origin of the $F$-quark masses, which can either originate from the PQ-scalar $\Phi$ or from the VEV of an additional spurion field $\Sigma$ without PQ charge.
The combination of the two possible charge assignments of $q$ and $F$ quarks results in four different models:
\begin{enumerate}
\setlength{\itemindent}{-0.2cm}
\item {\bf Use $\Phi$ in \eqref{eq:highenergy_Lag_appendix} and generate $F$ masses from $\braket{\Sigma}$.}
The additional mass term is 
\begin{equation}
\mathcal{L} \supset - \lambda \frac{\braket{\Sigma}}{\sqrt{2}} \bar{F}_L F_R + \mathrm{h.c.} \;,
\label{eq:spurion_mass_appendix}
\end{equation}
which implies
\begin{equation}
    \chi_{F_L} - \chi_{F_R} = 0  \quad \Rightarrow \quad \chi_{Q_L}-\chi_{q_R}=1.
\end{equation}
$F$ is vector-like with respect to the PQ symmetry and does not contribute to the anomaly. This is exactly the model which is analyzed in great detail in sections~\ref{sec:new_model} and \ref{sec:phenomenology}.

\item {\bf Use $\Phi^*$ in \eqref{eq:highenergy_Lag_appendix} and generate $F$ masses from $\braket{\Sigma}$.}
Using the same mass term as in~\eqref{eq:spurion_mass_appendix}, we get
\begin{equation}
    \chi_{F_L} - \chi_{F_R} = 0 \quad \Rightarrow \quad \chi_{Q_L}-\chi_{q_R}=-1 \;. 
\end{equation}
This option is of course equivalent to the first one after a redefinition of $\Phi\leftrightarrow\Phi^*$.

\item {\bf Use $\Phi$ in \eqref{eq:highenergy_Lag_appendix} and generate $F$ masses from $\braket{\Phi}$.}
The mass term of $F$ quarks in this case is given by
\begin{equation}
\mathcal{L} \supset - \lambda \frac{\braket{\phi}}{\sqrt{2}} \bar{F}_L F_R + \mathrm{h.c.} \;.
\label{eq:PQ_mass_appendix}
\end{equation}
And consequently,
\begin{equation}
    \chi_{F_L} - \chi_{F_R} = 1  \quad \Rightarrow \quad \chi_{Q_L}-\chi_{q_R}=0.
\end{equation}
Only $F$ carries an axial PQ charge, which means that no symmetry forbids a Yukawa term as in the SM,
    \begin{equation}
	\mathcal{L}\supset -\gamma^u \bar{Q}_L \tilde{H} u_R
	-\gamma^d \bar{Q}_L H d_R + \mathrm{h.c.}\; .
	\end{equation}
The existence of these couplings slightly modifies the diagonalization of the mass matrix.
Even though the $q$ fields are not charged under PQ in the UV model, they inherit axion couplings through the mixing with the heavy quarks during the diagonalization. These are however parametrically suppressed by $\epsilon^2=\frac{v^2}{\braket{\phi}^2}$.

\item {\bf Use $\Phi^*$ in \eqref{eq:highenergy_Lag_appendix} and generate $F$ masses from $\braket{\Phi}$.}
Using the same mass term as in~\eqref{eq:PQ_mass_appendix}, we get
\begin{equation}
    \chi_{F_L} - \chi_{F_R} = 1 \quad \Rightarrow \quad \chi_{Q_L}-\chi_{q_R}=-2 \;. 
\end{equation}
Both $q$ and $F$ quarks have axial PQ charges and therefore tree-level axion couplings in this variation of the model.
\end{enumerate}

In each of the cases above, all axial charges of all combinations of SM and $F$ quarks are fixed.
The only remaining freedom is a shift of all charges by an arbitrary constant, which can be used to fix one vector-like charge. 

Note that it only makes sense to work with the effective Lagrangian Eq.~\eqref{eq:new_model} in the first (or the equivalent second model) when $\braket{\phi}\ll m_{F_i}$.
Otherwise, the $F$ fields cannot be integrated out at any scale above PQ symmetry breaking as is done in the main text.
In models 3 and 4, a full diagonalization including all $q$ and $F$ quarks along the lines of what is done in appendix~\ref{app:diagonalization} is always necessary.

\section{Mass diagonalization and axion interactions}
\label{app:diagonalization}
This appendix contains the full diagonalization procedure of the quark mass matrix as well as the resulting couplings of the axion and light and heavy quarks. For convenience, this includes some repetitions of intermediate steps and results which are also included in section~\ref{sec:diagonalization}.
\subsection*{Absorbing the axion field into the quarks}
To have a field basis in which all fields are mass eigenstates, we have to diagonalize the mass matrix given by
\begin{equation}
\label{eq:MassMatr_appendix}
M = \frac{\braket{\Sigma}}{\sqrt{2}} \begin{pmatrix} 0 & \epsilon \epsilon'\alpha\\ \epsilon' \beta e^{i a/\langle \Phi \rangle}& \lambda\end{pmatrix}\,,
\end{equation}
with the two expansion parameters 
\begin{equation}
\epsilon = \frac{v}{\braket{\phi}}
    \quad \mathrm{and} \quad \epsilon' = \frac{\braket{\phi}}{\braket{\Sigma}}\ .
\end{equation}
As in the main text, we split the diagonalization into two parts, of which only the first one depends on the axion field. 
We start by absorbing the axion dependence of the quark mass matrix into the right-handed quarks,
\begin{align}
\label{eq:ChiralRotAx_appendix}
u_R\rightarrow e^{-\frac{i a}{\braket{\phi}}} u_R \, ,\quad d_R\rightarrow e^{-\frac{i a}{\braket{\phi}}} d_R\,.
\end{align}
This transformation removes the axion field $a$ from $M$ in Eq.~\eqref{eq:MassMatr_appendix} while the quark kinetic terms generate derivative couplings of the axion to right-handed quarks,
\begin{equation}
    \bar{q}_Ri\slashed{\partial}q_R \rightarrow \bar{q}_Ri\slashed{\partial}q_R + \frac{\partial_\mu a}{\braket{\phi}} \bar{q}_R\gamma^\mu q_R \,,
    \label{eq:axion_coupling_before_diag_appendix}
\end{equation}
where $q$ stands for both up- and down-type fields.
The path integral measure is not invariant under this transformation and anomalous interaction terms between the axion and gauge bosons arise.
These are~\cite{Adler:1969gk,Bell:1969ts,Bardeen:1969md,Peccei:1977hh,Peccei:1977ur}
\begin{align}
\begin{split}
\label{eq:AnomGluPhot_appendix}
  \mathcal{L} \supset &- 2N\cdot\frac{\alpha_s}{16\pi \braket{\phi}}\, a\, \epsilon^{\mu\nu\alpha\beta} G^a_{\mu\nu} G^a_{\alpha\beta} - E \cdot \frac{\alphaEM}{8\pi \braket{\phi}}\,a\,\epsilon^{\mu\nu\alpha\beta} F_{\mu\nu} F_{\alpha\beta}\\
 & + E \cdot \frac{\alphaEM}{4\pi} \frac{s_W}{c_W} \frac{a}{\braket{\phi}} \epsilon^{\mu\nu\alpha\beta} F_{\mu\nu} Z_{\alpha\beta} - E \cdot \frac{\alphaEM}{8\pi} \frac{s_W^2}{c_W^2} \frac{a}{\braket{\phi}} \epsilon^{\mu\nu\alpha\beta} Z_{\mu\nu} Z_{\alpha\beta}\,.
 \end{split}
\end{align}
With Eq.~\eqref{eq:axion_coupling_before_diag_appendix} we have chosen to apply axion-dependent rotations only to right-handed fields which do not couple to $SU(2)$ gauge fields. Therefore, $W$-couplings are absent.
$N$ and $E$ are the anomaly coefficients defined as 
\begin{align}
\label{eq:anomalycoeff_appendix}
N &= \sum_f  (\chi_{f_L}-\chi_{f_R}) T(R_f)=3, \\
E &= \sum_f  (\chi_{f_L}-\chi_{f_R}) Q_f^2.
\end{align}
$T(R_f)$ is the Dynkin index of the SU(3) representation and $Q_f$ are the electric charges.
In our model, $N=3$ and $E=5$~or~8 depending on whether the leptons are also charged under PQ.
We normalize the axion gluon coupling by defining
\begin{equation}
    f_a \equiv  \frac{1}{N_{\rm DW}}\braket{\phi}\,,
\end{equation}
where $N_\text{DW}\equiv 2N=6$ is the domain wall number counting the number of inequivalent vacua in the QCD-induced axion potential.
This means that we can write the gauge-boson coupling terms as
\begin{align}
\begin{split}
\mathcal{L} \supset &- \frac{\alpha_s}{16\pi f_a}\, a\, \epsilon^{\mu\nu\alpha\beta} G^a_{\mu\nu} G^a_{\alpha\beta} - \frac{E}{N} \cdot \frac{\alphaEM}{16\pi f_a}\,a\,\epsilon^{\mu\nu\alpha\beta} F_{\mu\nu} F_{\alpha\beta}\\
 & + \frac{E}{N} \cdot \frac{\alphaEM}{8\pi} \frac{s_W}{c_W} \frac{a}{f_a} \epsilon^{\mu\nu\alpha\beta} F_{\mu\nu} Z_{\alpha\beta} - \frac{E}{N} \cdot \frac{\alphaEM}{16\pi} \frac{s_W^2}{c_W^2} \frac{a}{f_a} \epsilon^{\mu\nu\alpha\beta} Z_{\mu\nu} Z_{\alpha\beta}\,.
 \label{eq:axion_boson_coup_app}
\end{split}
\end{align}
Comparing to the EFT setup defined in Eq.~\eqref{eq:EFT_Lagrangian}, we can identify $c_{gg} = 1$, $c_{WW} = 0$, $c_{\gamma\gamma} =E/N$, $c_{ZZ} =\tan(\theta_W)^2\ E/N$ and $c_{\gamma Z} =-2 \tan(\theta_W)\ E/N$.

We have eliminated the $a$ dependence of the mass matrix by the transformation in Eq.~\eqref{eq:ChiralRotAx_appendix}. The subsequent steps in the diagonalization procedure also contain axial transformations of the quark fields, which are however not axion dependent and only result in the usual shift of the theta angle of QCD $\theta_{\mathrm{QCD}}$
\begin{equation}
	\theta_{\mathrm{QCD}} \rightarrow \theta_{\mathrm{QCD}} + \arg (\det(\left. M \right|_{a=0})),
\end{equation}
therefore simply displacing the location of the minimum of the axion potential.

\subsection*{Mass diagonalization}
We continue with fully diagonalizing the matrices $M^u$ and $M^d$ by unitary transformations of left- and right handed fields. As before, we drop the labels $u/d$. At every point, we consider all three generations of SM and $F$ quarks even though the indices are also omitted.
	
Because $MM^\dagger$ is hermitian, it can be diagonalized by a unitary matrix $U$,
\begin{equation}
U^\dagger MM^\dagger U=   \Lambda^2 ,
\end{equation}
where $\Lambda^2$ is a diagonal matrix with only real positive eigenvalues. We then define the unitary matrix $S= M^\dagger U\Lambda^{-1}$ and a unitary transformation of fields
\begin{align}
\begin{pmatrix}q_L\\F_L\\\end{pmatrix} &\rightarrow U  \begin{pmatrix}q_L\\F_L\\\end{pmatrix}, \\
\begin{pmatrix}q_R\\F_R\\\end{pmatrix} &\rightarrow S  \begin{pmatrix}q_R\\F_R\\\end{pmatrix}.
\end{align}
Here, $q$ can be either $u$ or $d$. This transformation diagonalizes the mass matrix
\begin{equation}
	U^\dagger M S =	U^\dagger M M^\dagger U \Lambda^{-1} = \Lambda.
\end{equation}

We now need to perturbatively find $U$, which diagonalizes
\begin{equation}
M M^\dagger = \frac{\braket{\Sigma}^2}{2} \begin{pmatrix} \epsilon^2 \epsilon'^2 (\alpha\alpha^\dagger) & \epsilon \epsilon'(\alpha\lambda^\dagger)\\ \epsilon\epsilon'(\lambda\alpha^\dagger) & (\lambda\lambda^\dagger + \epsilon'^2\beta\beta^\dagger )	\end{pmatrix} \equiv \frac{\braket{\Sigma}^2}{2} \begin{pmatrix} (\epsilon\epsilon')^2 \delta & \epsilon\epsilon' \mu\\ 
\epsilon\epsilon' \mu^\dagger & \xi
\end{pmatrix}.
\end{equation}
In the last step we have defined the matrices $\delta$, $\mu$ and $\xi$ for notational convenience, of which $\delta$ and $\xi$ are hermitian. To quadratic order in $\epsilon$, $U$ is given by
\begin{equation}
\label{eq:UMatrixMod1}
U= \begin{pmatrix}
(-1+\frac{(\epsilon\epsilon')^2}{2}\mu \xi^{-2}\mu^\dagger)U_\delta & \epsilon\epsilon'\mu\xi^{-1}U_\xi \\ \epsilon\epsilon'\xi^{-1} \mu^\dagger U_\delta & (1-\frac{(\epsilon\epsilon')^2}{2}\xi^{-1}\mu^\dagger \mu \xi^{-1})U_\xi\\
\end{pmatrix} +\mathcal{O}(\epsilon^3)\,,
\end{equation}
where the unitary matrices $U_\delta$ and $U_\xi$ are defined by the property that they diagonalize hermitian matrices to give the $q$ and $F$ masses. Respectively,
\begin{align}
\label{eq:QMassMod1}
  M_q^2&=\text{diag}(m^2_{q_1},m^2_{q_2},m^2_{q_3})=\left[U_\delta^\dagger (\delta - \mu \xi^{-1}\mu^\dagger)U_\delta \; (\epsilon\epsilon')^2 + \mathcal{O}(\epsilon^3)\right] \frac{\braket{\Sigma}^2}{2}, \\
\label{eq:FMassMod1}
M_F^2&=\text{diag} (m^2_{F_1},m^2_{F_2},m^2_{F_3})=\left[U_\xi^\dagger(\xi +\frac{(\epsilon\epsilon')^2}{2}(\mu^\dagger \mu \xi^{-1}+\xi^{-1}\mu^\dagger\mu)) U_\xi+\mathcal{O}(\epsilon^3)\right] \frac{\braket{\Sigma}^2}{2} .
\end{align}
\vspace{-0.1cm}
In the region of interest for NA62, this expansion in $\epsilon$ can be done safely, as $\epsilon\lesssim 10^{-4}$. 

In order to map onto the effective description and to avoid sizable corrections to the effective description in~\eqref{eq:new_model}, we need to ensure that $m_{F_i} \gg \braket{\phi}$ such that the $F$ quarks can be integrated out at some scale above the PQ one. In other words, we need the scale separation between the mass of the $F$ quarks and the PQ scale to be sufficiently large.
This condition can be written as
\begin{align}
    &\min_i m_{F_i}^2 = \min \textrm{eig}(\xi) \Braket{\Sigma}^2 \gg  \Braket{\phi}^2\\
    \Rightarrow \quad& 1 \gtrsim \min\textrm{eig}(\xi)=\min\textrm{eig}(\lambda\lambda^\dagger +\epsilon'^2\beta \beta^\dagger) \gg \epsilon'^2\\
    \Rightarrow \quad& 1 \gtrsim \min\textrm{eig}(\lambda\lambda^\dagger)  \gg \epsilon'^2 \;.
\end{align}
\vspace{-0.1cm}
Here, $\min \textrm{eig}$ denotes the smallest eigenvalue of a matrix. In the second  and third lines, the size of the eigenvalues are constrained by perturbativity. We see that $\epsilon' \ll 1$ is only a necessary condition for the $F$ quarks to be much heavier than the PQ scale, while the last line is a sufficient condition. We define one more expansion parameter $\epsilon''$ as 
\begin{equation}
    \epsilon''^2 = \frac{\epsilon'^2}{\min \textrm{eig}(\lambda\lambda^\dagger)} \simeq  \frac{\Braket{\phi}^2}{\min_i m_{F_i}^2} \;.
\end{equation}
\vspace{-0.1cm}
When $\epsilon'' \ll 1$, we can expand  $\xi^{-1}$ as
\begin{align}
\label{eq:xi_inverse}
\xi^{-1}&=(\lambda\lambda^\dagger + \epsilon'^2 \beta\beta^\dagger)^{-1} \\
&=\lambda^{\dagger-1}\;(1 + \epsilon'^2 \lambda^{-1}\beta\beta^\dagger\lambda^{\dagger-1})^{-1}\;\lambda^{-1}\\
&=\lambda^{\dagger-1}\;\sum_n (-\epsilon'^2 \lambda^{-1}\beta\beta^\dagger\lambda^{\dagger-1})^n\;\lambda^{-1} \label{eq:xi_inverse_higherord}\\
&=\lambda^{\dagger-1}\;(1-\epsilon'^2\lambda^{-1}\beta\beta^\dagger\lambda^{\dagger-1}+ \mathcal{O}(\epsilon''^4))\;\lambda^{-1} \;,
\end{align}
where we have used a Neumann series from the second to the third line and assumed that eigenvalues of $\beta\beta^\dagger$ are at most of order 1.
Inserting the leading-order result in $\epsilon''$ into~\eqref{eq:QMassMod1}, we find for the light quark masses
\begin{equation}
\label{eq:appQMasses}
   M_q^2= \text{diag}(m^2_{q_1},m^2_{q_2},m^2_{q_3})\simeq
    U_\delta^\dagger\alpha\lambda^{-1}\beta\beta^\dagger\lambda^{\dagger -1} \alpha^\dagger U_\delta \; \;\epsilon^2\epsilon'^4 \frac{\braket{\Sigma}^2}{2} \simeq \mathcal{ABA^\dagger} \;\epsilon^2\epsilon'^4 \frac{\braket{\Sigma}^2}{2} \;.
\end{equation}
In the last step, we have defined the coupling matrices
\begin{equation}
\label{eq:mathcalABDef}
    \mathcal{A} = U^\dagger_\delta \mu \xi^{-1} U_\xi = U^\dagger_\delta \alpha \lambda^{-1} U_\xi + \mathcal{O}(\epsilon''^2) \quad \textrm{and}\quad \mathcal{B} = U_\xi^\dagger  \beta \beta^\dagger U_\xi\;.
\end{equation}
\newpage
\subsection*{Axion-quark interactions}
Because the axion field was absorbed entirely into the right-handed fields, the derivative terms with left-handed quark fields are not affected by the unitary transformation $U$. However, the derivative axion coupling to right-handed fields as in Eq.~\eqref{eq:axion_coupling_before_diag_appendix} does not transform trivially,
\begin{align}
\label{eq:quark-axion-int-app}
 \frac{\partial_\mu a}{\braket{\phi}}\begin{pmatrix}
\bar{q}_R & \bar{F}_R
\end{pmatrix}\;\gamma^\mu \begin{pmatrix}
\mathds{1} & 0 \\ 0 & 0
\end{pmatrix} \begin{pmatrix} q_R \\ F_R \end{pmatrix} \rightarrow
\frac{\partial_\mu a}{\braket{\phi}}\begin{pmatrix}
\bar{q}_R & \bar{F}_R
\end{pmatrix}\;\gamma^\mu S^\dagger\begin{pmatrix}
\mathds{1} & 0 \\ 0 & 0
\end{pmatrix}S \begin{pmatrix} q_R \\ F_R \end{pmatrix} \,.
\end{align}
To leading order in $\epsilon$, we obtain for $S$
\begin{align}
S &= M^\dagger U \Lambda^{-1} \\
\label{eq:SMatrixExpand}
&= \frac{\braket{\Sigma}}{\sqrt{2}} \left[\begin{pmatrix}
\epsilon \epsilon'^2\,\beta^\dagger \xi^{-1} \mu^\dagger U_\delta &  \epsilon' \, \beta^\dagger(1 - \frac{(\epsilon\epsilon')^2}{2} \xi^{-1} \mu^\dagger \mu \xi^{-1}) U_\xi \\ \epsilon \epsilon'\,(\lambda^{\dagger} \xi^{-1} \mu^\dagger -\alpha^\dagger) U_\delta   &  (\lambda^\dagger+(\epsilon \epsilon')^2 (\alpha^\dagger \mu \xi^{-1} -\frac{1}{2}\lambda^\dagger\xi^{-1}\mu^\dagger \mu \xi^{-1}))U_\xi\\
\end{pmatrix} +\mathcal{O}(\epsilon^3)\right] \Lambda^{-1}\,,
\end{align}
from which we can determine the relevant coupling matrix to leading order in $\epsilon$ as
\begin{align}
\frac{\slashed{\partial} a}{\braket{\phi}}S^\dagger\begin{pmatrix}
\mathds{1} & 0 \\ 0 & 0
\end{pmatrix}S
 =&\,\frac{\slashed{\partial}a}{\braket{\phi}}\,\Lambda^{-1}\frac{\braket{\Sigma}^2}{2}\,\begin{pmatrix}\epsilon^2 \epsilon'^4 U^\dagger_\delta \mu \xi^{-1} \beta \beta^\dagger \xi^{-1} \mu^\dagger U_\delta & \epsilon\epsilon'^3U^\dagger_\delta \mu \xi^{-1} \beta \beta^\dagger U_\xi\\ \epsilon\epsilon'^3U_\xi^\dagger  \beta \beta^\dagger \xi^{-1} \mu^{\dagger} U_\delta & \epsilon'^2 U_\xi^\dagger\beta \beta^\dagger U_\xi 
\end{pmatrix}\Lambda^{-1}
\\
\label{eq:AxQuarkIntNoExp_app}
=&\,\frac{\slashed{\partial}a}{\braket{\phi}}\, \Lambda^{-1}\frac{\braket{\Sigma}^2}{2}\begin{pmatrix}\epsilon^2\epsilon'^4 \mathcal{ABA^\dagger} & \epsilon\epsilon'^3 \mathcal{AB}\\ \epsilon\epsilon'^3 \mathcal{BA^\dagger} & \epsilon'^2\mathcal{B}\\ \end{pmatrix}\Lambda^{-1}\\
\simeq&\, (\slashed{\partial}a) \,\begin{pmatrix}
1 / \braket{\phi}& v\,\epsilon'\, M_q^{-1}\mathcal{A}\, \mathcal{B} M_F^{-1}/2\\ v\,\epsilon' M_F^{-1} \mathcal{B}\,\mathcal{A}^\dagger M_q^{-1}/2& \braket{\phi} M_F^{-1} \mathcal{B} M_F^{-1}/2\ \\
\end{pmatrix} \;,
\end{align}
where in the last line we have only kept the leading-order term in $\epsilon''$ as in Eq.~\eqref{eq:appQMasses}, which is of course only justified if $\epsilon''\ll 1$. So as one would expect, the coupling of SM quarks to axions is only strictly proportional to their masses if the heavy messenger fields are well separated from the PQ scale and can be integrated out. In the final step, we have also used that $\Lambda^{-1} = \text{diag}(M_q^{-1},M_F^{-1})$.  With this, we can finally write the axion couplings as
\begin{align}
\mathcal{L} &\supset 
	 \frac{\braket{\phi}}{2} \begin{pmatrix}\bar{q}_R & \bar{F}_R \\\end{pmatrix} (\slashed{\partial}a)\, \Lambda^{-1}\begin{pmatrix}
	(\epsilon\epsilon')^2 \mathcal{ABA^\dagger} & \epsilon\epsilon' \mathcal{AB}\\ \epsilon\epsilon' \mathcal{BA^\dagger} & \mathcal{B}\\
	\end{pmatrix}\Lambda^{-1} \begin{pmatrix} q_R \\ F_R\\ \end{pmatrix}
	\label{eq:AxDerivInt_appendix}\\
&\simeq\frac{1}{\braket{\phi}} \bar{q}_R \,(\slashed{\partial}a)\, q_R + \frac{\braket{\phi}}{2} \bar{F}_R(M_F^{-1} \mathcal{B} M_F^{-1} ) \,(\slashed{\partial}a)\, F_R + \left(\frac{v}{2}\,\epsilon' \bar{q}_R (M_q^{-1}\mathcal{A}\, \mathcal{B} M_F^{-1})\,(\slashed{\partial}a)\, F_R + \text{h.c.} \right),
\label{eq:AxDerivInt_appendix2}
\end{align}
in the form that has been used in the main body of this work.
\newpage
\subsection*{Tree-level contributions to \texorpdfstring{$s\rightarrow d + a$}{sda}}
The previous leading-order expansion only induces flavour-diagonal tree-level couplings between the axion and SM quarks. In this case the $s\rightarrow d + a$ process is only induced at loop level as we have discussed in the main text. However, in general we can not exclude the possibility that higher orders in the expansion of both $\epsilon$ and $\epsilon''$, which induce non-diagonal coupling structures, are relevant. These flavour-violating couplings would trigger the $s\rightarrow d + a$ decay already at tree-level. It therefore becomes important to analyze whether these tree-level couplings can in fact spoil our analysis or, at the very least, whether they restrict the range of values of our expansion parameters.

Before going into the details, let us briefly summarize how large the axial-vector coupling, irrespective of whether it is induced at tree- or loop-level, is allowed to be without being in conflict with the bound $\text{BR}(K^+\rightarrow\pi^+ + a) < 7.3\cdot 10^{-11}$~\cite{Adler:2008zza}. Generally, a coupling of the form
\begin{align}
\mathcal{H}_{s\rightarrow da} = \partial_\mu a\, \bar{d} h^S_{ds} \gamma^\mu (1+\gamma_5) s + \text{h.c.}
\end{align}
results in the decay width of Eq.~\eqref{eq:Kpia_width}
\begin{align}
\Gamma (K^+ \rightarrow \pi^+ a) = \frac{|h^S_{ds}|^2}{16 \pi m_{K^+}^3} ( m_{K^+}^2 -  m_{\pi^+}^2 )^2 \lambda^{1/2}(m_{K^+}^2,\,m_{\pi^+}^2,\,m_{a}^2)\, f^2_+(m_a^2)\,,
\end{align}
which leads to
\begin{align}
\label{eq:GeneralhSBound}
    |h^S_{ds}| \lesssim 1.5 \cdot 10^{-12}\,\frac{1}{\mathrm{GeV}}\;.
\end{align}

From this we can readily conclude that the $\epsilon$ expansion is safe: The parameter $\epsilon = v/\braket{\phi}$ is $\mathcal{O}(10^{-4})$ for $f_a \sim 10^{6}\,\mathrm{GeV}$, which is the region where NA62 is sensitive. The leading-order axial-vector coupling between quarks and the axion in Eq.~\eqref{eq:AxDerivInt_appendix2} therefore corresponds to $1/(2\braket{\phi}) \sim 1/(12 \cdot f_a) \sim 1/(10^{7}\,\mathrm{GeV})$. The NLO contribution to the axial-vector coupling between the axion and SM quarks would be suppressed by an additional factor of $\epsilon^2$ with respect to this leading-order coupling if we were to insert higher orders terms in $S$ in Eq.~\eqref{eq:SMatrixExpand}. This therefore means that the NLO coupling is parametrically suppressed by $10^{-7} \cdot 10^{-8}\,\mathrm{GeV}^{-1} = 10^{-15}\,\mathrm{GeV}^{-1}$, which is sufficiently far away from the E787 bound quoted in Eq.~\eqref{eq:GeneralhSBound} and is also out of reach for NA62.

Next, we want to quantify how small $\epsilon''$ has to be in order for significant flavour-violating axion couplings to be avoided at tree level.
By expanding $\xi^{-1}$ in $\epsilon''$ as in Eq.~\eqref{eq:xi_inverse}, we can write $S$ schematically as
\begin{equation}
    S=\begin{pmatrix}
    \textrm{unitary} + \mathcal{O}(\epsilon''^2) & \mathcal{O}(\epsilon'')\\ \mathcal{O}(\epsilon'') & \textrm{unitary} + \mathcal{O}(\epsilon''^2) 
    \end{pmatrix}\;,
\end{equation}
and estimate the parametric size of the coupling structure in Eq.~\eqref{eq:quark-axion-int-app} as
\begin{equation}
     S^\dagger\begin{pmatrix}
\mathds{1} & 0 \\ 0 & 0
\end{pmatrix}S = \begin{pmatrix}
\mathds{1} + \mathcal{O}(\epsilon''^2) & \mathcal{O}(\epsilon'')\\
\mathcal{O}(\epsilon'') & \mathcal{O}(\epsilon''^2)
\end{pmatrix} \; .
\end{equation}
Hence, we can put an upper bound on the tree-level flavour violating axion couplings to the SM quarks,
\begin{equation}
 |h^{S,\mathrm{tree level}}_{ds}| = \frac{1}{2\braket{\phi}}\left[ S^\dagger\begin{pmatrix}
\mathds{1} & 0 \\ 0 & 0
\end{pmatrix}S\right]_{ds} \lesssim \frac{\epsilon''^2}{2\Braket{\phi}}. 
\end{equation}
To avoid the constraint in~\eqref{eq:GeneralhSBound}, it is sufficient to require $\epsilon''\lesssim 10^{-3}$. By further restricting to $\epsilon''\lesssim 10^{-4}$, the tree-level effect becomes negligible compared to the loop-induced effect. This is exactly the setup which is considered in the main text.

Two more comments are in order. First, note that flavour-violating axion couplings at tree-level are not necessarily a problem but could also be considered an interesting feature of our model. However, since we initially set out to UV complete the model in \eqref{eq:new_model} and not to build an axion-flavour model, we choose to restrict ourselves to the case of large scale separations, where the tree-level flavour violation is negligible and flavour violation is induced only by loop processes.

Second, tree-level flavour violation can also be suppressed by the coupling matrices without requiring the masses of the additional fields to be much larger than the PQ scale. For instance, when $\epsilon'=0.2$, we can set $\beta = \lambda=\mathds{1}$ and $\alpha = Y\sqrt{\frac{1+0.2^2}{0.2^2}}$, with $Y$ being the SM Yukawa couplings. This results in vanishing flavour off-diagonal axion couplings to SM quarks in Eq.~\eqref{eq:AxQuarkIntNoExp_app} at all orders in $\epsilon''$, as the inversion of $\xi$ in~\eqref{eq:xi_inverse} is trivial in this case and thus $\mathcal{ABA^\dagger}$ becomes proportional to $M_q^2$, which is exactly diagonal, without expanding in $\epsilon''$:
\begin{align}
\label{eq:explicit_matrices}
    M_q^2&=\left[U_\delta^\dagger (\delta - \mu \xi^{-1}\mu^\dagger)U_\delta \; (\epsilon\epsilon')^2 + \mathcal{O}(\epsilon^3)\right] \frac{\braket{\Sigma}^2}{2}\\
    &=\left[U_\delta^\dagger (\alpha \alpha^\dagger - \frac{1}{1+\epsilon'^2} \alpha \alpha^\dagger )  U_\delta \; (\epsilon\epsilon')^2 + \mathcal{O}(\epsilon^3)\right] \frac{\braket{\Sigma}^2}{2}\\
    &= \frac{1}{1+\epsilon'^2}\mathcal{ABA^\dagger} \epsilon^2\epsilon'^4\frac{\braket{\Sigma}^2}{2} + \mathcal{O}(\epsilon^3).
\end{align}
Moreover, we also see that corrections to the relation in Eq.~\eqref{eq:appQMasses}, crucial to obtain the effective coefficient Eq.~\eqref{eq:hdsUVCoeffWqF}, are of the order $\epsilon''^2=\epsilon'^2=4\%$ such that our discussion of the $K^+ \rightarrow \pi^+ + a$ decay is still valid. Note however that $\alpha$ in this explicit realization is already very close to the perturbativity limit of Yukawa couplings.

\subsection*{Electroweak interactions}
Right-handed $q$- and $F$-quarks are in identical representations of the SM gauge group (considering up- and down-type separately). Hence, the interactions of this chiral components with gauge bosons are unchanged under our transformation. This is not the case for left-handed fields. It is useful to write the 6x6 matrix $U$ as a block matrix
\begin{equation}
	U= \begin{pmatrix}
	A & B \\C & D \\
	\end{pmatrix},
\end{equation}   
where each block is a 3x3 matrix.
Note, however, that unitarity of $U$ does not imply unitarity of any of the blocks.
Because the transformation above mixes different representations of $\text{SU(2)}\times\text{U(1)}_\text{Y}\text{/U(1)}_\text{EM}$, we have to check how the interactions with $W$ and $Z$ bosons are modified. 
Let us start with the $Z$ bosons,
\begin{align}
\begin{split}
	&\begin{pmatrix}\bar{q} \\ \bar{F}\end{pmatrix}^T \gamma_\mu Z^\mu \frac{-g}{\cos(\theta_W)} \left[
	\pm\frac{1}{2}\begin{pmatrix}\mathds{1} & 0 \\ 0 & 0 \\\end{pmatrix}P_L
	-Q \sin^2(\theta_W)   \begin{pmatrix}\mathds{1} & 0 \\ 0 & \mathds{1} \\\end{pmatrix}
	\right] \begin{pmatrix}
	q \\ F\\
	\end{pmatrix} \\
	\label{eq:Zint}
	\rightarrow &\begin{pmatrix}\bar{q} \\ \bar{F}\end{pmatrix}^T \gamma_\mu Z^\mu \frac{-g}{\cos(\theta_W)} \left[
	\pm\frac{1}{2}\begin{pmatrix}A^\dagger A & A^\dagger B \\ B^\dagger A & B^\dagger B \\\end{pmatrix}P_L
	-Q \sin^2(\theta_W) \begin{pmatrix}\mathds{1} & 0 \\ 0 & \mathds{1} \\\end{pmatrix}
	\right] \begin{pmatrix}
	q \\ F\\
	\end{pmatrix}\;.
	\end{split}
\end{align}
In this expression, the upper (lower) sign refers to up- (down-) type quarks and $Q$ is the electromagnetic charge.  We see that $Z$ can in principle couple to all available neutral currents, including ones involving light SM quarks of different flavour because $A$ does not have to be unitary. By identifying the blocks $A$ and $B$ in our perturbative result of $U$ in \eqref{eq:UMatrixMod1}, we find the $Z$-interactions at leading order in $\epsilon$ to be
	\begin{equation}
\label{eq:ZBosInt_appendix}
    \mathcal{L} \supset\begin{pmatrix}\bar{q} \\ \bar{F}\end{pmatrix}^T \gamma_\mu Z^\mu \frac{-g}{\cos(\theta_W)} \left[
	\pm\frac{1}{2}\begin{pmatrix} \mathds{1} & -\epsilon\epsilon'\mathcal{A} \\-\epsilon\epsilon'\mathcal{A}^\dagger & (\epsilon\epsilon')^2\mathcal{A}^\dagger \mathcal{A} \\\end{pmatrix}P_L
	-Q \sin^2(\theta_W)   \begin{pmatrix}\mathds{1} & 0 \\ 0 & \mathds{1} \\\end{pmatrix}
	\right] \begin{pmatrix}
	q \\ F\\
	\end{pmatrix}.
\end{equation}
Tree-level flavour-changing couplings to SM quarks only appear at order $\epsilon^2$.

In the next step, we look at the $W$ interactions. Because these mix up- and down-type quarks, we reintroduce the corresponding labels to write
\begin{align}
	& \frac{-g}{\sqrt{2}}\begin{pmatrix}\bar{u}_L \\ \bar{d}_L \\ \bar{F}^u_L \\ \bar{F}^d_L\end{pmatrix}^T \gamma^\mu
	\left[W^+_\mu \begin{pmatrix}
	0 &  \mathds{1}  & 0 & 0 \\
	0 & 0 & 0 & 0 \\
	0 & 0 & 0 & 0 \\
	0 & 0 & 0 & 0 \\
	\end{pmatrix}+W^-_\mu \begin{pmatrix}
	0 & 0 & 0 & 0 \\
	\mathds{1} & 0 & 0 & 0 \\
	0 & 0 & 0 & 0 \\
	0 & 0 & 0 & 0 \\
	\end{pmatrix}\right]
		\begin{pmatrix}
		u_L \\ d_L \\ F^u_L \\  F^d_L \\
		\end{pmatrix} \nonumber \\
		\label{eq:WInt}
	\rightarrow &  \frac{-g}{\sqrt{2}}\begin{pmatrix}\bar{u}_L \\ \bar{d}_L \\ \bar{F}^u_L \\ \bar{F}^d_L\end{pmatrix}^T  \gamma^\mu
	\left[W^+_\mu \begin{pmatrix}
	0 &  A^{u\dagger}A^d  & 0 &A^{u\dagger}B^d  \\
    0 & 0 & 0 & 0 \\
	0 &  B^{u\dagger}A^d  & 0 &   B^{u\dagger}B^d \\
	0 & 0 & 0 & 0 \\
	\end{pmatrix}+ W^-_\mu\begin{pmatrix}
	0 & 0 & 0 & 0 \\
	A^{d\dagger}A^u & 0 & A^{d\dagger}B^u & 0 \\
	0 & 0 & 0 & 0 \\
	B^{d\dagger}A^u & 0 &  B^{d\dagger}B^u & 0 \\
	\end{pmatrix} \right]
	\begin{pmatrix}
	u_L \\ d_L \\ F^u_L \\  F^d_L \\
	\end{pmatrix}\;.
\end{align}
The $W$ boson couples to all available charged currents. We identify the CKM matrix $V$ as 
\begin{equation}
   V= A^{u\dagger}A^d = U_\delta^{u\dagger}U_\delta^d + \mathcal{O}(\epsilon^2)\;,
\end{equation}
which unlike in the SM does not have to be unitary, but non-unitarity only appears at order $\epsilon^2$. When we again insert the perturbative results for $A$ and $B$, we arrive at
\begin{align}
\begin{split}
	\mathcal{L} \supset  \frac{-g}{\sqrt{2}} \begin{pmatrix}\bar{u}_L \\ \bar{d}_L \\ \bar{F}^u_L \\ \bar{F}^d_L\end{pmatrix}^T \gamma^\mu
	&\left[    W^+_\mu \begin{pmatrix}
	0 &  V  & 0 &-\epsilon\epsilon'V\mathcal{A}_d  \\
    0 & 0 & 0 & 0 \\
	0 &  -\epsilon \epsilon' \mathcal{A}_u^\dagger V\phantom{^\dagger}  & 0 & \phantom{^\dagger}(\epsilon \epsilon')^2 \mathcal{A}^\dagger_u V \mathcal{A}_d \\
	0 & 0 & 0 & 0 \\
	\end{pmatrix}\right. \\
 + & \phantom{\Bigg[}\left.  W^-_\mu\begin{pmatrix}
	0 & 0 & 0 & 0 \\
	V^\dagger & 0 & -\epsilon\epsilon' V^\dagger\mathcal{A}_u & 0 \\
	0 & 0 & 0 & 0 \\
	-\epsilon\epsilon'\mathcal{A}_d^\dagger V^\dagger & 0 &  (\epsilon \epsilon')^2 \mathcal{A}^\dagger_d V^\dagger \mathcal{A}_u & 0 \\
	\end{pmatrix} \right]
	\begin{pmatrix}
	u_L \\ d_L \\ F^u_L \\  F^d_L \\
	\end{pmatrix}\;.
	\label{eq:quark-W-coupling_appendix}
	\end{split}
\end{align}

\subsection*{Radial modes of $H$ and $\Phi$}

So far we have not included the radial modes of the Higgs field $H$ and of the PQ field $\Phi$ in our discussion. To capture the impact of the diagonalization procedure on their couplings, it is convenient to write
    \begin{equation}
\mathcal{L}\supset -\begin{pmatrix}
\bar{q}_{L} & \bar{F}_{L}
\end{pmatrix}
 M_\text{rad}
\begin{pmatrix}
q_{R} \\
F_{R}
\end{pmatrix} + \mathrm{h.c.},
\end{equation}
where
\begin{align}
     M_\text{rad} = 
    \begin{pmatrix}
    0 & \alpha \frac{h}{\sqrt{2}}\\
    \beta \frac{\phi}{\sqrt{2}} & 0
    \end{pmatrix}
    \;.
\end{align}
The unitary transformation then results in
\begin{align}
    M_\text{rad} \rightarrow U^\dagger M_\text{rad} S\;.
\end{align}
As the expressions are lengthy, we quote each 3x3 blocks separatel 
\begin{align}
\notag
    [U^\dagger M_\text{rad} S]_{qq} &= \frac{\Braket{\Sigma}}{2} \left(\Big(\epsilon^2 \epsilon'^3 \mathcal{A} \mathcal{B} \mathcal{A}^\dagger \phi+ \mathcal{O}(\epsilon^4)\Big) - \Big(\epsilon \epsilon' U_\delta^\dagger \alpha(\lambda^\dagger \xi^{-1} \mu^\dagger - \alpha^\dagger ) U_\delta\, h + \mathcal{O}(\epsilon^3)\Big)\right) M_q^{-1}\\ 
    &= \frac{\Braket{\Sigma}}{2} \left(\epsilon^2 \epsilon'^3 \mathcal{A} \mathcal{B} \mathcal{A}^\dagger \phi + \frac{2}{\epsilon \epsilon' \Braket{\Sigma}^2}  M_q^2 \,h \right) M_q^{-1}\\
    &\simeq \frac{M_q}{\braket{\phi}}\,\phi + \frac{M_q}{v}\, h\;,\\
    [U^\dagger M_\text{rad} S]_{qF} &= \frac{\Braket{\Sigma}}{2} \left(\Big(\epsilon \epsilon'^2 \mathcal{A} \mathcal{B} \phi + \mathcal{O}(\epsilon^3)\Big) - \Big(\mathcal{C}\,h + \mathcal{O}(\epsilon^2)\Big) \right) M_F^{-1}\;,\\
    \notag
    [U^\dagger M_\text{rad} S]_{Fq} &= \frac{\Braket{\Sigma}}{2} \left( \Big(\epsilon \epsilon'^2 \mathcal{B} \mathcal{A}^\dagger \phi + \mathcal{O}(\epsilon^3) \Big)+ \Big( \epsilon^2 \epsilon'^2 \mathcal{A}^\dagger U_\delta^\dagger \alpha(\lambda^\dagger \xi^{-1} \mu^\dagger - \alpha^\dagger ) U_\delta h+ \mathcal{O}(\epsilon^3) \Big) \right) M_q^{-1} \\
    &= \frac{\Braket{\Sigma}}{2} \left( \epsilon \epsilon'^2 \mathcal{B} \mathcal{A}^\dagger \phi  - \frac{2}{\Braket{\Sigma}^2} \mathcal{A}^\dagger M_q^2\,h  \right)M_q^{-1}\;,\label{eq:unsuppressed_coupling}\\
    [U^\dagger M_\text{rad} S]_{FF} &= \frac{\Braket{\Sigma}}{2} \left(\Big(\epsilon' \mathcal{B}\,\phi + \mathcal{O}(\epsilon^2)\Big) + \Big(\epsilon \epsilon' \mathcal{A}^\dagger \mathcal{C}\,h + \mathcal{O}(\epsilon^3)\Big)\right) M_F^{-1}\;,
\end{align}
where we have defined $\mathcal{C} = U_\delta^\dagger \alpha \lambda^\dagger U_\xi  \simeq 2\,\mathcal{A}\,M_F^2/\braket{\Sigma}^2$. As we can see, modifications of the Higgs coupling to SM quarks always appear together with additional factors of $\epsilon^2$ and are thus negligible. 
A contribution of $\phi$ to $s\to d+a$, on the other hand, must involve internal $F$-quarks and is therefore proportional to $\epsilon''^2$. Consequently, any contributions from $\phi$ loops, in particular the ones related to the unsuppressed coupling in Eq.~\eqref{eq:unsuppressed_coupling}, become negligible in the $\epsilon'' \ll 1$ limit (as long as this suppression also compensates for possible enhancements originating from the hierarchical coupling matrices). But even if $\epsilon''$ is not small, the $\phi$-mediated flavor violation can be suppressed given an explicit form for the $\alpha$ and $\beta$ coupling matrices. For instance, in the realisation proposed above Eq.~\eqref{eq:explicit_matrices}, the $\phi$-induced loop processes become either flavour-diagonal (because the amplitude has the same flavour structure as the mass matrix) or numerically negligible. This is also true for Higgs-induced loops.

\section{Counterterm contribution to the kaon decay width}

\label{app:Counterterms}

As mentioned in the main text, counterterm contributions are relevant for the computation of the $s\rightarrow d + a$ transition as a result of the renormalization of quark fields (see the right diagram in figure~\ref{fig:diagramssketch}). Most important in this context is the renormalization of SM quark fields, while contributions related to heavy $F$-fields are of higher order in $\epsilon''$. Note that in the latter case one also encounters divergent loop integrals, which are parametrically suppressed by $\mathcal{O}(\epsilon''^2)$. Cancelling all of these divergences (and, more generally, all one-loop divergences in our UV model) requires a more complete renormalization discussion. However, in the following, we focus on the leading order processes that are not suppressed by $\epsilon''^2$, and therefore avoid this complication.

\begin{figure}[t]
	\center
	\includegraphics{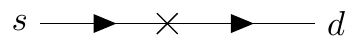} + \includegraphics{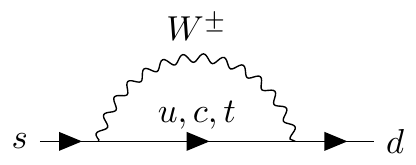} + ... = 0
	\caption{Relevant diagrams for determining the renormalization constants contributing to $s \rightarrow d$. The cross marks the insertion of a counterterm. The dots stand for higher order diagrams in our UV model, such as $Z$ boson-induced flavour changes or diagrams with heavy $F$ quarks.}
	\label{fig:RenCond}
\end{figure}

We start by parameterizing the renormalized down-type quark fields as~\cite{Bobeth:2001sq}
\begin{align}
\label{eq:RenDFields}
q^b_L= \begin{pmatrix}d\\s\\b\end{pmatrix}^{\hspace{-0.1cm}\text{bare}}_{\hspace{-0.1cm}L} \hspace{-0.3cm}= \left( 1 + \frac{1}{2}\delta Z^L\right) \underbrace{\begin{pmatrix}d\\s\\b\end{pmatrix}_{\hspace{-0.1cm}L}}_{q_L}\,,\hspace{1cm}q^b_R=\begin{pmatrix}d\\s\\b\end{pmatrix}^{\hspace{-0.1cm}\text{bare}}_{\hspace{-0.1cm}R} \hspace{-0.3cm}= \left( 1 + \frac{1}{2}\delta Z^R\right) \underbrace{\begin{pmatrix}d\\s\\b\end{pmatrix}_{\hspace{-0.1cm}R}}_{q_R}\,.
\end{align}
The renormalization constants of relevance for the $s\rightarrow d + a$ transition are determined by demanding that the one-loop $W^\pm$ contribution to $s \rightarrow d$ and the counterterm cancel each other~\cite{Hall:1981bc}, as depicted in figure~\ref{fig:RenCond}. The renormalization constants are therefore $\mathcal{O}(g^2)$ as a result of canceling the $W^\pm$ loop.

\begin{figure}[t]
	\center
	\includegraphics{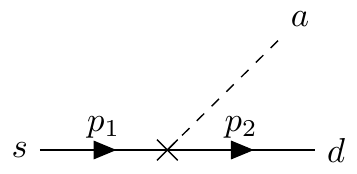} +\includegraphics{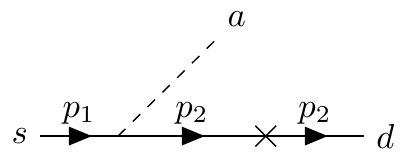} + \includegraphics{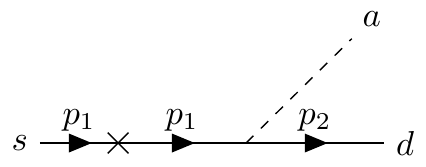} = 0\,,
	\caption{Sum of counterterm-induced diagrams for $s\rightarrow d + a$ adding up to zero.}    \label{fig:SumCTDiagr}
\end{figure}

\begin{figure}[!t]
	\center
	\notag
	\includegraphics{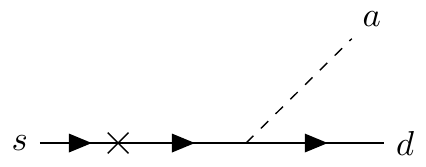} = --\includegraphics{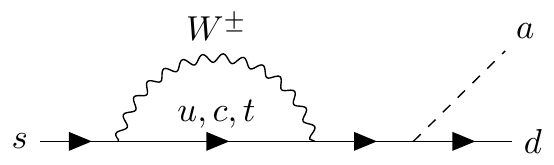}
	\\
	\includegraphics{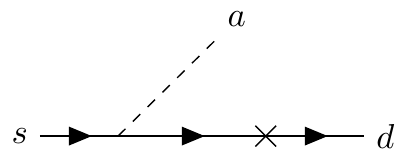} = --\includegraphics{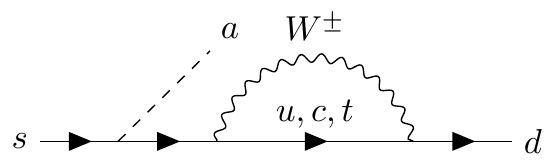}
	\caption{Relation between counterterm insertions and $W^\pm$ loops for $s\rightarrow d + a$.}\label{fig:RelCTWLoop}
\end{figure}

\begin{figure}[t]
	\center
	\notag
	\includegraphics{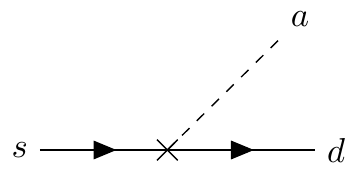} = \includegraphics{Figures/RenormalizationDiagrams/s2dAdWLoopNoLabel}+\includegraphics{Figures/RenormalizationDiagrams/sAs2dWLoopNoLabel}
	\caption{Relation between $s \rightarrow d + a$ counterterm contribution and ALP emission from external legs.}\label{fig:RelCTALPWLoop}
\end{figure}

Inserting the renormalized quark fields from Eq.~\eqref{eq:RenDFields} into the right-handed derivative axion-fermion interactions, we obtain the terms
\begin{align}
\mathcal{L} = \frac{1}{\braket{\phi}} \bar{q}_R \,(\slashed{\partial}a)\, q_R + \frac{1}{2\braket{\phi}} \bar{d}_R {\delta Z^R_{sd}}^* \,(\slashed{\partial}a)\,  s_R  + \frac{1}{2\braket{\phi}} \bar{d}_R \,(\slashed{\partial}a)\,{\delta Z^R_{ds}}s_R + \text{h.c.}\,,
\end{align}
where the quark fields now denote the renormalized ones. These new terms involving the renormalization constants therefore induce  $s\rightarrow d + a$ at tree-level. Note that this process is of order $g^2/\braket{\phi}$, just like the one-loop FCNCs in Eq.~\eqref{eq:hdsUVCoeffWqF}. Hence, there is no reason to neglect this counterterm contribution at this stage. As is known in literature, the explicit renormalization calculation does not need to be performed. For this, as will be proven below,
the diagrammatic equation in figure~\ref{fig:SumCTDiagr} can be shown to hold without knowing the precise expressions for the renormalization constants.
Moreover, from the renormalization condition shown in figure~\ref{fig:RenCond}, we can also conclude the two relations showcased in figure~\ref{fig:RelCTWLoop}\,.

Combining all the relations above, we end up with the key relation depicted in figure~\ref{fig:RelCTALPWLoop}, which allows us to exchange the counterterm calculation with the computation of additional self-energy diagrams where the ALP is emitted from the external down-type quark legs~\cite{Logan:2000iv}, i.e.~the diagrams on the right-hand side. From this point it is most straightforward to simply compute the additional loop diagrams instead of taking a detour to compute renormalization constants. Upon performing the calculations, one notices that the diagrams on the right-hand side of figure~\ref{fig:RelCTALPWLoop} add up to zero at linear order in the down- and strange-quark masses. Therefore, the counterterm contribution is subdominant and can safely be neglected for all practical purposes.

\subsection*{Proof of the relation in figure~\ref{fig:SumCTDiagr}}

\label{eq:ProofFigSumCTDiagr}
The derivative axion-quarks interaction can be generally written as
\begin{align}
 \mathcal{L} \supset \partial_\mu a\,\bar{q}_R^b\,G_Q \gamma^\mu q_R^b\,,
\end{align}
where $q^b = (d^b,\,s^b,\,b^b)^T$ denote the bare down-type quark fields and $G_Q$ is a general coupling matrix of diagonal structure, i.e. $G_Q = \text{diag}(g_d,\,g_s,\,g_b)$. Inserting now the renormalized quark fields from Eq.~\eqref{eq:RenDFields}, we obtain
\begin{align} \label{eq:AxCT}
\mathcal{L} &\supset \partial_\mu a\, \bar{q}_R \left( 1 + \frac{1}{2} \delta Z^{R\,\dagger} \right) G_Q \left( 1 + \frac{1}{2} \delta Z^R \right)\gamma^\mu  q_R\\
&= \partial_\mu a\, \bar{q}_R\, G_Q \gamma^\mu  q_R + \partial_\mu a\, \bar{q}_R\, \frac{1}{2} \delta Z^{R\,\dagger} \,G_Q \gamma^\mu  q_R +\partial_\mu a\, \bar{q}_R\, G_Q\,\frac{1}{2} \delta Z^R \, \gamma^\mu  q_R\,.
\end{align}
We also need to trace the influence on the kinetic and mass terms,
\begin{align}
\mathcal{L} \supset&\,i \bar{q}^b \slashed{\partial} q^b - \bar{q}^b M q^b\\
=& i \bar{q}^b_R \slashed{\partial} q^b_R + i \bar{q}^b_L \slashed{\partial} q^b_L - \bar{q}^b_R M q^b_L - \bar{q}^b_L M q^b_R\\
=&\,i \bar{q}_R \slashed{\partial} q_R + i \bar{q}_L \slashed{\partial} q_L - \bar{q}_R M q_L - \bar{q}_L M q_R + \mathcal{L}_C\,,
\end{align}
where $M=\text{diag}(m_d, m_s, m_b)$ and the counterterm Lagrangian $\mathcal{L}_C$ is given by 
\begin{align}
\begin{split}
\label{eq:FreeLagrCT}
\mathcal{L}_C =& \frac{i}{2} \bar{q}_R (\delta Z^{R})^{\dagger} \slashed{\partial} q_R + \frac{i}{2} \bar{q}_R \slashed{\partial}\,\delta Z^{R} q_R + \frac{i}{2} \bar{q}_L (\delta Z^{L})^{\dagger} \slashed{\partial} q_L + \frac{i}{2} \bar{q}_L  \slashed{\partial} \,\delta Z^{L} q_L \\
 &-\frac{1}{2} \bar{q}_R (\delta Z^{R})^{\dagger} M q_L -\frac{1}{2} \bar{q}_R M\,\delta Z^{L} q_L - \frac{1}{2} \bar{q}_L (\delta Z^{L})^{\dagger} M q_R - \frac{1}{2} \bar{q}_L M \delta Z^{R} q_R\,.
 \end{split}
\end{align}
For the left diagram in figure~\ref{fig:SumCTDiagr}, we use Eq.~\eqref{eq:AxCT} to obtain
\begin{align}
\label{eq:FirstDiagrCTSum}
i \mathcal{M}_1 = \frac{i (p_1 - p_2)_\mu}{2} \bar{u}_d(p_2) \left( i g_s\, \delta Z^{R*}_{sd} + i g_d\, \delta Z^{R}_{ds} \right)  \gamma^\mu  P_R u_s(p_1)\,,
\end{align}
where $Z^{R*}_{sd}$ is a specific matrix element from $\delta Z^{R\,\dagger} $ and $^*$ denotes complex conjugation.

For the diagram in the middle, we can use Eqs.~\eqref{eq:AxCT} and \eqref{eq:FreeLagrCT} to find
\allowdisplaybreaks
\begin{align}
i \mathcal{M}_2 =& \bar{u}_d(p_2) \frac{i}{2} \bigg\{ \left[ \left( \delta Z^{R*}_{sd} + \delta Z^R_{ds}\right) \slashed{p}_2  - \left(m_d \delta Z^R_{ds} + m_s \delta Z^{L*}_{sd} \right)\right] P_R \notag\\
&\qquad \qquad + \left[\left(\delta Z^{L*}_{sd} + \delta Z^L_{ds} \right)\slashed{p}_2 - \left( m_d \delta Z^L_{ds} + m_s \delta Z^{R*}_{sd} \right) \right] P_L \bigg\} \notag\\
&\times \frac{i (\slashed{p}_2 + m_s)}{m_d^2 - m_s^2}\, i (p_1 - p_2)_\mu i g_s \gamma^\mu  P_R u_s(p_1)\\
=& \frac{-i (p_1-p_2)_\mu}{2} \,i g_s \,\bar{u}_d(p_2) \left\{ \left[\delta Z^{R*}_{sd} m_d  - m_s \delta Z^{L*}_{sd}\right] P_R  + \left[\delta Z^{L*}_{sd} m_d - m_s \delta Z^{R*}_{sd} \right]P_L \right\}\notag\\ & \times\frac{\slashed{p}_2 +m_s}{m_d^2 - m_s^2} \gamma^\mu  P_R u_s(p_1)\,,
\end{align}
 \interdisplaylinepenalty=10000
where we have used the equation of motion $\bar{u}_d(p_2) \slashed{p}_2 = \bar{u}_d(p_2) m_d$. Using the equation of motion once more for $\slashed{p}_2$ in the propagator yields
\begin{align}
\notag
i \mathcal{M}_2 =& \frac{-i (p_1-p_2)_\mu}{2} \,i g_s\, \bar{u}_d(p_2) \bigg\{ \Big(\left[ \delta Z^{R*}_{sd} m_d - m_s \delta Z^{L*}_{sd}\right] P_R +\left[\delta Z^{L*}_{sd} m_d -m_s \delta Z^{R*}_{sd}\right] P_L \Big)\frac{m_s}{m_d^2 - m_s^2}\\
&+ \Big(\left[\delta Z^{R*}_{sd} m_d - m_s \delta Z^{L*}_{sd} \right]P_L +\left[\delta Z^{L*}_{sd} m_d -m_s \delta Z^{R*}_{sd} \right]P_R\Big)\frac{m_d}{m_d^2 - m_s^2}\bigg\}\gamma^\mu P_R u_s(p_1)\\
&= \frac{-i (p_1-p_2)_\mu}{2}\,i g_s\,\bar{u}_d(p_2) \Big\{ -m_s^2 \delta Z ^{L*}_{sd} P_R - m_s^2 \delta Z^{R*}_{sd} P_L\notag\\
&\hspace{3.5cm} + \delta Z^{R*}_{sd} m_d^2 P_L + \delta Z^{L*}_{sd} m_d^2 P_R \Big\}\frac{1}{m_d^2-m_s^2}\gamma^\mu \,P_R\,u_s(p_1)\\
&= \frac{-i (p_1-p_2)_\mu}{2}\,i g_s\,\bar{u}_d(p_2) \,\delta Z^{R*}_{sd}\,\gamma^\mu  P_R u_s(p_1)\,,
\end{align}
\vspace*{-0.1cm}
which cancels the $g_s$ term in Eq.~\eqref{eq:FirstDiagrCTSum}. Note that the terms involving $\delta Z^L$ cancel due to $P_R P_L = 0$. 

We can proceed in a similar fashion for the right diagram in figure~\ref{fig:SumCTDiagr},
\begin{align}
\notag
i \mathcal{M}_3 =& \bar{u}_d(p_2) i (p_1 - p_2)_\mu \,i g_d\,\gamma^\mu  P_R \frac{i (\slashed{p}_1 + m_d)}{m_s^2 - m_d^2}\frac{i}{2} \bigg\{ \left[ \left( \delta Z^{R*}_{sd} + \delta Z^R_{ds}\right) \slashed{p}_1- \left(m_d \delta Z^R_{ds} + m_s \delta Z^{L*}_{sd} \right)\right] P_R \\
&\quad  + \left[\left(\delta Z^{L*}_{sd} + \delta Z^L_{ds} \right)\slashed{p}_1 - \left( m_d \delta Z^L_{ds} + m_s \delta Z^{R*}_{sd} \right) \right] P_L \bigg\} u_s(p_1)\\
=&\frac{-i (p_1-p_2)_\mu}{2}\,i g_d\,\bar{u}_d(p_2) \gamma^\mu  P_R \frac{ \slashed{p}_1 + m_d}{m_s^2 - m_d^2} \notag \\    
&\quad\times\Big\{\delta Z^R_{ds} m_s P_L - m_d \delta Z^R_{ds} P_R + \delta Z^L_{ds} m_s P_R - m_d \delta Z^L_{ds} P_L\Big\} u_s(p_1)\\
\notag
=& \frac{-i (p_1-p_2)_\mu}{2}\,i g_d\,\bar{u}_d(p_2) \gamma^\mu  P_R  \bigg\{\frac{ m_d}{m_s^2 - m_d^2}\left(\delta Z^R_{ds} m_s P_L - m_d \delta Z^R_{ds} P_R + \delta Z^L_{ds} m_s P_R - m_d \delta Z^L_{ds} P_L \right) \notag\\
&\quad + \frac{ m_s}{m_s^2 - m_d^2}\left(\delta Z^R_{ds} m_s P_R - m_d \delta Z^R_{ds} P_L + \delta Z^L_{ds} m_s P_L - m_d \delta Z^L_{ds} P_R \right) \bigg\} u_s(p_1)\\
=& \frac{-i (p_1-p_2)_\mu}{2}\,i g_d\,\bar{u}_d(p_2) \gamma^\mu  \,\delta Z^R_{ds}\,P_R u_s(p_1)\,,
\end{align}
\vspace*{-0.1cm}
which cancels the $g_d$ term in Eq.~\eqref{eq:FirstDiagrCTSum}. We therefore conclude that the sum of all three diagrams vanishes,
\begin{align}
i \mathcal{M}_1 + i \mathcal{M}_2 +i \mathcal{M}_3 = 0\,,
\end{align}
\vspace*{-0.1cm}
independently of the precise expression of the renormalization constants.

\section{EDM calculations}\label{app:EDM}

\subsection*{General considerations}
\begin{figure}
\begin{minipage}{0.3\textwidth}
	\centering
	\includegraphics[width=1.0\linewidth]{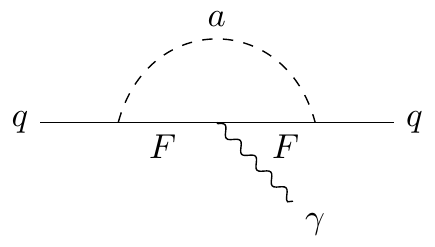}  \end{minipage}\hfill
\begin{minipage}{0.3\textwidth}
	\centering
	\vspace*{-0.15cm}
	\includegraphics[width=1.0\linewidth]{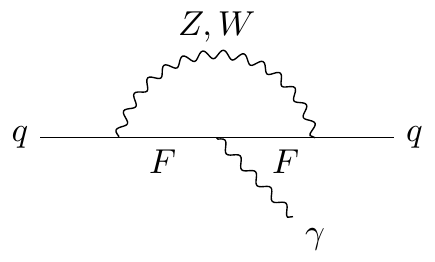}  \end{minipage}\hfill
\begin{minipage}{0.3\textwidth}
	\centering
	\vspace*{-1.2cm}
	\includegraphics[width=1.0\linewidth]{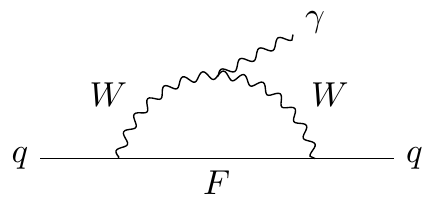}
	\end{minipage}
\caption{One-loop contributions to photon interactions with quarks in our UV model, which are of relevance for the EDM discussion.  }
	\label{fig:EDMdiagrams}	
\end{figure}

\begin{figure}
\hspace*{-0.3cm}
\begin{minipage}{0.3\textwidth}
	\centering
	\includegraphics[width=0.97\linewidth]{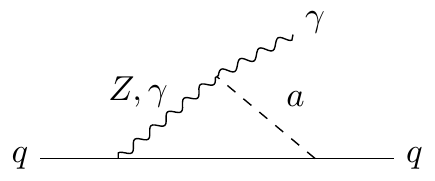}  \end{minipage}\hfill
\begin{minipage}{0.3\textwidth}
	\centering
	\vspace*{0.35cm}
	\includegraphics[width=1.1\linewidth]{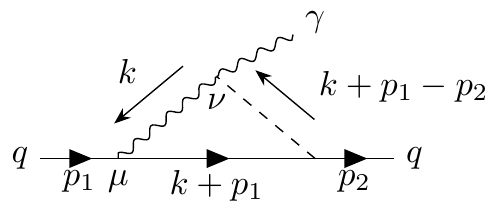}  \end{minipage}\hfill
\begin{minipage}{0.3\textwidth}
	\centering
	\vspace*{0.35cm}
	\includegraphics[width=1.1\linewidth]{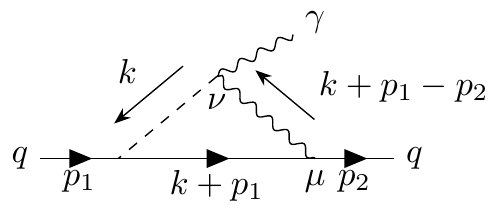}
	\end{minipage}
\hspace{0.2cm}
\caption{Barr-Zee diagrams relevant for the quark EDM computation in our UV model.}
	\label{fig:EDMBZdiagrams}	
\end{figure}

In this appendix, we assess whether the one-loop contributions shown in Figs.~\ref{fig:EDMdiagrams} and~\ref{fig:EDMBZdiagrams} can generate quark EDMs and hence also a neutron EDM. Before computing the relevant diagrams explicitly, it is worthwhile to remember the discussion of appendix~\ref{app:CPVInt}. Similar to the case of weak interactions, CP-violating interactions can be parameterized by assigning complex phases to the corresponding operators. Therefore, once a specific operator is inserted twice in a given diagram, i.e.~a specific vertex and its hermitian conjugate appear together, the CP phase necessarily drops out.

To be more specific, consider the diagram on the left in figure~\ref{fig:EDMdiagrams}. Let us first write the corresponding $a-q-F$ vertices generally as $(\partial_\mu a) \bar{q}_i \gamma^\mu\,X_{ij}\,P_R F_j + \text{h.c.}$ with an arbitrary $3\times 3$ flavour matrix $X$, and as usual use $q$ ($F$) to denote the up- or down-type quark triplets in the SM (F-quark) sector. Note that the right-handed coupling structure is fixed by our diagonalization procedure, see appendix~\ref{app:diagonalization}. This operator violates CP if $X$ has complex-valued entries. Note now that the external quarks are fixed in the Feynman diagram and hence the same $a-q-F$ vertex appears twice in the one-loop diagram because the photon only couples flavour-diagonally to all orders in $\epsilon$ and $\epsilon''$. If we rewrite the relevant entry $X_{ql}$ as $x_{ql} e^{i \theta_{ql}}$, where $l$ denotes the internal $F$-quark flavour and $\theta_{ql}$ denotes the CP violating phase, we get $e^{i \theta_{ql}} \cdot e^{-i \theta_{ql}} = 1 $ in the Feynman amplitude and hence the CP violating phase drops out. As a result, the diagram on the left in figure~\ref{fig:EDMdiagrams} cannot produce any net CP-violating quark EDM. This is also consistent with a description based on Jarlskog invariants, discussed in ref.~\cite{DiLuzio:2020oah}, upon taking the right-handed coupling structure into account.

An analogous argument can be applied to the remaining diagrams in figure~\ref{fig:EDMdiagrams}, and also to the ones where the photon is radiated off an external quark line in those diagrams. None of these diagrams can therefore induce an EDM at any order in $\epsilon$ and $\epsilon''$.
This is true even if further CP-violating coupling structures were to occur at higher orders, which could `allow' for an internal SM quark $q' \neq q$ to appear in the loops (if $q' = q$, no CP violating vertex is possible as discussed in appendix~\ref{app:CPVInt}).

Thus, only the Barr-Zee diagram on the left of figure~\ref{fig:EDMBZdiagrams} and its `conjugate' with the internal gauge boson and axion line interchanged remain.\footnote{Formally, by counting the coupling constant insertions, these diagrams are of the same order as other two-loop diagrams. We still consider them explicitly as only a single loop integration is involved.} For these diagrams the reasoning is not as straightforward since no vertex appears twice in the diagram. For an internal photon line, however, one notices that no CP violating vertex appears in the diagram. As the photon only couples flavour-diagonally, the internal quark line has to be the same quark $q$ as the external one. In this case, the axion-quark vertex cannot violate CP (see appendix~\ref{app:CPVInt}) and consequently this diagram is fully CP conserving.
This reasoning does not apply to the Barr-Zee diagram with an internal Z-boson and hence a contribution to the quark EDM $d_q$ given by  $- (d_q/2) F^{\mu\nu} \bar{q} \sigma_{\mu\nu} i \gamma_5 q$ can be expected. For the explicit computation below, we follow the momentum flow as shown in the middle and right panels of figure~\ref{fig:EDMBZdiagrams}.

    \subsection*{Z-induced Barr-Zee diagram}
    We now calculate the EDM contribution based on the Barr-Zee diagram with an internal $Z$ boson in figure~\ref{fig:EDMBZdiagrams}. Considering the order in $\epsilon$ and $\epsilon''$ to which we have expanded, to obtain a net CP violating contribution we have to focus on an internal $F$-quark for the fermionic line. We parameterize the flavour matrix structure of the $a-q-F$ and $Z-q-F$ vertices by $X_{ql} = \epsilon' v (M_q^{-1} \mathcal{A} \mathcal{B} M_F^{-1})_{ql}/2 $ and $Y_{ql} = \frac{g}{2 c_W} \epsilon \epsilon' \mathcal{A}_{ql}$, respectively, as dictated by Eqs.~\eqref{eq:AxDerivInt} and~\eqref{eq:ZBosInt}. For both diagrams and for each individual $F$-quark with index $l$, the amplitude reads
    \begin{align}
    \begin{split}
    \label{eq:BarrZeeZFAmpl}
    i \mathcal{M} \simeq &(\pm)\int \frac{\text{d}^4k}{(2\pi)^4} \Big[\bar{u}_q(p_2) \,i X_{ql} \,i (\slashed{k} + \slashed{p}_1 - \slashed{p}_2 )\, P_R\, \frac{i (\slashed{k} + \slashed{p}_1 + m_{F_l})}{(k+p_1)^2 - m_{F_l}^2}\,i \gamma_\mu Y^\dagger_{lq} P_L u_q(p_1)\Big]\\
    & \qquad \times \frac{-i\left(g^{\mu\nu} - \frac{k^\mu k^\nu}{m_Z^2}\right)}{k^2 - m_Z^2} \,\frac{e^2}{(4\pi)^2f_a}\,i \epsilon^{\beta\nu\lambda\alpha} (i k_\beta) (i (p_1-p_2)_\lambda) \epsilon^*_\alpha \,\frac{i}{(k+p_1-p_2)^2 -m_a^2}\\
    &(\pm) \int \frac{\text{d}^4k}{(2\pi)^4} \Big[\bar{u}_q(p_2)\, i \gamma_\mu Y_{ql} P_L\, \frac{i (\slashed{k} + \slashed{p}_1 + m_{F_l})}{(k+p_1)^2 - m_{F_l}^2}\,i X^\dagger_{lq} \,(-i\slashed{k}) P_R \,u_q(p_1)\Big] \\
     & \qquad\times \frac{-i\left(g^{\mu\nu} - \frac{(k+p_1-p_2)^\mu (k+p_1-p_2)^\nu}{m_Z^2}\right)}{(k+p_1-p_2)^2 - m_Z^2} \frac{i}{k^2 -m_a^2} \, \frac{e^2}{(4\pi)^2f_a}\,
     \\
     & \qquad\times i \epsilon^{\beta\nu\lambda\alpha} (-i) (k+p_1-p_2)_\beta (i (p_1-p_2)_\lambda) \epsilon^*_\alpha\;.
    \end{split}
    \end{align}
   Here, we have schematically used the Feynman rule due to $(e/4\pi)^2 (a/f_a)F^{\mu\nu} \tilde{Z}_{\mu\nu}$ resulting from the chiral rotation discussed in section~\ref{subsec:AnomTerms}. The global $\pm$ depends on whether $q$ is an up- or down-type quark, see Eq.~\eqref{eq:ZBosInt}, and $\epsilon^*_\alpha$ is the photon polarization vector. We can write the amplitude in a more compact form as
    \begin{align}
   \begin{split}
     i \mathcal{M} \simeq &(\pm)(-i)\frac{e^2}{(4\pi)^2f_a}\int \frac{\text{d}^4k}{(2\pi)^4}\,X_{ql}\,Y^\dagger_{lq}\,\epsilon^{\beta\nu\lambda\alpha}\,k_\beta\, (p_1-p_2)_\lambda\,\epsilon^*_\alpha\,\Big[\bar{u}_q(p_2) (\slashed{k} + \slashed{p}_1 - \slashed{p}_2 )\, P_R\\
     &\times \frac{m_{F_l}}{(k+p_1)^2 - m_{F_l}^2}\,\gamma_\mu\,u_q(p_1)\Big]\,\frac{g^{\mu\nu} - \frac{k^\mu k^\nu}{m_Z^2}}{k^2 - m_Z^2}\,\,\frac{1}{(k+p_1-p_2)^2 -m_a^2}\\
     &(\pm) (-i)\frac{e^2}{(4\pi)^2f_a}\int \frac{\text{d}^4k}{(2\pi)^4} \,Y_{ql}\,X^\dagger_{lq}\,\epsilon^{\beta\nu\lambda\alpha}\,(k+p_1-p_2)_\beta\,(p_1-p_2)_\lambda\,\epsilon^*_\alpha\,\Big[\bar{u}_q(p_2)\,\gamma_\mu P_L\\
     &\times \frac{m_{F_l}}{(k+p_1)^2 - m_{F_l}^2}\,\slashed{k}\,u_q(p_1)\Big]\,\frac{g^{\mu\nu} - \frac{(k+p_1-p_2)^\mu (k+p_1-p_2)^\nu}{m_Z^2}}{(k+p_1-p_2)^2 - m_Z^2} \frac{1}{k^2 -m_a^2}\,.
      \end{split}
    \end{align}
    We then use \texttt{Package-X}~\cite{Patel:2015tea} to compute the loop integrals and map out the relevant Lorentz structure for the quark EDM. Both of the Barr-Zee diagrams give the same loop functions $f(m_{F_l},\,m_a,\,m_Z,\,m_q) \equiv f(m_{F_l},\,m_q)$ for the EDM operator, differing only in their sign. The relevant amplitude structure therefore reads
    \begin{align}
        i\mathcal{M} \simeq (-i)\frac{e^2}{(4\pi)^2f_a}(X_{ql}\,Y^\dagger_{lq} - Y_{ql}\,X^\dagger_{lq})\, \frac{i}{16\pi^2}\,m_{F_l}\,f(m_{F_l},\,m_q)\, \bar{u}_q(p_2)\,\frac{i\sigma^{\alpha\beta}\gamma_5}{2m_q}\,\epsilon^*_\alpha\,(p_1 - p_2)_\beta\,u_q(p_1).
    \end{align}
    Inserting the LO expressions for $X$ and $Y$, we get
    \begin{align}
    \begin{split}
        i\mathcal{M} \simeq & (-i)\frac{e^2}{(4\pi)^2 f_a}\,\epsilon\epsilon'^2\, \frac{v}{2}\,\frac{g}{2 c_W}\,\frac{1}{m_q}\,((\mathcal{A} \mathcal{B})_{ql}\,\mathcal{A}^\dagger_{lq} - \mathcal{A}_{ql}\,(\mathcal{B} \mathcal{A}^\dagger)_{lq})\\
        &\qquad\times\frac{i}{16\pi^2}\,f(m_{F_l},\,m_q)\, \bar{u}_q(p_2)\,\frac{i\sigma^{\alpha\beta}\gamma_5}{2m_q}\,\epsilon^*_\alpha\,(p_1 - p_2)_\beta\,u_q(p_1).
        \end{split}
    \end{align}
    In the case considered in the main text, where all $F$ quarks are assumed to be equally heavy, performing the sum over all internal $F$-quarks, (i.e.~over the index $l$) leads to a vanishing amplitude. Higher orders in the expansion of the relevant $Z-q-F$ and $a-q-F$ vertices can in principle again lead to a non-zero result, but one that would be suppressed by higher powers of $\epsilon$ and would therefore be negligible.
    
    As the Z-induced Barr-Zee diagram is formally of two-loop order and to motivate that the estimate for the upper limit of the neutron EDM in Eq.~\eqref{eq:NEDMestimate} is reasonable, we can, for the sake of an estimate, assume that all $F$ quarks have different mass and hence no cancellation takes place. The loop function $f(m_{F_l},\,m_q)$ is UV divergent and hence exhibits a scale dependence $\mu$. This dependence would of course disappear if further contributions were taken into account. Expanding in inverse powers of $m_{F_l}$, we find
    \begin{align}
    f(m_{F_l},\,m_q) \approx \frac{ i m_q}{4} \left(-3 + 2 \log\left(\frac{\mu^2}{m_{F_l}^2}\right)\right) +...\;.
    \end{align}
    Without performing a full two-loop analysis, we conservatively assume the term in the brackets to be an $\mathcal{O}(1)$ factor. The remaining finite terms in $f(m_{F_l})$, which are not expected to fully cancel, would actually be significantly smaller for $m_F \geq 1\,\mathrm{TeV}$. Moreover, if we parameterize the entries of $(\mathcal{AB})_{ql}$ and $\mathcal{A}_{ql}$ by $x_{ql} \,e^{i\theta_{ql}} $ and $a_{ql}\,e^{i \phi_{ql}}$, respectively, we obtain
    \begin{align}
    \label{eq:BarrZeeZCalc}
    \begin{split}
         i\mathcal{M} \simeq & (-i)\frac{e^2}{(4\pi)^2 f_a}\,\epsilon\epsilon'^2\, \frac{v}{2}\,\frac{g}{2 c_W}\,\frac{1}{m_q}\,\left(x_{ql} \,e^{i\theta_{ql}}\,a_{ql}\,e^{-i \phi_{ql}} - a_{ql}\,e^{i \phi_{ql}}\,x_{ql} \,e^{-i\theta_{ql}} \right)\\
         &\qquad\times \frac{i}{16\pi^2}\,\frac{i m_q}{4} \bar{u}_q(p_2)\,\frac{i\sigma^{\alpha\beta}\gamma_5}{2m_q}\,\epsilon^*_\alpha\,(p_1 - p_2)_\beta\,u_q(p_1)\;.
    \end{split}
    \end{align}
    The EDM operator to map onto is $-(d_q/2)\,F^{\mu\nu} \bar{q} \sigma_{\mu\nu} i \gamma_5 q$, where $d_q$ has to be real-valued for the operator to be hermitian. By comparing with Eq.~\eqref{eq:BarrZeeZCalc} and including the contributions from all three $F$ quarks, we have
    \begin{align}
        d_q & \sim 3\cdot\frac{1}{128\pi^2}\,\frac{e^2}{(4\pi)^2 f_a}\,\epsilon\epsilon'^2 \,\frac{v}{2}\, \frac{g}{2 c_W}\,\frac{1}{m_q}\,x_{ql}\,a_{ql}\,2 \sin(\theta_{ql}-\phi_{ql})\\
        &\sim 3\cdot\frac{1}{128\pi^2}\,\frac{\alpha}{4\pi f_a} \,\epsilon\epsilon'^2 \,\frac{v}{2}\, \frac{e}{2 s_W c_W}\,\frac{1}{m_q}\,\frac{2\,m_q^2}{\epsilon^2 \epsilon'^4 \braket{\Sigma}^2}\,2 \sin(\theta_{ql}-\phi_{ql})\\
        &\sim 3\cdot\frac{1}{128\pi^2}\,\,\frac{\alpha}{4\pi f_a}\,\frac{1}{6 f_a} \,\frac{e}{2 s_W c_W}\,m_q\,2 \sin(\theta_{ql}-\phi_{ql})\;,
    \end{align}
    where we have identified $x_{ql} a_{ql} \sim 2\,m_q^2/(\epsilon^2 \epsilon'^4 \braket{\Sigma}^2)$
    in the second line based on Eq.~\eqref{eq:appQMasses}. For the up- and down-quarks respectively, fixing $f_a = 4 \cdot 10^6\,\mathrm{GeV}$ and assuming $\sin(\theta_{ql}-\phi_{ql}) \approx 1$, we get
    \begin{align}
    \label{eq:dEDMestimate}
        d_d \sim 3 \cdot 10^{-36}\,e\cdot\mathrm{cm}\,,\qquad  d_u \sim 1 \cdot 10^{-36}\,e\cdot\mathrm{cm}\,.
    \end{align}
    This result is in agreement with the naive estimate for the upper limit on the neutron EDM in Eq.~\eqref{eq:NEDMestimate} for $d_n^\text{\,UV} \sim (4/3) d_d - (1/3) d_u$~\cite{Dar:2000tn}. Of course, it should be kept in mind that the result in Eq.~\eqref{eq:dEDMestimate} is only an estimate as well. However, given that the sensitivity of current neutron EDM experiments is away by several orders of magnitude, it is an estimate that is sufficient for our purposes.
\bibliographystyle{utphys}
\bibliography{references} 

\end{document}